\documentclass[a4paper]{article}
\usepackage{amssymb}
\usepackage{amsfonts}
\usepackage{amsmath}
\usepackage[margin=2cm]{geometry}
\usepackage{graphicx}
\usepackage{caption}
\usepackage{subfig}
\usepackage{multirow}
\usepackage{multicol}
\usepackage{makecell}
\usepackage{hhline}
\usepackage[T1]{fontenc}

\allowdisplaybreaks

\input{tcilatex}
\begin{document}

\title{Fractional Burgers wave equation on a finite domain}
\author{Sla\dj an Jeli\'{c}\thanks{
Faculty of Technical Sciences, University of Novi Sad, Trg D. Obradovi\'{c}a
6, 21000 Novi Sad, Serbia, df.sladjan@uns.ac.rs}, Du\v{s}an Zorica\thanks{
Department of Physics, Faculty of Sciences, University of Novi Sad, Trg D.
Obradovi\'{c}a 4, 21000 Novi Sad, Serbia and Mathematical Institute, Serbian
Academy of Arts and Sciences, Kneza Mihaila 36, 11000 Belgrade, Serbia,
dusan.zorica@df.uns.ac.rs}}
\maketitle

\begin{abstract}
\noindent Dynamic response of the one-dimensional viscoelastic rod of finite
length, that has one end fixed and the other subject to prescribed either
displacement or stress, is analyzed by the analytical means of Laplace
transform, yielding the displacement and stress of an arbitrary rod's point
as a convolution of the boundary forcing and solution kernel.
Thermodynamically consistent Burgers models are adopted as the constitutive
equations describing mechanical properties of the rod. Short-time
asymptotics implies the finite wave propagation speed in the case of the
second class models, contrary to the case of the first class models.
Moreover, Burgers model of the first class yield quite classical shapes of
displacement and stress time profiles resulting from the boundary forcing
assumed as the Heaviside function, while model of the second class yield
responses that resemble to the sequence of excitation and relaxation
processes.

\end{abstract}

\section{Introduction}

The fractional Burgers wave equation is considered in \cite{OZO} for the
Cauchy initial value problem on the unbounded domain, and here the aim is to
solve and analyze the initial-boundary value problem in space $x\in \left[
0,L\right] $ during time $t>0$, i.e., to consider the wave propagation in a
viscoelastic rod of finite length $L$ fixed at one of its ends and free on
the other, that has either prescribed displacement $u,$ or it is subject to
a given stress $\sigma .$ The particular interest is the behavior of
displacement and stress, obtained as a response to the boundary conditions
assumed as the Heaviside step function, since the stress for prescribed
displacement of rod's free end correspond to the relaxation modulus, while
displacement for prescribed stress acting on rod's free end correspond to
the creep compliance, that are studied in \cite{OZ-2} for the
thermodynamically consistent Burgers models. Writing the constitutive
equation of viscoelastic body in terms of relaxation modulus found
application in proving the dissipativity properties of the hereditary
fractional wave equations using a priori energy estimates in \cite{ZO}.
Note, the relaxation modulus represents the time-evolution of stress,
obtained from the constitutive equation for strain prescribed as the step
function, while the creep compliance represents the time-evolution of
strain, obtained from the constitutive equation for stress prescribed as the
step function.

Therefore, in order to model the wave propagation in one-dimensional
deformable viscoelastic body, the equation of motion and strain $\varepsilon 
$ for small local deformations 
\begin{equation}
\frac{\partial }{\partial x}\sigma (x,t)=\rho \,\frac{\partial ^{2}}{%
\partial t^{2}}u(x,t)\;\;\text{and}\;\;\varepsilon (x,t)=\frac{\partial }{%
\partial x}u(x,t)  \label{eq-motion}
\end{equation}%
are coupled with the thermodynamically consistent fractional Burgers model
either of the first class%
\begin{equation}
\left( 1+a_{1}\,{}_{0}\mathrm{D}_{t}^{\alpha }+a_{2}\,{}_{0}\mathrm{D}%
_{t}^{\beta }+a_{3}\,{}_{0}\mathrm{D}_{t}^{\gamma }\right) \sigma \left(
x,t\right) =\left( b_{1}\,{}_{0}\mathrm{D}_{t}^{\mu }+b_{2}\,{}_{0}\mathrm{D}%
_{t}^{\mu +\eta }\right) \varepsilon \left( x,t\right) ,  \label{UCE-1-5}
\end{equation}%
or of the second class 
\begin{equation}
\left( 1+a_{1}\,{}_{0}\mathrm{D}_{t}^{\alpha }+a_{2}\,{}_{0}\mathrm{D}%
_{t}^{\beta }+a_{3}\,{}_{0}\mathrm{D}_{t}^{\beta +\eta }\right) \sigma
\left( x,t\right) =\left( b_{1}\,{}_{0}\mathrm{D}_{t}^{\beta }+b_{2}\,{}_{0}%
\mathrm{D}_{t}^{\beta +\eta }\right) \varepsilon \left( x,t\right) ,
\label{UCE-6-8}
\end{equation}%
where ${}_{0}\mathrm{D}_{t}^{\xi }$ denotes the operator of
Riemann-Liouville fractional differentiation of order $\xi \in \left[ n,n+1%
\right] ,$ $n\in 
\mathbb{N}
_{0},$ defined by%
\begin{equation*}
{}_{0}\mathrm{D}_{t}^{\xi }y\left( t\right) =\frac{\mathrm{d}^{n+1}}{\mathrm{%
d}t^{n+1}}\left( \frac{t^{-\left( \xi -n\right) }}{\Gamma \left( 1-\left(
\xi -n\right) \right) }\ast y\left( t\right) \right) ,\;\;t>0
\end{equation*}%
through the convolution in time: $f\left( t\right) \ast _{t}g\left( t\right)
=\int_{0}^{t}f\left( t^{\prime }\right) g\left( t-t^{\prime }\right) \mathrm{%
d}t^{\prime },$ $t>0,$ see \cite{TAFDE}.

Fractional Burgers wave equation, represented by the system of equations (%
\ref{eq-motion}) and either (\ref{UCE-1-5}) or (\ref{UCE-6-8}), is subject
to zero initial conditions%
\begin{equation}
u(x,0)=0,\;\;\frac{\partial }{\partial t}u(x,0)=0,\;\;\sigma (x,0)=0,\;\;%
\frac{\partial }{\partial t}\sigma (x,0)=0,\;\;\varepsilon (x,0)=0,\;\;\frac{%
\partial }{\partial t}\varepsilon (x,0)=0,\;\;x\in \left[ 0,L\right] ,
\label{ic}
\end{equation}%
as well as to the boundary conditions 
\begin{equation}
u(0,t)=0\;\;\text{and either}\;\;u(L,t)=\Upsilon (t),\;\;\text{or}\;\;\sigma
(L,t)=\Sigma (t),\;\;t>0,  \label{bc}
\end{equation}%
corresponding to a rod fixed at one end and forced on the other. Wave
propagation in a rod of finite length, i.e., the initial-boundary value
problem (\ref{eq-motion}), subject to initial and boundary conditions (\ref%
{ic}) and (\ref{bc}), is considered in \cite{APZ-4,APZ-3} for the case of
viscoelastic material modeled by the fractional distributed-order equation
with power type constitutive function, while in \cite{AKOZ} a fluid-like
model of viscoelastic body is employed.

Thermodynamical consistency analysis of the fractional Burgers model%
\begin{equation}
\left( 1+a_{1}\,{}_{0}\mathrm{D}_{t}^{\alpha }+a_{2}\,{}_{0}\mathrm{D}%
_{t}^{\beta }+a_{3}\,{}_{0}\mathrm{D}_{t}^{\gamma }\right) \sigma \left(
x,t\right) =\left( b_{1}\,{}_{0}\mathrm{D}_{t}^{\mu }+b_{2}\,{}_{0}\mathrm{D}%
_{t}^{\nu }\right) \varepsilon \left( x,t\right) ,  \label{fbm}
\end{equation}%
containing model parameters: $a_{1},a_{2},a_{3},b_{1},b_{2}>0,$ $\alpha
,\beta ,\mu \in \left[ 0,1\right] ,$ with $\alpha \leq \beta ,$ and $\gamma
,\nu \in \left[ 1,2\right] $, performed in \cite{OZ-1}, implied two classes
of thermodynamically consistent models, represented by (\ref{UCE-1-5}) and (%
\ref{UCE-6-8}). In the case of models belonging to the first class, the
highest differentiation order of strain $\mu +\eta \in \left[ 1,2\right] ,$
with $\eta \in \left\{ \alpha ,\beta \right\} ,$ is greater than the highest
differentiation order of stress, that is either $\gamma \in \left[ 0,1\right]
$ in the case of Model I, in addition to $0\leq \alpha \leq \beta \leq
\gamma \leq \mu \leq 1$ and $\eta \in \left\{ \alpha ,\beta ,\gamma \right\}
,$ or $\gamma \in \left[ 1,2\right] $ in the case of Models II - V, in
addition to $0\leq \alpha \leq \beta \leq \mu \leq 1$ and $\left( \eta
,\gamma \right) \in \left\{ \left( \alpha ,2\alpha \right) ,\left( \alpha
,\alpha +\beta \right) ,\left( \beta ,\alpha +\beta \right) ,\left( \beta
,2\beta \right) \right\} ,$ while for models belonging to the second class
differentiation orders of stress $\beta \in \left[ 0,1\right] $ and $\beta
+\eta \in \left[ 1,2\right] $ coincide with the highest differentiation
orders of strain in addition to $0\leq \alpha \leq \beta \leq 1,$ so that $%
\eta =\alpha ,$ in the case of Model VI; $\eta =\beta $ in the case of Model
VII; and $\alpha =\eta =\beta ,$ $\bar{a}_{1}=a_{1}+a_{2},$ and $\bar{a}%
_{2}=a_{3}$ in the case of Model VIII. Similar forms of the fractional
Burgers models are checked for the thermodynamical consistency in \cite%
{AJP,BazhlekovaTsocheva}, while the classical and different variants of
fractional Burgers models, describing the flow of viscoelastic fluids in
various geometries, are considered in \cite%
{HyderAli,HyderAli-1,HyderAliQi,JamilFetecau,KangLiuXia,KhanAnjumFetecauQi,KhanHyderAliQi,KhanHyderAliQi-1}%
.

In \cite{Heymans,Kim,ZhouYanMasudaKuriyagawa}, the classical Burgers model
is used for description of polymer dynamics, viscoelastic material behavior
of asphalts, and molding of glass, while in \cite%
{AbbasMasadPapagiannakisHarman,LiuDaiYou} the micromechanical approach is
adopted for asphalt mixtures modeling. Fractional version of the Burgers
constitutive equation is used in \cite{XuJiang} for modeling polymers, while 
\cite{CelauroFecarottiPirrottaCollop,OeserPellinenScarpasKasbergen,Zbiciak}
found optimal model parameters in fractional Burgers model by using data
from creep and creep-recovery experiments, performed on asphalt concrete
mixtures. Experimental data from creep and stress relaxation of biological
tissues is also used in \cite{DemirciTonuk,grah}. On the other hand,
theoretical investigation of the creep compliance and relaxation modulus,
corresponding to fractional viscoelastic models having differentiation
orders below the first order, is presented in \cite%
{BazhlekovaBazhlekov,Mai-10,MainardiSpada}, where it is found that creep
compliance is a Bernstein function and relaxation modulus is a completely
monotonic function, while in \cite{OZ-2} thermodynamically consistent
Burgers models (\ref{UCE-1-5}) and (\ref{UCE-6-8}) proved to have the same
properties of creep compliance and relaxation modulus if the thermodynamical
requirements are narrowed. The creep compliances corresponding to the
classical models of viscoelasticity are reviewed in \cite{Makris}.

The damped oscillations and wave propagation problems on bounded and
semi-bounded domain are considered in \cite{R-S1,R-S-2001,R-S,R-S2,R-S-2008}
by modeling viscoelastic materials using Zener, modified Zener, and modified
Maxwell constitutive equations. The question of wave propagation speed,
asymptotics of solution near the wavefront, wave dispersion and attenuation
properties, accounted for by the fractional wave equation, are considered in 
\cite{Hanyga2013,Han7,Han8,Han6,Hanyga2019}, while in \cite%
{ColombaroGiustiMainardi2,ColombaroGiustiMainardi1} Buchen-Mainardi
wavefront solution expansion, introduced in \cite{BuchenMainardi}, is used
in examining fractional wave equations in media modeled by the Bessel,
integer and fractional order Maxwell and Kelvin-Voigt constitutive
equations. More on the Bessel model can be found in \cite%
{ColombaroGiustiMainardi,GiustiMainardi}. Wave propagation for a class of
thermodynamically consistent fractional models of viscoelastic body is
accounted for in \cite{KOZ19}. Solution's peak propagation speed is
considered in \cite%
{LuchkoMainardi-1,LuchkoMainardi-2,LuchkoMainardiPovstenko} as the wave
propagation speed. Fractional wave equations found their applications in
modeling seismic wave propagation, see \cite{Wu}, as well as in the
acoustics of complex media, see \cite{Cai2018}. The overview of fractional
order models of viscoelastic materials, wave propagation problems including
dispersion and attenuation processes are found in \cite%
{APSZ-1,APSZ-2,Holm-book,Mai-10,R-S-2010}.

\section{Solution of fractional Burgers wave equation}

The system of governing equations (\ref{eq-motion}) with either (\ref%
{UCE-1-5}) or (\ref{UCE-6-8}), representing the fractional Burgers wave
equation, subject to initial and boundary conditions (\ref{ic}) and (\ref{bc}%
), transforms into 
\begin{gather}
\frac{\partial }{\partial x}\sigma (x,t)=\frac{\partial ^{2}}{\partial t^{2}}%
u(x,t),\;\;\varepsilon (x,t)=\frac{\partial }{\partial x}u(x,t),\;\;\text{%
with either}  \label{btj-0} \\
\left( 1+a_{1}\,{}_{0}\mathrm{D}_{t}^{\alpha }+a_{2}\,{}_{0}\mathrm{D}%
_{t}^{\beta }+a_{3}\,{}_{0}\mathrm{D}_{t}^{\gamma }\right) \sigma \left(
x,t\right) =\left( {}_{0}\mathrm{D}_{t}^{\mu }+b\,{}_{0}\mathrm{D}_{t}^{\mu
+\eta }\right) \varepsilon \left( x,t\right) ,\;\;\text{or}  \label{btj-1} \\
\left( 1+a_{1}\,{}_{0}\mathrm{D}_{t}^{\alpha }+a_{2}\,{}_{0}\mathrm{D}%
_{t}^{\beta }+a_{3}\,{}_{0}\mathrm{D}_{t}^{\beta +\eta }\right) \sigma
\left( x,t\right) =\left( {}_{0}\mathrm{D}_{t}^{\beta }+b\,{}_{0}\mathrm{D}%
_{t}^{\beta +\eta }\right) \varepsilon \left( x,t\right) ,  \label{btj-2}
\end{gather}%
subject to 
\begin{gather}
u(x,0)=0,\;\;\frac{\partial }{\partial t}u(x,0)=0,\;\;\sigma (x,0)=0,\;\;%
\frac{\partial }{\partial t}\sigma (x,0)=0,\;\;\varepsilon (x,0)=0,\;\;\frac{%
\partial }{\partial t}\varepsilon (x,0)=0,\;\;x\in \left[ 0,1\right] ,
\label{ic-bd} \\
u(0,t)=0\;\;\text{and either}\;\;u(1,t)=\Upsilon (t),\;\;\text{or}\;\;\sigma
(1,t)=\Sigma (t),\;\;t>0,  \label{bc-bd}
\end{gather}%
after introducing dimensionless quantities%
\begin{gather*}
\bar{x}=\frac{x}{L},\;\;\bar{t}=\frac{t}{T},\;\;\bar{u}=\frac{u}{L},\;\;\bar{%
\Upsilon}=\frac{\Upsilon }{L},\;\;\bar{\sigma}=\sigma \frac{T^{\xi }}{b_{1}}%
,\;\;\bar{\Sigma}=\Sigma \frac{T^{\xi }}{b_{1}},\;\;\bar{\varepsilon}%
=\varepsilon , \\
\bar{a}_{1}=\frac{a_{1}}{T^{\alpha }},\;\;\bar{a}_{2}=\frac{a_{2}}{T^{\beta }%
},\;\;\bar{a}_{3}=\frac{a_{3}}{T^{\zeta }},\;\;\bar{b}=\frac{b_{2}}{%
b_{1}T^{\eta }},\;\;\text{with}\;\;T=\left( \frac{\rho L^{2}}{b_{1}}\right)
^{\frac{1}{2-\xi }},
\end{gather*}%
where $\left( \xi ,\zeta \right) =\left( \mu ,\gamma \right) $ for the first
and $\left( \xi ,\zeta \right) =\left( \beta ,\beta +\eta \right) ,$ $\eta
\in \left\{ \alpha ,\beta ,\gamma \right\} ,$ for the second class of
Burgers models and after omitting bars over dimensionless quantities. By
applying the Laplace transform, defined by%
\begin{equation*}
\tilde{f}\left( s\right) =\tciLaplace \lbrack f\left( t\right) ]\left(
s\right) =\int_{0}^{\infty }f\left( t\right) \mathrm{e}^{-st}\mathrm{d}t,
\end{equation*}%
and by taking into account zero initial conditions (\ref{ic-bd}), the
governing equations, i.e., either (\ref{btj-0}), (\ref{btj-1}), or (\ref%
{btj-0}), (\ref{btj-2}), become%
\begin{equation}
\frac{\partial }{\partial x}\tilde{\sigma}\left( x,s\right) =s^{2}\tilde{u}%
\left( x,s\right) ,\;\;\tilde{\varepsilon}\left( x,s\right) =\frac{\partial 
}{\partial x}\tilde{u}\left( x,s\right) ,\;\;\tilde{\sigma}\left( x,s\right)
=\tilde{G}\left( s\right) \tilde{\varepsilon}\left( x,s\right) ,
\label{btj-lt-1}
\end{equation}%
where the complex modulus is 
\begin{align}
& \tilde{G}\left( s\right) =\frac{\phi _{\varepsilon }\left( s\right) }{\phi
_{\sigma }\left( s\right) },\;\;\text{with either}  \label{g-tilda} \\
& \phi _{\sigma }\left( s\right) =1+a_{1}s^{\alpha }+a_{2}s^{\beta
}+a_{3}s^{\gamma },\;\;\phi _{\varepsilon }\left( s\right) =s^{\mu }+bs^{\mu
+\eta },\;\;\text{or}  \label{fi-prva-klasa} \\
& \phi _{\sigma }\left( s\right) =1+a_{1}s^{\alpha }+a_{2}s^{\beta
}+a_{3}s^{\beta +\eta },\;\;\phi _{\varepsilon }\left( s\right) =s^{\beta
}+bs^{\beta +\eta },  \label{fi-druga-klasa}
\end{align}%
for the first, respectively second class of Burgers models, so that system
of equations (\ref{btj-lt-1}) solved with respect to $\tilde{u}$ reduce to
the ordinary differential equation with constant coefficients%
\begin{equation*}
\frac{\partial ^{2}}{\partial x^{2}}\tilde{u}\left( x,s\right) -\frac{s^{2}}{%
\tilde{G}\left( s\right) }\tilde{u}\left( x,s\right) =0,
\end{equation*}%
whose solution is%
\begin{eqnarray}
\tilde{u}\left( x,s\right) &=&C_{1}\left( s\right) \mathrm{e}^{\frac{xs}{%
\sqrt{\tilde{G}\left( s\right) }}}+C_{2}\left( s\right) \mathrm{e}^{-\frac{xs%
}{\sqrt{\tilde{G}\left( s\right) }}},\;\;\text{i.e.,}  \notag \\
\tilde{u}\left( x,s\right) &=&C\left( s\right) \sinh \frac{xs}{\sqrt{\tilde{G%
}\left( s\right) }},  \label{u preko c}
\end{eqnarray}%
since the first boundary condition in (\ref{bc-bd}), corresponding to fact
that rod's end is fixed, ensures that $2C=C_{1}=-C_{2},$ while the Laplace
transform of the displacement (\ref{u preko c}) combined with (\ref{btj-lt-1}%
)$_{2,3}$ yields the Laplace transform of stress in the form 
\begin{equation}
\tilde{\sigma}\left( x,s\right) =C\left( s\right) s\sqrt{\tilde{G}\left(
s\right) }\cosh \frac{xs}{\sqrt{\tilde{G}\left( s\right) }}.
\label{sigma preko c}
\end{equation}

\subsection{Solution for prescribed displacement of rod's free end}

Displacement and stress in the Laplace domain for the prescribed
displacement of rod's free end, according to (\ref{u preko c}) and (\ref%
{sigma preko c}), take the following forms%
\begin{equation}
\tilde{u}\left( x,s\right) =\tilde{\Upsilon}\left( s\right) \frac{\sinh 
\frac{xs}{\sqrt{\tilde{G}\left( s\right) }}}{\sinh \frac{s}{\sqrt{\tilde{G}%
\left( s\right) }}}\;\;\text{and}\;\;\tilde{\sigma}\left( x,s\right) =\tilde{%
\Upsilon}\left( s\right) s\sqrt{\tilde{G}\left( s\right) }\frac{\cosh \frac{%
xs}{\sqrt{\tilde{G}\left( s\right) }}}{\sinh \frac{s}{\sqrt{\tilde{G}\left(
s\right) }}},  \label{u-i-sigma-tilda}
\end{equation}%
since the function $C$ is determined from the Laplace transform of
displacement (\ref{u preko c}) and boundary condition (\ref{bc-bd})$_{2}$ as%
\begin{equation*}
C\left( s\right) =\frac{\tilde{\Upsilon}\left( s\right) }{\sinh \frac{s}{%
\sqrt{\tilde{G}\left( s\right) }}}.
\end{equation*}

\subsubsection{Displacement for given $\Upsilon $}

Displacement in the Laplace domain, given by (\ref{u-i-sigma-tilda})$_{1}$,
can be expressed either through the solution kernel image $\tilde{P}$ in the
case of Burgers models of the first class, or through the regularized
solution kernel image $\tilde{P}_{\mathrm{reg}}$ in the case of models
belonging to the second class, that are defined as%
\begin{equation}
\tilde{P}\left( x,s\right) =\frac{\sinh \frac{xs}{\sqrt{\tilde{G}\left(
s\right) }}}{\sinh \frac{s}{\sqrt{\tilde{G}\left( s\right) }}}\;\;\text{and}%
\;\;\tilde{P}_{\mathrm{reg}}\left( x,s\right) =\frac{1}{s}\tilde{P}\left(
x,s\right) .  \label{pe-tilda}
\end{equation}

Considering the asymptotics of solution kernel image $\tilde{P}$ and its
regularized version $\tilde{P}_{\mathrm{reg}}$ as $s\rightarrow \infty ,$
one obtains%
\begin{equation}
\tilde{P}\left( x,s\right) =\mathrm{e}^{-\left( 1-x\right) \frac{s}{\sqrt{%
\tilde{G}\left( s\right) }}}\frac{1-\mathrm{e}^{-2\frac{xs}{\sqrt{\tilde{G}%
\left( s\right) }}}}{1-\mathrm{e}^{-2\frac{s}{\sqrt{\tilde{G}\left( s\right) 
}}}}\sim \mathrm{e}^{-\sqrt{\frac{a_{3}}{b}}\left( 1-x\right) s^{1-\frac{%
\delta }{2}}}\;\;\text{and}\;\;\tilde{P}_{\mathrm{reg}}\left( x,s\right)
\sim \frac{1}{s}\mathrm{e}^{-\sqrt{\frac{a_{3}}{b}}\left( 1-x\right) s},
\label{pe-s-besk}
\end{equation}%
because of the asymptotics of complex modulus $\tilde{G},$ given by (\ref%
{g-tilda}), that yields%
\begin{equation}
\tilde{G}\left( s\right) \sim \left\{ 
\begin{tabular}{ll}
$\frac{b}{a_{3}}s^{\delta },\smallskip $ & for models of the first class,
with $\delta =\mu +\eta -\gamma ,$ \\ 
$\frac{b}{a_{3}},$ & for models of the second class,%
\end{tabular}%
\right. \;\;\text{as}\;\;s\rightarrow \infty .  \label{ge-tilde-besk}
\end{equation}%
Therefore, the short-time asymptotics of solution kernel $P$ for models of
the first class is obtained as 
\begin{equation}
P\left( x,t\right) \sim \frac{1}{\pi }\int_{0}^{\infty }\sin \left( \sqrt{%
\frac{a_{3}}{b}}\left( 1-x\right) \rho ^{1-\frac{\delta }{2}}\sin \frac{%
\delta \pi }{2}\right) \mathrm{e}^{-\rho t+\sqrt{\frac{a_{3}}{b}}\left(
1-x\right) \rho ^{1-\frac{\delta }{2}}\cos \frac{\delta \pi }{2}}\mathrm{d}%
\rho ,\;\;\text{as}\;\;t\rightarrow 0,  \label{pe-te(zi)-nula}
\end{equation}%
by inverting the Laplace transform of (\ref{pe-s-besk})$_{1}$ using the
definition and integration in the complex plane, while the short-time
asymptotics of regularized solution kernel $P_{\mathrm{reg}},$ corresponding
to models of the second class, yields%
\begin{equation}
P_{\mathrm{reg}}\left( x,t\right) \sim H\left( t-\sqrt{\frac{a_{3}}{b}}%
\left( 1-x\right) \right) ,\;\;\text{as}\;\;t\rightarrow 0.
\label{pe-reg-te(zi)-nula}
\end{equation}

On the other hand, the asymptotics of regularized solution kernel image $%
\tilde{P}_{\mathrm{reg}}$ as $s\rightarrow 0,$ yields%
\begin{equation}
\tilde{P}_{\mathrm{reg}}\left( x,s\right) =\frac{1}{s}\frac{\left( 1+xs^{1-%
\frac{\xi }{2}}+\ldots \right) -\left( 1-xs^{1-\frac{\xi }{2}}+\ldots
\right) }{\left( 1+s^{1-\frac{\xi }{2}}+\ldots \right) -\left( 1-s^{1-\frac{%
\xi }{2}}+\ldots \right) }\sim \frac{1}{s}\,x,\;\;\text{implying}\;\;P_{%
\mathrm{reg}}\left( x,t\right) \sim x\,H\left( t\right) =x,\;\;\text{as}%
\;\;t\rightarrow \infty ,  \label{pe-reg-te(zi)-besk}
\end{equation}%
since the asymptotics of complex modulus $\tilde{G},$ given by (\ref{g-tilda}%
), yields%
\begin{equation}
\tilde{G}\left( s\right) \sim s^{\xi },\;\;\text{as}\;\;s\rightarrow 0,\;\;%
\text{with}\;\;\xi \in \left\{ \mu ,\beta \right\} .  \label{ge-tilde-nula}
\end{equation}

Using the inverse Laplace transform of the derivative of function in (\ref%
{pe-tilda})$_{2}$ one obtains%
\begin{equation}
P(x,t)=\frac{\partial }{\partial t}P_{\mathrm{reg}}\left( x,t\right) +P_{%
\mathrm{reg}}\left( x,0\right) \,\delta \left( t\right) =\frac{\partial }{%
\partial t}P_{\mathrm{reg}}\left( x,t\right) ,  \label{pe-pe-reg}
\end{equation}%
since the asymptotics of $\tilde{P}_{\mathrm{reg}}$ as $s\rightarrow \infty $
yields $s\tilde{P}_{\mathrm{reg}}\left( x,s\right) \sim \mathrm{e}^{-\sqrt{%
\frac{a_{3}}{b}}\left( 1-x\right) s}\rightarrow 0,$ see (\ref{pe-s-besk})$%
_{2},$ and by the initial value Tauber theorem $P_{\mathrm{reg}}\left(
x,0\right) =\lim_{s\rightarrow \infty }s\tilde{P}_{\mathrm{reg}}\left(
x,s\right) =0.$ More precisely, the solution kernel is 
\begin{eqnarray}
P(x,t) &=&\frac{\partial }{\partial t}\left( P_{\mathrm{reg}}\left(
x,t\right) \,H\left( t-\sqrt{\frac{a_{3}}{b}}\left( 1-x\right) \right)
\right)  \notag \\
&=&\frac{\partial }{\partial t}P_{\mathrm{reg}}\left( x,t\right) \,H\left( t-%
\sqrt{\frac{a_{3}}{b}}\left( 1-x\right) \right) +P_{\mathrm{reg}}\left(
x,t\right) \,\delta \left( t-\sqrt{\frac{a_{3}}{b}}\left( 1-x\right) \right)
,  \label{pe-reg}
\end{eqnarray}%
since the regularized solution kernel $P_{\mathrm{reg}}$ is zero up to $t=%
\sqrt{\frac{a_{3}}{b}}\left( 1-x\right) $ and non-zero afterwards, according
to (\ref{pe-reg-te(zi)-nula}).

The solution kernel $P$ is calculated by the definition of inverse Laplace
transform in Section \ref{kalk-pe} using the Cauchy residues theorem, since
complex valued function $\tilde{P}$ has infinite number of poles, each of
them of the first order, that are obtained as zeros of its denominator,
i.e., as solutions of the equation%
\begin{equation}
\sinh \frac{s}{\sqrt{\tilde{G}\left( s\right) }}=0\;\;\text{implying}\;\;%
\frac{s}{\sqrt{\tilde{G}\left( s\right) }}=-\mathrm{i}k\pi ,\;\;\text{i.e.,}%
\;\;\frac{s^{2}}{\tilde{G}\left( s\right) }+\left( k\pi \right)
^{2}=0,\;\;k=0,\pm 1,\pm 2,....  \label{sinus}
\end{equation}%
More precisely, as proved in Section \ref{polovi}, there is a pair of
complex conjugated poles $s_{k}$ and $\bar{s}_{k}$ for each $k\in 
\mathbb{N}
_{0}$ lying in the left complex half-plane. In addition to poles, function $%
\tilde{P}$ may have branch points other than $s=0,$ due to the square root
of function $\tilde{G}$, since its denominator $\phi _{\sigma }$ has either
one negative real zero or a pair of complex conjugated zeros with negative
real part, while in the case when function $\phi _{\sigma }$ does not have
zeros, then function $\tilde{P}$ has no branch points other than $s=0$. The
explicit form of solution kernel $P$ and its regularized form $P_{\mathrm{reg%
}}$ in the case when function $\tilde{P}$ either has no branch points other
then $s=0$ or has one negative real branch point are given by%
\begin{eqnarray}
&&P\left( x,t\right) =-\frac{1}{\pi }\int_{0}^{\infty }\func{Im}\left( \frac{%
\sinh \frac{x\rho \mathrm{e}^{\mathrm{i}\pi }}{\sqrt{\tilde{G}\left( \rho 
\mathrm{e}^{\mathrm{i}\pi }\right) }}}{\sinh \frac{\rho \mathrm{e}^{\mathrm{i%
}\pi }}{\sqrt{\tilde{G}\left( \rho \mathrm{e}^{\mathrm{i}\pi }\right) }}}%
\right) \mathrm{e}^{-\rho t}\mathrm{d}\rho  \notag \\
&&\quad \quad \quad \quad \quad \quad +2\sum_{k=1}^{\infty }\left( -1\right)
^{k}\frac{\sin \left( k\pi x\right) }{k\pi }\mathrm{e}^{-\rho
_{k}t\left\vert \cos \varphi _{k}\right\vert }\func{Re}\left( \frac{s_{k}}{%
1+\left( k\pi \right) ^{2}\frac{\tilde{G}^{\prime }\left( s_{k}\right) }{%
2s_{k}}}\mathrm{e}^{\mathrm{i}\rho _{k}t\sin \varphi _{k}}\right) ,
\label{pe} \\
&&P_{\mathrm{reg}}\left( x,t\right) =x-\frac{1}{\pi }\int_{0}^{\infty }\func{%
Im}\left( \frac{1}{\rho \mathrm{e}^{\mathrm{i}\pi }}\frac{\sinh \frac{x\rho 
\mathrm{e}^{\mathrm{i}\pi }}{\sqrt{\tilde{G}\left( \rho \mathrm{e}^{\mathrm{i%
}\pi }\right) }}}{\sinh \frac{\rho \mathrm{e}^{\mathrm{i}\pi }}{\sqrt{\tilde{%
G}\left( \rho \mathrm{e}^{\mathrm{i}\pi }\right) }}}\right) \mathrm{e}%
^{-\rho t}\mathrm{d}\rho  \notag \\
&&\quad \quad \quad \quad \quad \quad +2\sum_{k=1}^{\infty }\left( -1\right)
^{k}\frac{\sin \left( k\pi x\right) }{k\pi }\mathrm{e}^{-\rho
_{k}t\left\vert \cos \varphi _{k}\right\vert }\func{Re}\left( \frac{1}{%
1+\left( k\pi \right) ^{2}\frac{\tilde{G}^{\prime }\left( s_{k}\right) }{%
2s_{k}}}\mathrm{e}^{\mathrm{i}\rho _{k}t\sin \varphi _{k}}\right) ,
\label{pe-regular}
\end{eqnarray}%
while the solution kernel $P$ and its regularized form take the form%
\begin{align}
& P\left( x,t\right) =\frac{1}{\pi }\int_{0}^{\infty }\func{Im}\left( \frac{%
\sinh \frac{x\rho \mathrm{e}^{\mathrm{i}\varphi _{0}}}{\sqrt{\tilde{G}\left(
\rho \mathrm{e}^{\mathrm{i}\varphi _{0}}\right) }}}{\sinh \frac{\rho \mathrm{%
e}^{\mathrm{i}\varphi _{0}}}{\sqrt{\tilde{G}\left( \rho \mathrm{e}^{\mathrm{i%
}\varphi _{0}}\right) }}}\mathrm{e}^{\mathrm{i}\left( \varphi _{0}+\rho t%
\mathrm{\sin }\varphi _{0}\right) }\right) \mathrm{e}^{-\rho t\left\vert 
\mathrm{\cos }\varphi _{0}\right\vert }\mathrm{d}\rho  \notag \\
& \quad \quad \quad \quad \quad \quad +2\sum_{k=1}^{\infty }\left( -1\right)
^{k}\frac{\sin \left( k\pi x\right) }{k\pi }\mathrm{e}^{-\rho
_{k}t\left\vert \cos \varphi _{k}\right\vert }\func{Re}\left( \frac{s_{k}}{%
1+\left( k\pi \right) ^{2}\frac{\tilde{G}^{\prime }\left( s_{k}\right) }{%
2s_{k}}}\mathrm{e}^{\mathrm{i}\rho _{k}t\sin \varphi _{k}}\right) ,
\label{peee} \\
& P_{\mathrm{reg}}\left( x,t\right) =x\frac{\varphi _{0}}{\pi }+\frac{1}{\pi 
}\int_{0}^{\infty }\func{Im}\left( \frac{1}{\rho \mathrm{e}^{\mathrm{i}%
\varphi _{0}}}\frac{\sinh \frac{x\rho \mathrm{e}^{\mathrm{i}\varphi _{0}}}{%
\sqrt{\tilde{G}\left( \rho \mathrm{e}^{\mathrm{i}\varphi _{0}}\right) }}}{%
\sinh \frac{\rho \mathrm{e}^{\mathrm{i}\varphi _{0}}}{\sqrt{\tilde{G}\left(
\rho \mathrm{e}^{\mathrm{i}\varphi _{0}}\right) }}}\mathrm{e}^{\mathrm{i}%
\left( \varphi _{0}+\rho t\mathrm{\sin }\varphi _{0}\right) }\right) \mathrm{%
e}^{-\rho t\left\vert \mathrm{\cos }\varphi _{0}\right\vert }\mathrm{d}\rho 
\notag \\
& \quad \quad \quad \quad \quad \quad +2\sum_{k=1}^{\infty }\left( -1\right)
^{k}\frac{\sin \left( k\pi x\right) }{k\pi }\mathrm{e}^{-\rho
_{k}t\left\vert \cos \varphi _{k}\right\vert }\func{Re}\left( \frac{1}{%
1+\left( k\pi \right) ^{2}\frac{\tilde{G}^{\prime }\left( s_{k}\right) }{%
2s_{k}}}\mathrm{e}^{\mathrm{i}\rho _{k}t\sin \varphi _{k}}\right) ,
\label{peee-regular}
\end{align}%
in the case when function $\tilde{P}$ has a pair of complex conjugated
branch points with negative real part $s_{0}=\rho _{0}\mathrm{e}^{\mathrm{i}%
\varphi _{0}}$ and $\bar{s}_{0}$ in addition to $s=0$. Note, the form (\ref%
{peee}) of solution kernel $P$ is more general, since it reduces to (\ref{pe}%
) for $\varphi _{0}=\pi $.

The solution kernel $P,$ according to either (\ref{pe}) or (\ref{peee}),
consist of two terms: the first is at most non-monotonic in both space and
time and the second one is a superposition of standing waves oscillating in
time with angular frequency $\omega _{k}=\rho _{k}\sin \varphi _{k}$ and
amplitude decreasing in time. Note, values of $x\in \left( 0,1\right) $ and $%
t>0$ are not independent in the case of the second model class, since $P_{%
\mathrm{reg}}\left( x,t\right) \neq 0$ for $t>\sqrt{\frac{a_{3}}{b}}\left(
1-x\right) ,$ according to (\ref{pe-reg-te(zi)-nula}), implying the finite
velocity of disturbance propagation, which is not the case for the first
model class, due to the short-time asymptotics of solution kernel $P$, see (%
\ref{pe-te(zi)-nula}).

Having the solution kernel calculated either by (\ref{pe}) and (\ref{peee})
in the case of the first model class, or by (\ref{pe-reg}) in the case of
the second model class, the displacement in the case of prescribed
displacement of rod's free end is%
\begin{equation}
u\left( x,t\right) =\Upsilon \left( t\right) \ast P\left( x,t\right) ,
\label{u-ipsilon-pe}
\end{equation}%
by the inverse Laplace transform of (\ref{u-i-sigma-tilda})$_{1}$, with the
solution kernel image $\tilde{P}$ defined by (\ref{pe-tilda}).

\subsubsection{Stress for given $\Upsilon $}

Stress in the Laplace domain, given by (\ref{u-i-sigma-tilda})$_{2}$, can be
expressed either through the solution kernel image $\tilde{R}$ in the case
of Burgers models of the first class, or through the regularized solution
kernel image $\tilde{R}_{\varepsilon }$ in the case of models belonging to
the second class, that are defined as%
\begin{equation}
\tilde{R}\left( x,s\right) =s\sqrt{\tilde{G}\left( s\right) }\frac{\cosh 
\frac{xs}{\sqrt{\tilde{G}\left( s\right) }}}{\sinh \frac{s}{\sqrt{\tilde{G}%
\left( s\right) }}}\;\;\text{and}\;\;\tilde{R}_{\varepsilon }\left(
x,s\right) =\tilde{R}\left( x,s\right) \mathrm{e}^{-\varepsilon \sqrt{s}}.
\label{er-tilda}
\end{equation}

Considering the asymptotics of solution kernel image $\tilde{R}$ and its
regularized version $\tilde{R}_{\varepsilon }$ as $s\rightarrow \infty ,$
one obtains%
\begin{eqnarray}
\tilde{R}\left( x,s\right) &=&s\sqrt{\tilde{G}\left( s\right) }\mathrm{e}%
^{-\left( 1-x\right) \frac{s}{\sqrt{\tilde{G}\left( s\right) }}}\frac{1+%
\mathrm{e}^{-2\frac{xs}{\sqrt{\tilde{G}\left( s\right) }}}}{1-\mathrm{e}^{-2%
\frac{s}{\sqrt{\tilde{G}\left( s\right) }}}}\sim \sqrt{\frac{b}{a_{3}}}s^{1+%
\frac{\delta }{2}}\mathrm{e}^{-\sqrt{\frac{a_{3}}{b}}\left( 1-x\right) s^{1-%
\frac{\delta }{2}}}\;\;\text{and}  \label{er-s-besk-1} \\
\tilde{R}_{\varepsilon }\left( x,s\right) &\sim &\sqrt{\frac{b}{a_{3}}}s%
\mathrm{e}^{-\sqrt{\frac{a_{3}}{b}}\left( 1-x\right) s}\mathrm{e}%
^{-\varepsilon \sqrt{s}},  \label{er-s-besk-2}
\end{eqnarray}%
because of the asymptotics of complex modulus $\tilde{G},$ given by (\ref%
{ge-tilde-besk}), so that the short-time asymptotics of solution kernel $R$
for models of the first class is obtained as 
\begin{equation}
R\left( x,t\right) \sim -\frac{1}{\pi }\sqrt{\frac{b}{a_{3}}}%
\int_{0}^{\infty }\sin \left( \sqrt{\frac{a_{3}}{b}}\left( 1-x\right) \rho
^{1-\frac{\delta }{2}}\sin \frac{\delta \pi }{2}-\frac{\delta \pi }{2}%
\right) \mathrm{e}^{-\rho t+\sqrt{\frac{a_{3}}{b}}\left( 1-x\right) \rho ^{1-%
\frac{\delta }{2}}\cos \frac{\delta \pi }{2}}\rho ^{1+\frac{\delta }{2}}%
\mathrm{d}\rho ,\;\;\text{as}\;\;t\rightarrow 0,  \label{er-te-tezi-nula}
\end{equation}%
by inverting the Laplace transform of (\ref{er-s-besk-1}) using the
definition and integration in the complex plane, while the short-time
asymptotics of regularized solution kernel $R_{\varepsilon },$ corresponding
to models of the second class, reads%
\begin{eqnarray}
R_{\varepsilon }\left( x,t\right) &\sim &\sqrt{\frac{b}{a_{3}}}\delta \left(
t-\sqrt{\frac{a_{3}}{b}}\left( 1-x\right) \right) \ast \frac{\mathrm{d}}{%
\mathrm{d}t}\left( \frac{\varepsilon }{2t\sqrt{\pi t}}\mathrm{e}^{-\frac{%
\varepsilon ^{2}}{4t}}\right)  \notag \\
&\sim &\sqrt{\frac{b}{a_{3}}}\frac{\mathrm{d}}{\mathrm{d}t}\left( \frac{%
\varepsilon }{2\tau \sqrt{\pi \tau }}\mathrm{e}^{-\frac{\varepsilon ^{2}}{%
4\tau }}\right) _{\tau =t-\sqrt{\frac{a_{3}}{b}}\left( 1-x\right) },\;\;%
\text{as}\;\;t\rightarrow 0,  \label{er-eps-te-tezi-nula}
\end{eqnarray}%
since $\mathcal{L}^{-1}\left[ \mathrm{e}^{-\varepsilon \sqrt{s}}\right]
\left( t\right) =\frac{\varepsilon }{2t\sqrt{\pi t}}\mathrm{e}^{-\frac{%
\varepsilon ^{2}}{4t}}\rightarrow 0,$ when $t\rightarrow 0.$

Applying the inverse Laplace transform to the regularized solution kernel
image $\tilde{R}_{\varepsilon },$ given by (\ref{er-tilda})$_{2},$ one has%
\begin{equation*}
R_{\varepsilon }(x,t)=R\left( x,t\right) \ast \left( \frac{\varepsilon }{2t%
\sqrt{\pi t}}\mathrm{e}^{-\frac{\varepsilon ^{2}}{4t}}\right) ,\;\;\text{%
i.e.,}\;\;\lim_{\varepsilon \rightarrow 0}R_{\varepsilon }(x,t)=R\left(
x,t\right) \ast \delta \left( t\right) =R\left( x,t\right) ,
\end{equation*}%
since $\lim_{\varepsilon \rightarrow 0}\frac{\varepsilon }{2t\sqrt{\pi t}}%
\mathrm{e}^{-\frac{\varepsilon ^{2}}{4t}}=\delta \left( t\right) ,$ because
of $\left. \mathcal{L}\left[ \frac{\varepsilon }{2t\sqrt{\pi t}}\mathrm{e}^{-%
\frac{\varepsilon ^{2}}{4t}}\right] \left( s\right) \right\vert
_{\varepsilon =0}=\left. \mathrm{e}^{-\varepsilon \sqrt{s}}\right\vert
_{\varepsilon =0}=1=\mathcal{L}\left[ \delta \left( t\right) \right] .$

Similarly as done in Section \ref{kalk-pe}, the calculation of regularized
solution kernel $R_{\varepsilon }$ is also performed by the inverse Laplace
transform formula, so that it takes the following form%
\begin{align}
& R_{\varepsilon }\left( x,t\right) =\frac{1}{\pi }\int_{0}^{\infty }\func{Im%
}\left( \sqrt{\tilde{G}\left( \rho \mathrm{e}^{\mathrm{i}\varphi
_{0}}\right) }\frac{\cosh \frac{x\rho \mathrm{e}^{\mathrm{i}\varphi _{0}}}{%
\sqrt{\tilde{G}\left( \rho \mathrm{e}^{\mathrm{i}\varphi _{0}}\right) }}}{%
\sinh \frac{\rho \mathrm{e}^{\mathrm{i}\varphi _{0}}}{\sqrt{\tilde{G}\left(
\rho \mathrm{e}^{\mathrm{i}\varphi _{0}}\right) }}}\mathrm{e}^{\mathrm{i}%
\left( 2\varphi _{0}+\rho t\mathrm{\sin }\varphi _{0}\right) }\mathrm{e}%
^{-\varepsilon \sqrt{\rho }\mathrm{e}^{\mathrm{i}\frac{\varphi _{0}}{2}%
}}\right) \mathrm{e}^{-\rho t\left\vert \mathrm{\cos }\varphi
_{0}\right\vert }\rho \mathrm{d}\rho  \notag \\
& \quad \quad \quad \quad \quad \quad =2\sum_{k=1}^{\infty }\left( -1\right)
^{k+1}\frac{\cos \left( k\pi x\right) }{\left( k\pi \right) ^{2}}\mathrm{e}%
^{-\rho _{k}t\left\vert \cos \varphi _{k}\right\vert }\func{Re}\left( \frac{%
s_{k}^{3}}{1+\left( k\pi \right) ^{2}\frac{\tilde{G}^{\prime }\left(
s_{k}\right) }{2s_{k}}}\mathrm{e}^{\mathrm{i}\rho _{k}\sin \varphi
_{k}}\right) \mathrm{e}^{-\varepsilon \sqrt{s_{k}}},  \label{er}
\end{align}%
in the case when function $\tilde{R}_{\varepsilon }$, given by (\ref%
{er-tilda})$_{2},$ has a pair of complex conjugated branch points $%
s_{0}=\rho _{0}\mathrm{e}^{\mathrm{i}\varphi _{0}}$ and $\bar{s}_{0}$ in
addition to $s=0,$ while by putting $\varphi _{0}=\pi $ in (\ref{er}) one
obtains the form of solution kernel $R_{\varepsilon },$ analogous to (\ref%
{pe}), corresponding to case when function $\tilde{R}_{\varepsilon },$
except for $s=0,$ has either no branch points or has one negative real
branch point. Note, the form of solution kernel $R,$ corresponding to the
Burgers models of the first class, is obtained by putting $\varepsilon =0$
into expression (\ref{er}) for $R_{\varepsilon }$, since regularization is
not required.

Having the solution kernel $R_{\varepsilon }$ calculated by (\ref{er}),
either with $\varepsilon =0$ in the case of the first model class, or with $%
\varepsilon \neq 0$ in the case of the second model class, the stress in the
case of prescribed displacement of rod's free end is%
\begin{equation}
\sigma \left( x,t\right) =\Upsilon \left( t\right) \ast R\left( x,t\right) ,
\label{sigma-ipsilon-er}
\end{equation}%
by the inverse Laplace transform of (\ref{u-i-sigma-tilda})$_{2}$, with the
solution kernel image $\tilde{R}$ defined by (\ref{er-tilda}).

\subsubsection{Numerical examples}

\begin{figure}[h]
\begin{center}
\begin{minipage}{0.46\columnwidth}
  \subfloat[Case of Model V.]{
   \includegraphics[width=\columnwidth]{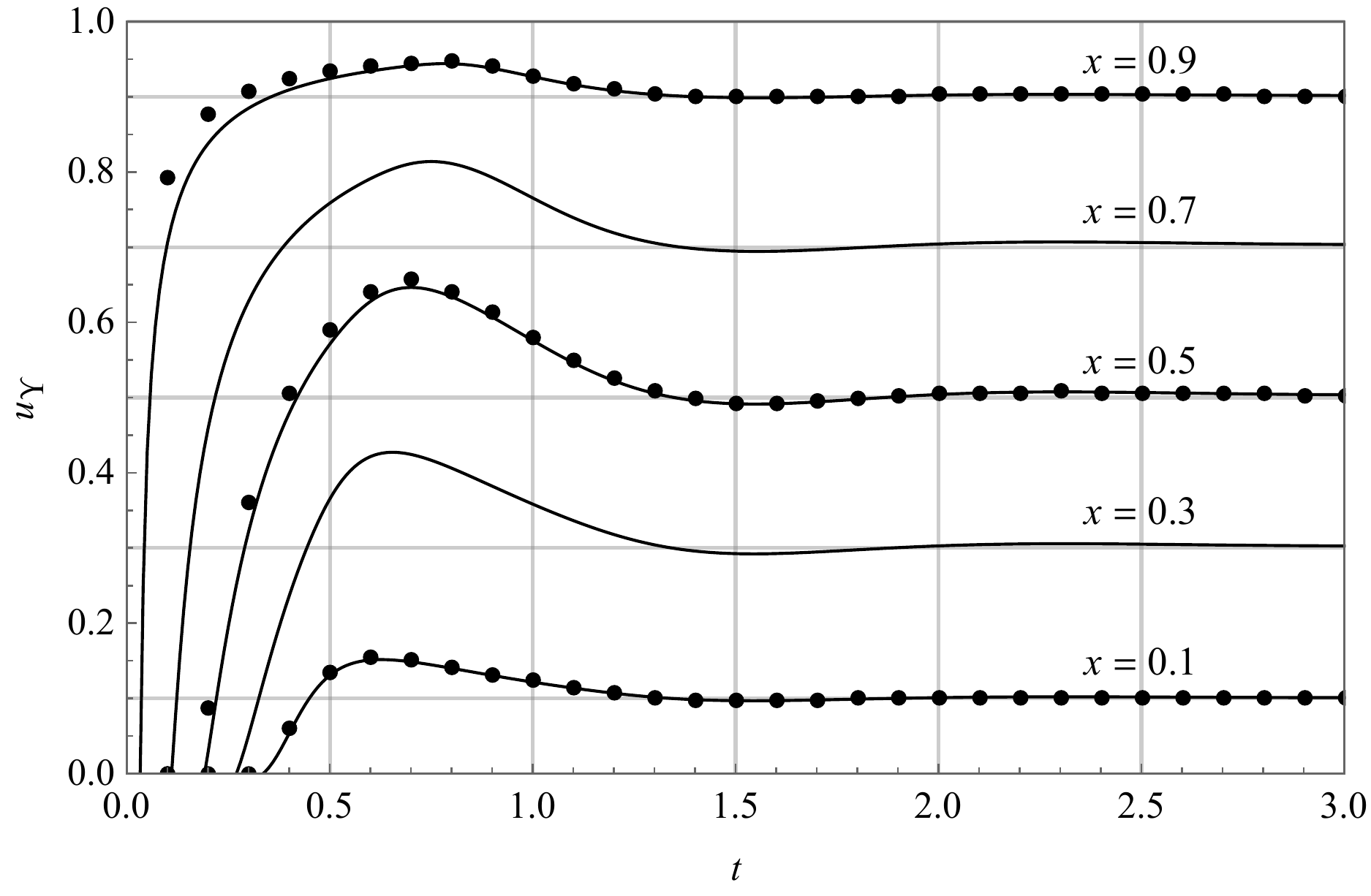}
   \label{sr-m5-u}}
  \end{minipage}
\hfil
\begin{minipage}{0.46\columnwidth}
  \subfloat[Case of Model VII.]{
   \includegraphics[width=\columnwidth]{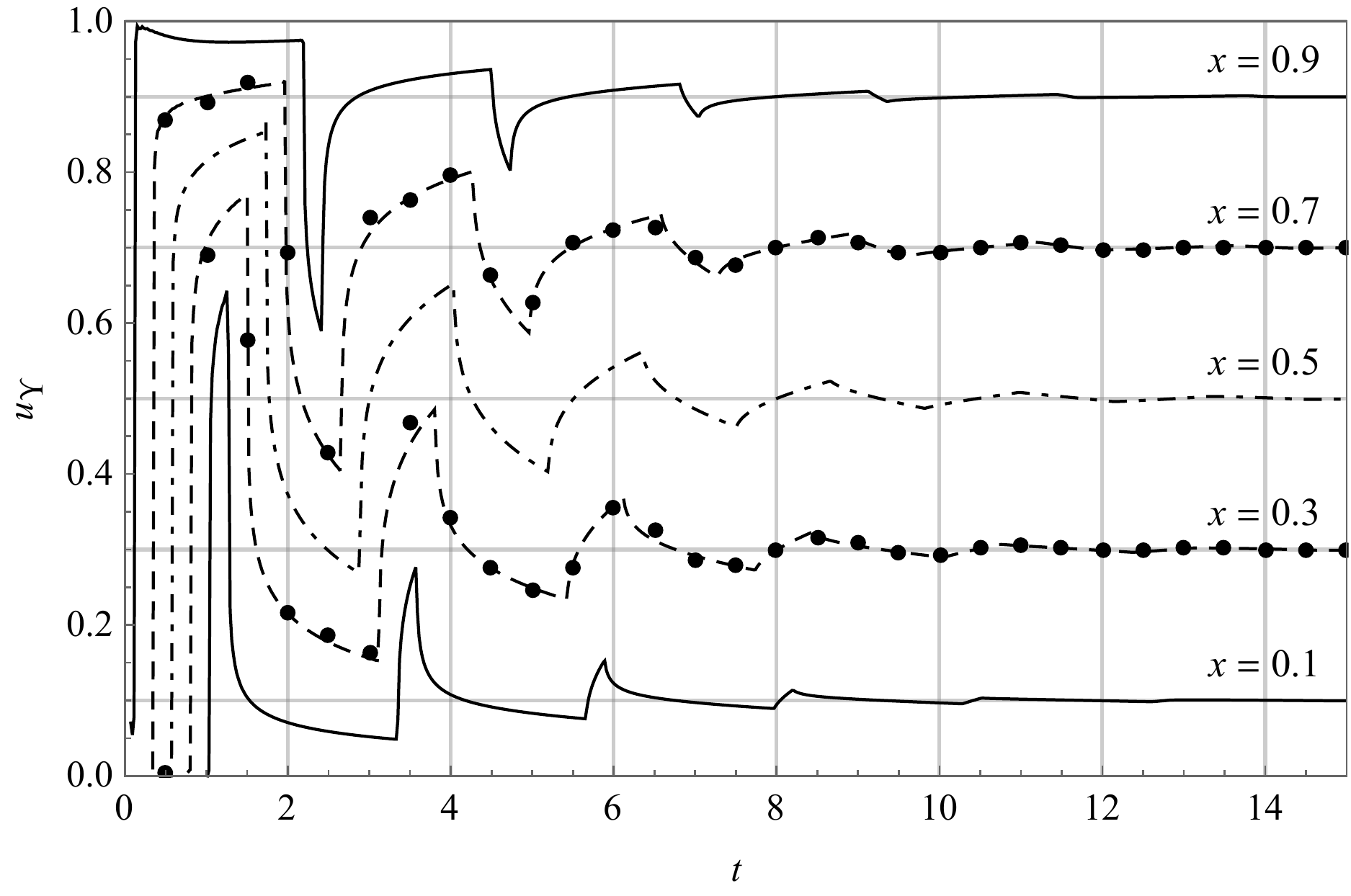}
   \label{sr-m7-u}}
  \end{minipage}
\end{center}
\caption{Displacement of a rod when the displacement of its free end is
assumed as the Heaviside function, i.e., $\Upsilon=H$, obtained according to
analytical expression (lines) and by numerical Laplace transform inversion
(dots). }
\label{sr-u}
\end{figure}

\begin{figure}[h]
\begin{center}
\includegraphics[width=0.6\columnwidth]{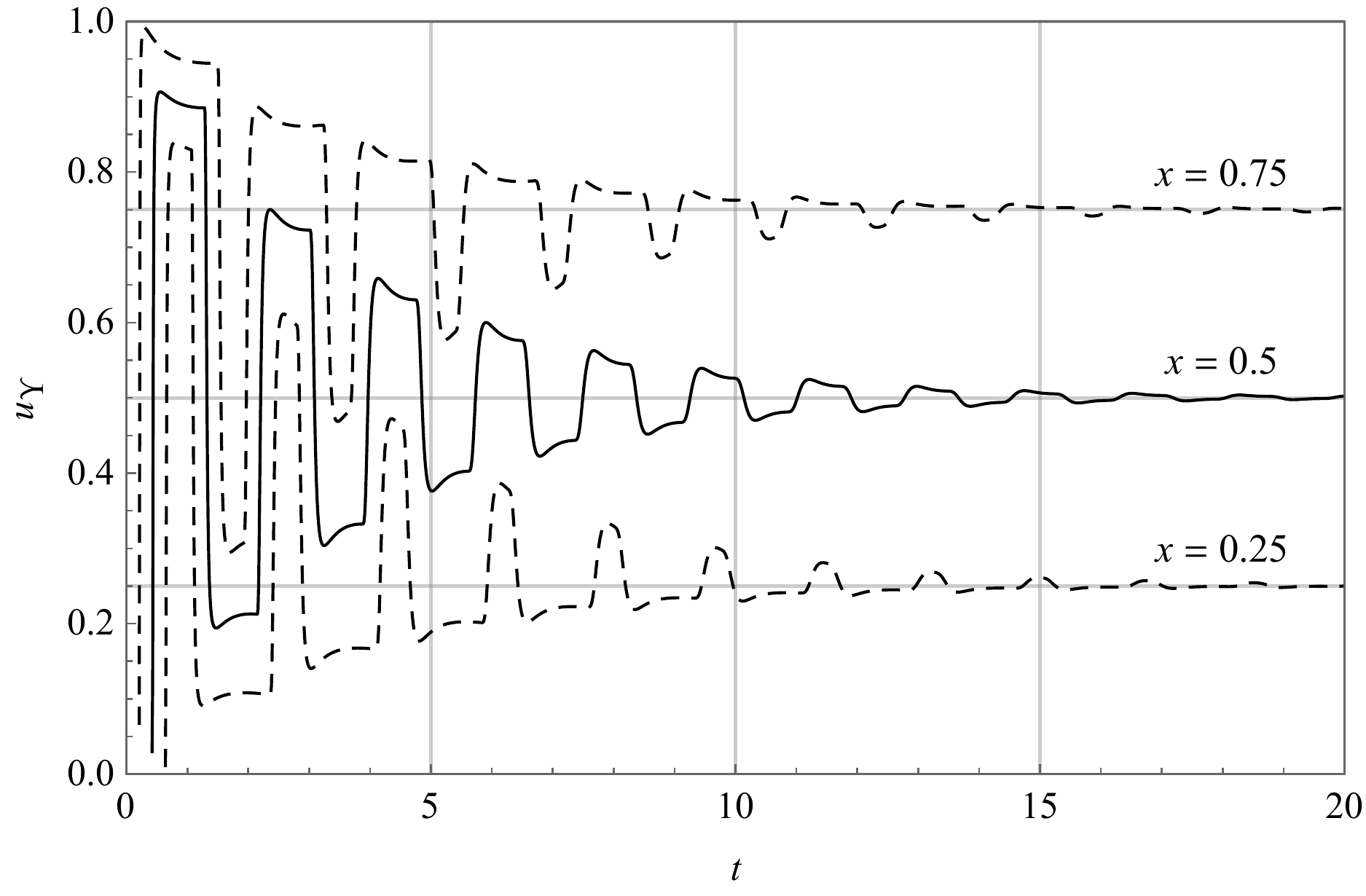}
\end{center}
\caption{Displacement of a rod when the displacement of its free end is
assumed as the Heaviside function, i.e., $\Upsilon =H$, for Model V in the
case of complex conjugated branch points.}
\label{sr-m5-u-kompl-nule}
\end{figure}

Figures \ref{sr-u} and \ref{sr-m5-u-kompl-nule} present displacements of
several points of the rod for displacement of rod's free end given as the
Heaviside step function, i.e., for boundary condition (\ref{bc-bd})$_{2}$
taken as $\Upsilon =H.$ The regularized solution kernel $P_{\mathrm{reg}}$
actually represents the step response, due to defining relation (\ref%
{pe-tilda})$_{2}$ for regularized solution kernel image $\tilde{P}_{\mathrm{%
reg}},$ that yields%
\begin{equation*}
u_{\Upsilon }\left( x,t\right) =P_{\mathrm{reg}}\left( x,t\right) =H\left(
t\right) \ast P\left( x,t\right)
\end{equation*}%
after performing the inverse Laplace transform, see also (\ref{u-ipsilon-pe}%
).

The step response displays damped oscillatory behavior that settles at the
value of point's position, i.e.,%
\begin{equation}
\lim_{t\rightarrow \infty }u_{\Upsilon }\left( x,t\right) =x,
\label{u-ipsilon-besk}
\end{equation}%
as predicted by the large-time asymptotics of regularized solution kernel $%
P_{\mathrm{reg}},$ given by (\ref{pe-reg-te(zi)-besk}). The time profiles of
step response in the case of Model V have quite classical shapes of the
oscillatory behavior with pronounced damping, see Figure \ref{sr-m5-u}. On
the other hand, the profiles in the case of Model VII, being also damped
oscillatory, resemble to the sequence of excitation and relaxation
processes, since profiles repeatedly change their convexity from concave to
convex, as clearly visible from Figure \ref{sr-m7-u}. Nevertheless, curves
obtained by analytical expressions are consistent with curves represented by
dots, that are obtained by numerical Laplace transform inversion using fixed
Talbot method, see \cite{AbateValko}. Regarding the short-time asymptotics,
step response differs for Burgers models of the first and second class,
since in the case of first class models time profiles continuously increase
from zero, obtaining non-zero values depending on point's position, see
short-time asymptotics (\ref{pe-te(zi)-nula}) of solution kernel $P$ and
Figure \ref{sr-m5-u}, while in the case of second class models time profiles
jump from zero depending on point's position, since the short-time
asymptotics is represented by the Heaviside function, see (\ref%
{pe-reg-te(zi)-nula}) and Figure \ref{sr-m7-u}. 

Plots from Figure \ref{sr-u} correspond to the case when solution kernel
image $\tilde{P}$ and its regularization $\tilde{P}_{\mathrm{reg}}$ have no
branch points except for $s=0,$ so that the step responses are obtained
using (\ref{pe-regular}) with model parameters as in Table \ref{tbl}, while
time profiles from Figure \ref{sr-m5-u-kompl-nule} correspond to the case
when kernel image additionally has a pair of complex conjugated branch
points with negative real part, hence the step responses are calculated by (%
\ref{peee-regular}) using model parameters from Table \ref{tbl}. Although
model parameters in the case of complex conjugated branch points do not
satisfy narrowed thermodynamical restrictions, required in the proof that
kernel image has a pair of complex conjugated poles $s_{k},$ $k\in \mathbb{N}%
,$ this requirement is checked numerically, so as the fact that the argument
of branch point is greater than the arguments of poles. Time profiles of the
step response from Figure \ref{sr-m5-u-kompl-nule} are peculiarly shaped, as
if two vibrations with different frequencies are superposed, since there is
a relaxation process instead of peak, that is followed by another faster
relaxation process, appearing after two successive excitation processes
having different speeds. Moreover, responses have an envelope, that is
typical for damped oscillations. 
\begin{table}[h]
\begin{center}
\begin{tabular}{c|c|c|c|c|c|c|c|c|c|c}
Model & Branch points & $a_1$ & $a_2$ & $a_3$ & $b$ & $\alpha$ & $\beta$ & $%
\gamma$ & $\mu$ & $\eta$ \\ 
\hhline{=|=|=|=|=|=|=|=|=|=|=} \multirow{2}{*}{Model V} & $s=0$ & $0.005$ & $%
0.8$ & $0.115$ & $0.376$ & $0.6$ & $0.61$ & \multirow{2}{*}{$2\beta$} & $0.8$
& \multirow{3}{*}{$\beta$} \\ \cline{2-8}\cline{10-10}
& $s=0$, $s_0$, $\bar{s}_0$ & $0.075$ & $0.8$ & $1.14$ & $1.39$ & $0.4$ & $%
0.685$ &  & $0.7$ &  \\ \cline{1-10}
Model VII & $s=0$ & $0.01$ & $4.5$ & $4$ & $3$ & $0.7$ & $0.845$ & - & - & 
\end{tabular}%
\end{center}
\caption{Model parameters.}
\label{tbl}
\end{table}

\begin{figure}[p]
\begin{center}
\begin{minipage}{0.5\columnwidth}
			\subfloat[Case of Model V - line corresponds to analytical expression and geometrical shapes to numerical Laplace transform inversion.]{
			\includegraphics[width=\columnwidth]{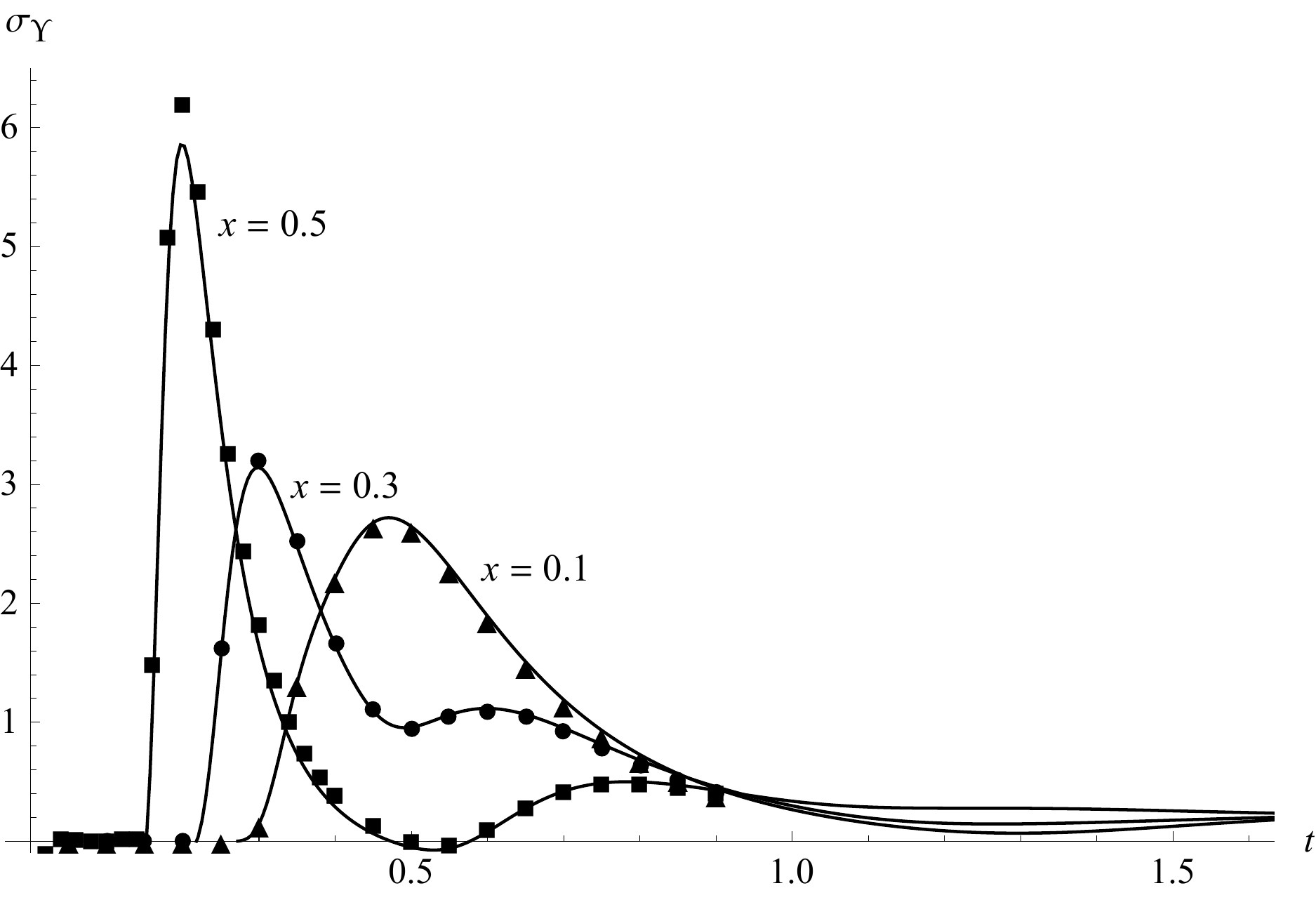}
			\label{sr-m5-sigma-1}}
		\end{minipage}
\vfil
\begin{minipage}{0.46\columnwidth}
				\subfloat[Case of Model V.]{
				\includegraphics[width=\columnwidth]{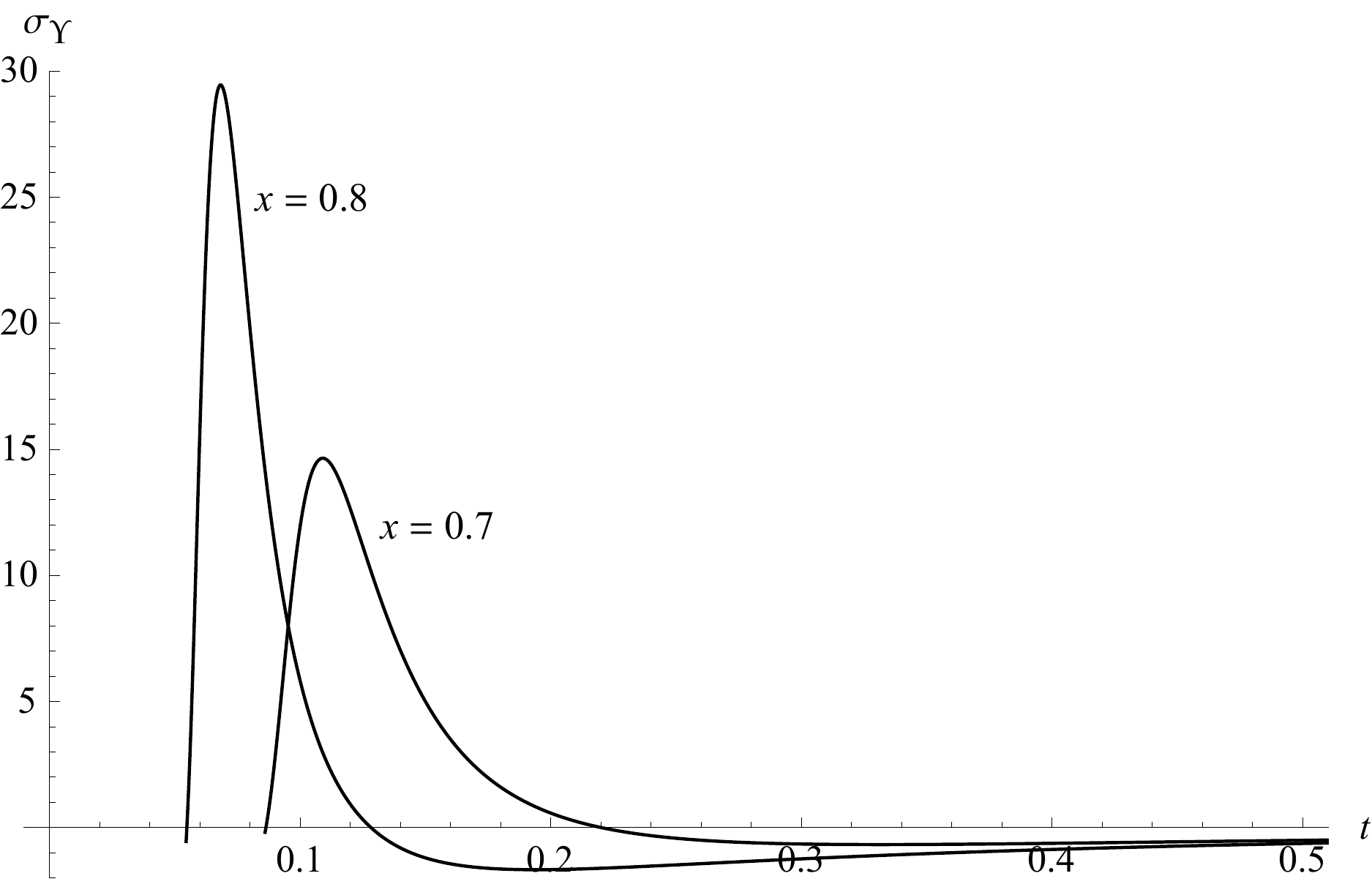}
				\label{sr-m5-sigma-2}}
		\end{minipage}
\hfill 
\begin{minipage}{0.46\columnwidth}
				\subfloat[Case of Model V.]{
				\includegraphics[width=\columnwidth]{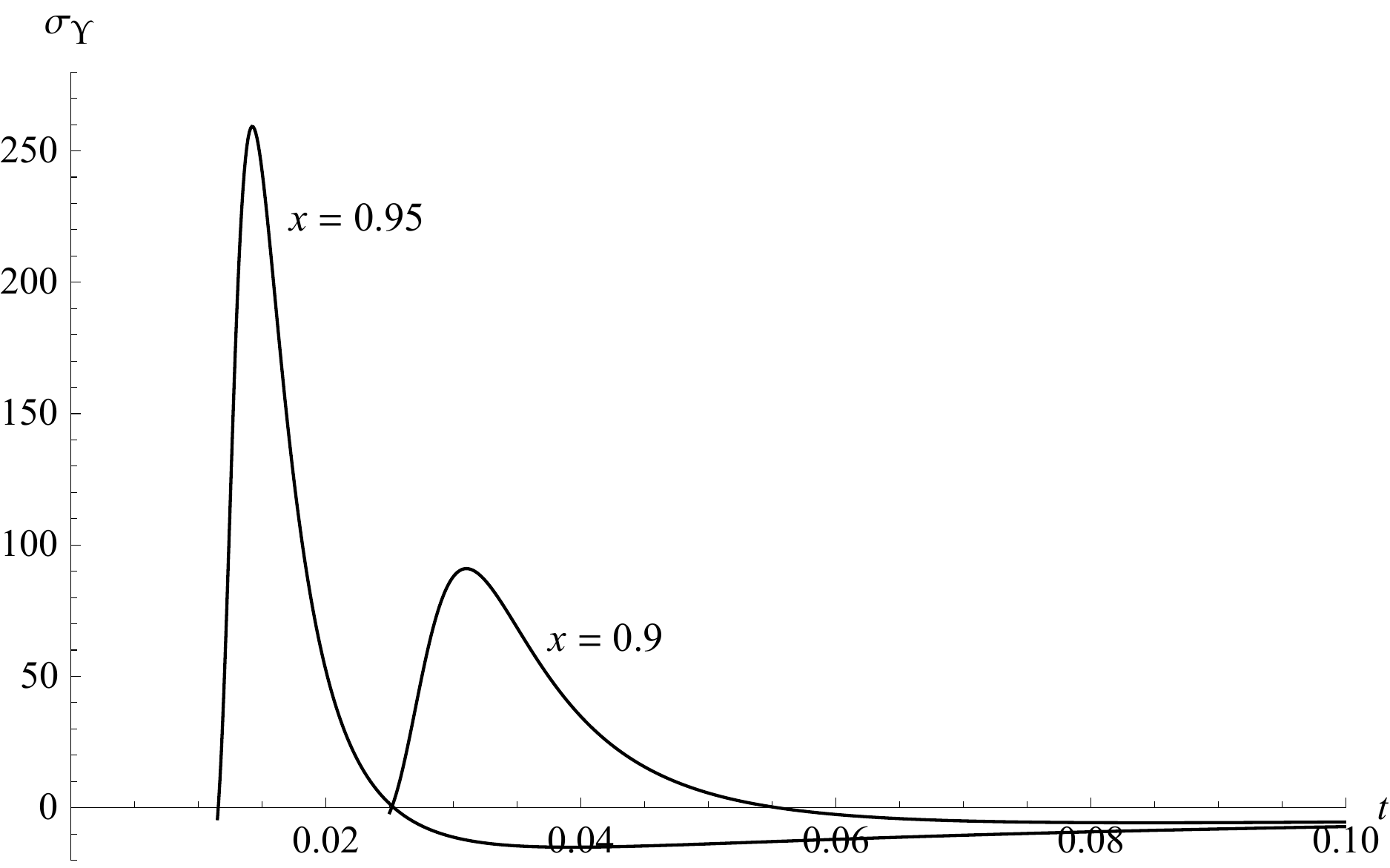}
				\label{sr-m5-sigma-3}}
		\end{minipage}
\vfil
\begin{minipage}{0.46\columnwidth}
  \subfloat[Case of Model VII - dotted, solid, and dashed lines correspond to $x\in\{0.1,0.3,0.5\}$, respectively.]{
   \includegraphics[width=\columnwidth]{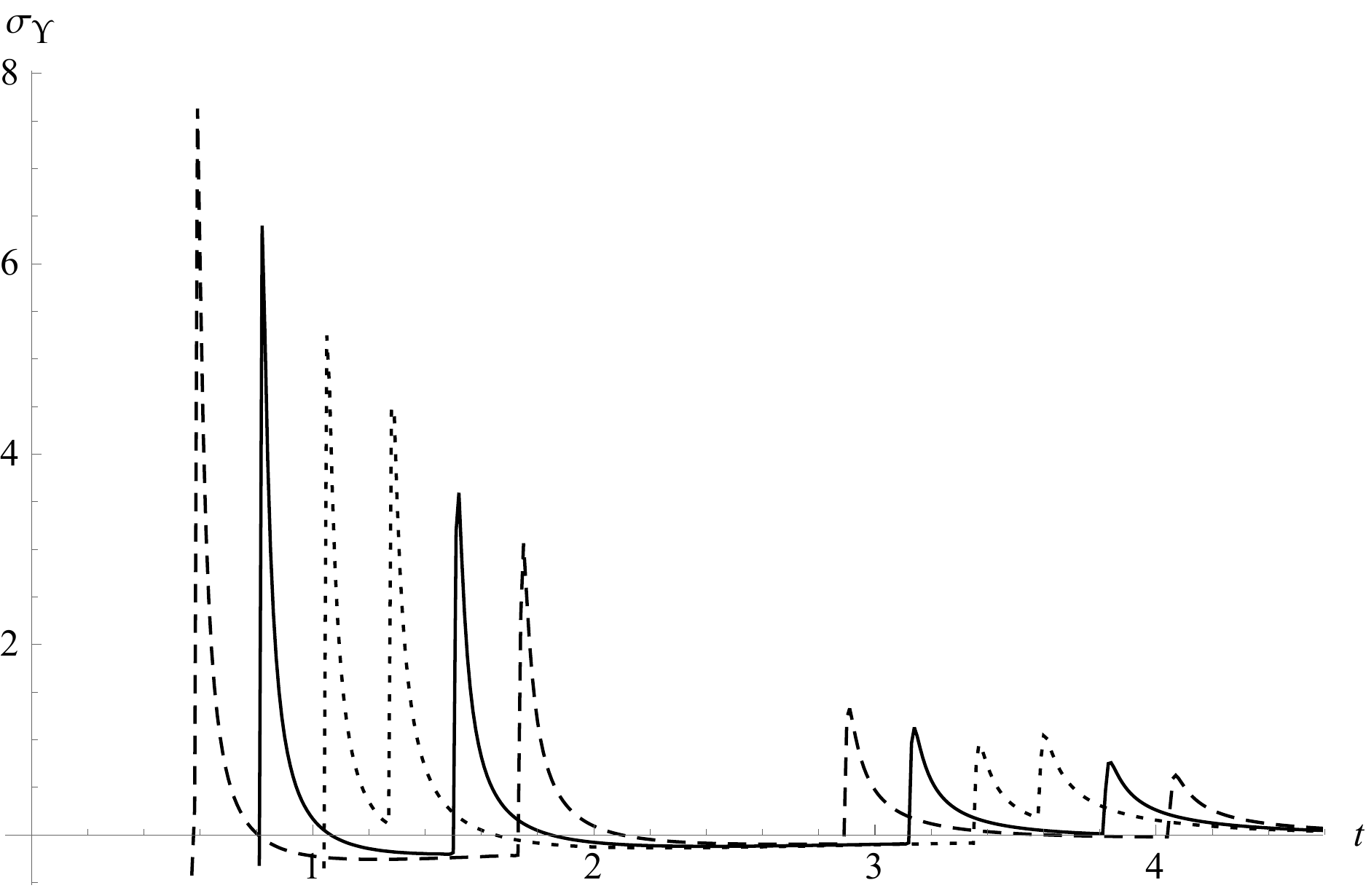}
   \label{sr-m7-sigma-1}}
\end{minipage}
\hfil
\begin{minipage}{0.46\columnwidth}
  \subfloat[Case of Model VII - dotted, solid, and dashed lines correspond to $x\in\{0.7,0.8,0.9\}$, respectively.]{
   \includegraphics[width=\columnwidth]{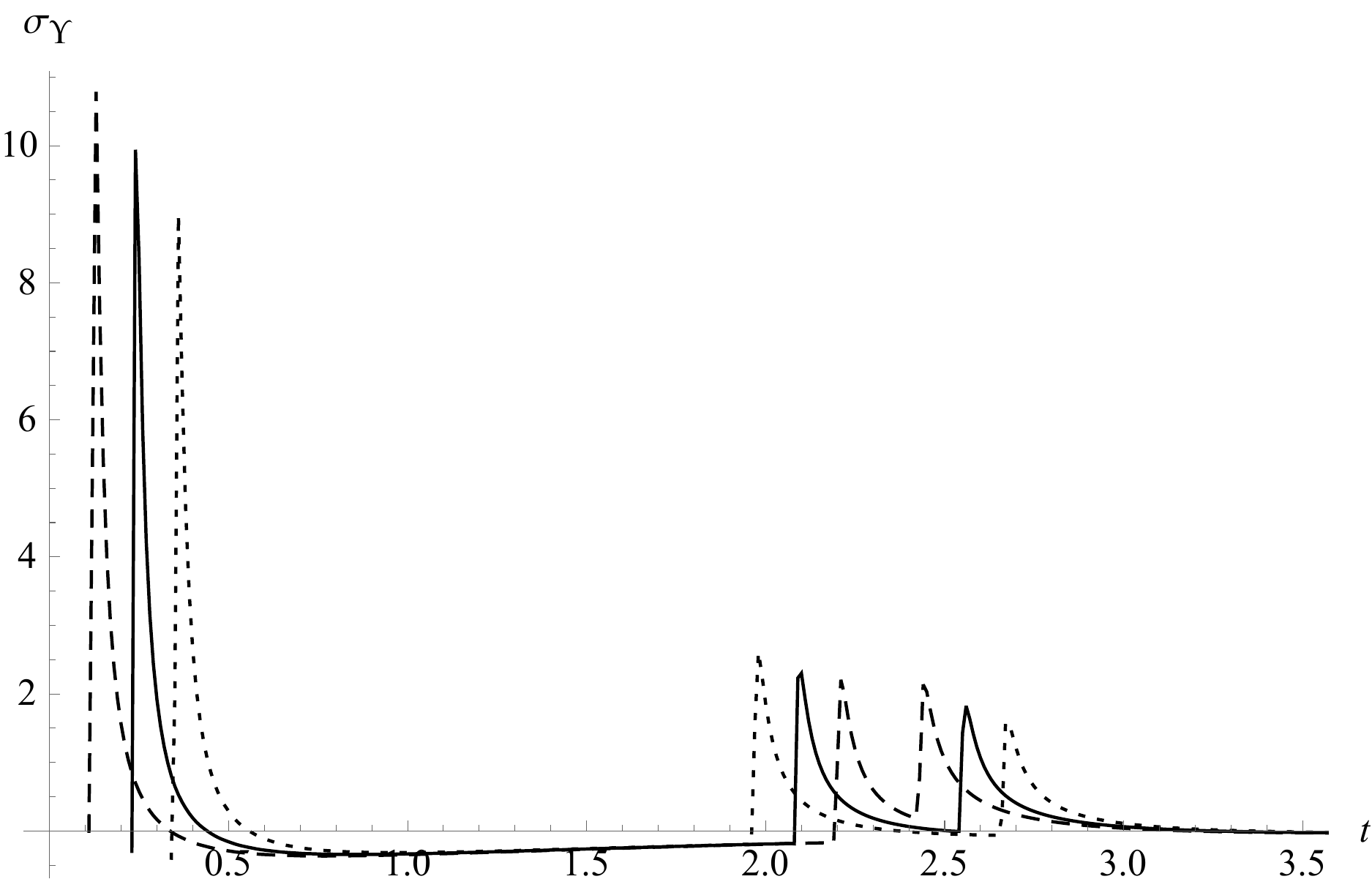}
   \label{sr-m7-sigma-2}}
\end{minipage}
\end{center}
\caption{Stress in a rod when the displacement of its free end is assumed as
the Heaviside function, i.e., $\Upsilon=H$, obtained according to analytical
expression (lines) and by numerical Laplace transform inversion (geometrical
shapes).}
\label{sr-sigma}
\end{figure}
\begin{figure}[h]
\begin{center}
\begin{minipage}{0.46\columnwidth}
				\subfloat[]{
				\includegraphics[width=\columnwidth]{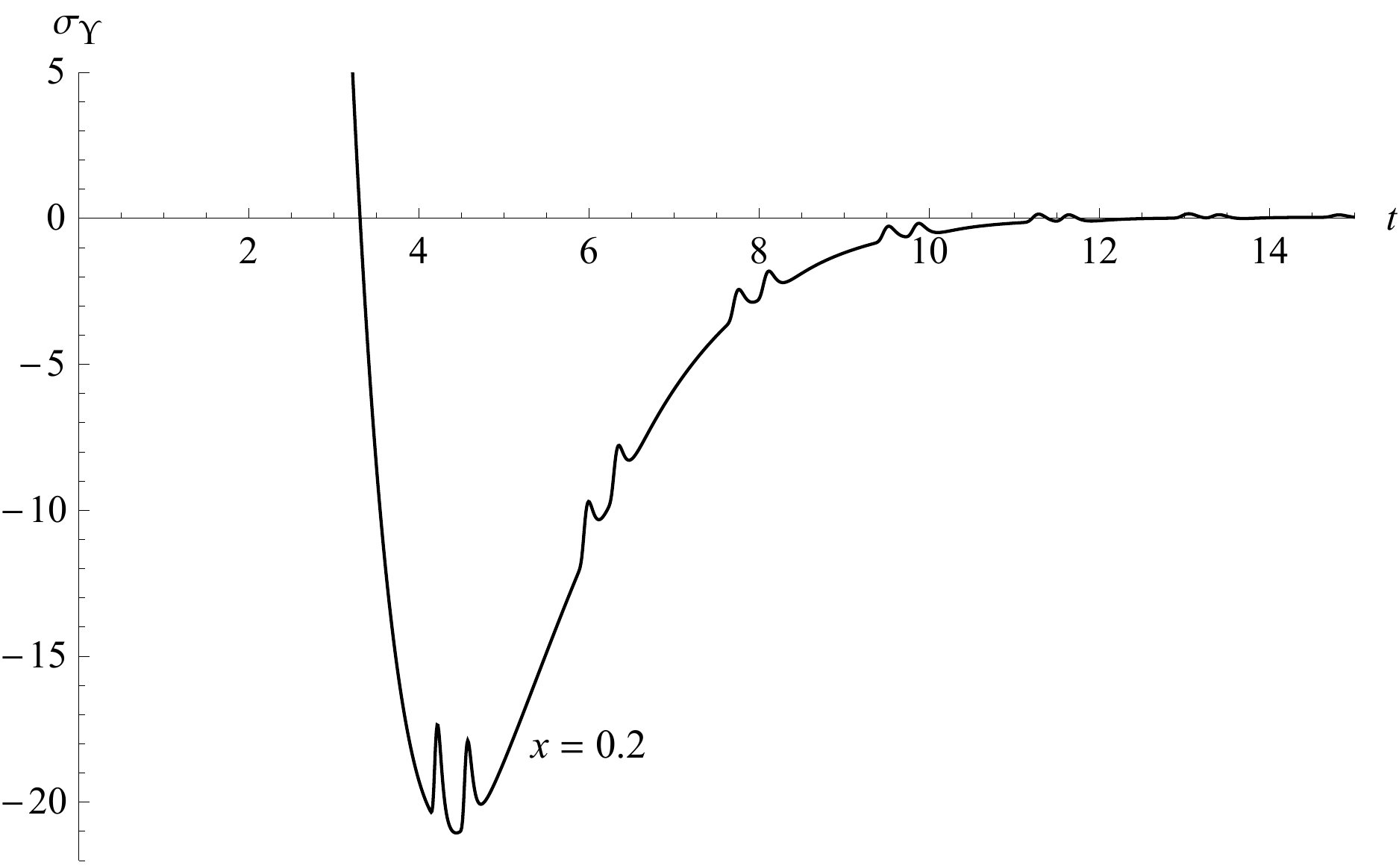}
				\label{sr-m5-sigma1-kompl-nule}}
		\end{minipage}
\hfill 
\begin{minipage}{0.46\columnwidth}
				\subfloat[]{
				\includegraphics[width=\columnwidth]{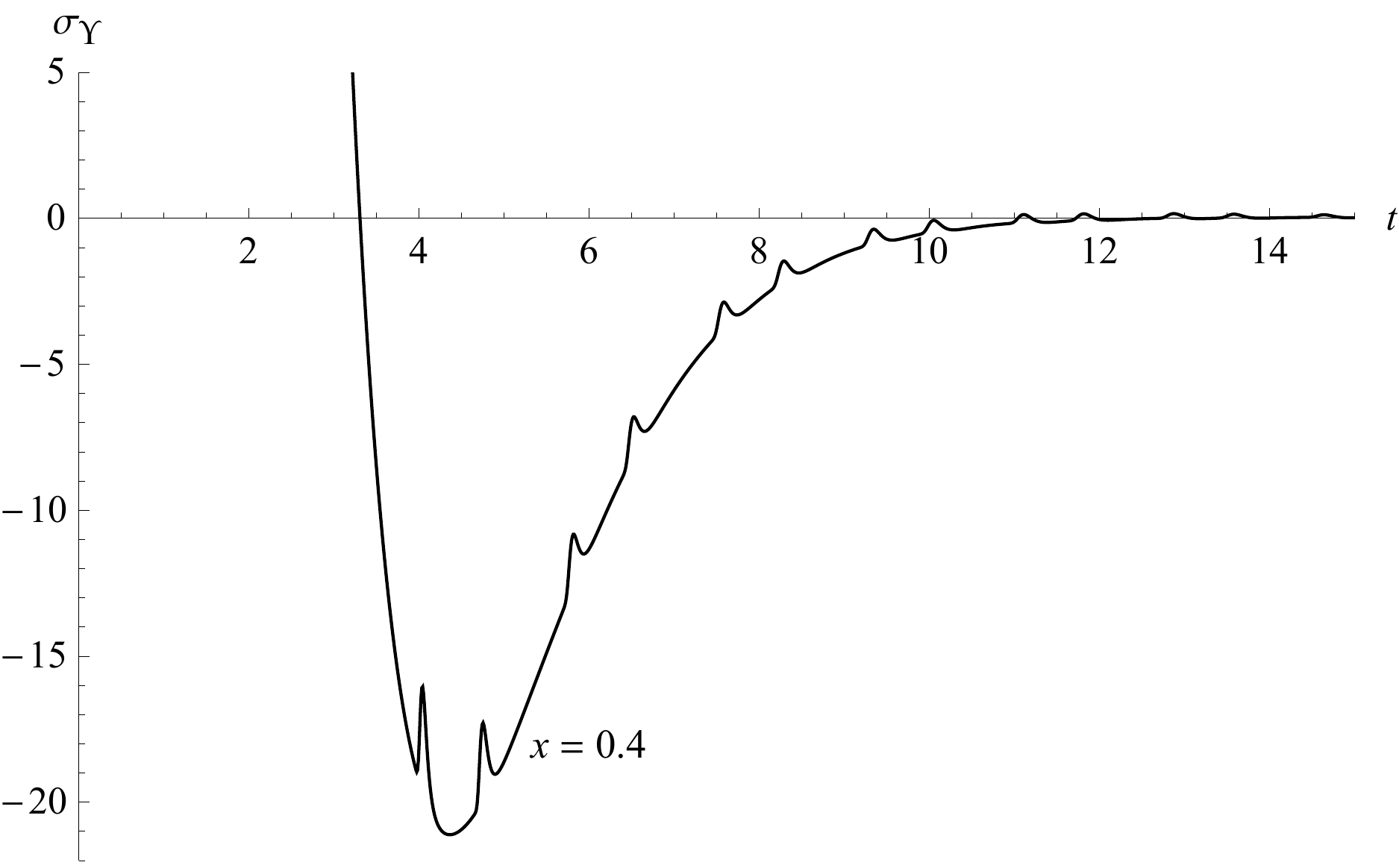}
				\label{sr-m5-sigma2-kompl-nule}}
		\end{minipage}
\vfil
\begin{minipage}{0.46\columnwidth}
  \subfloat[]{
   \includegraphics[width=\columnwidth]{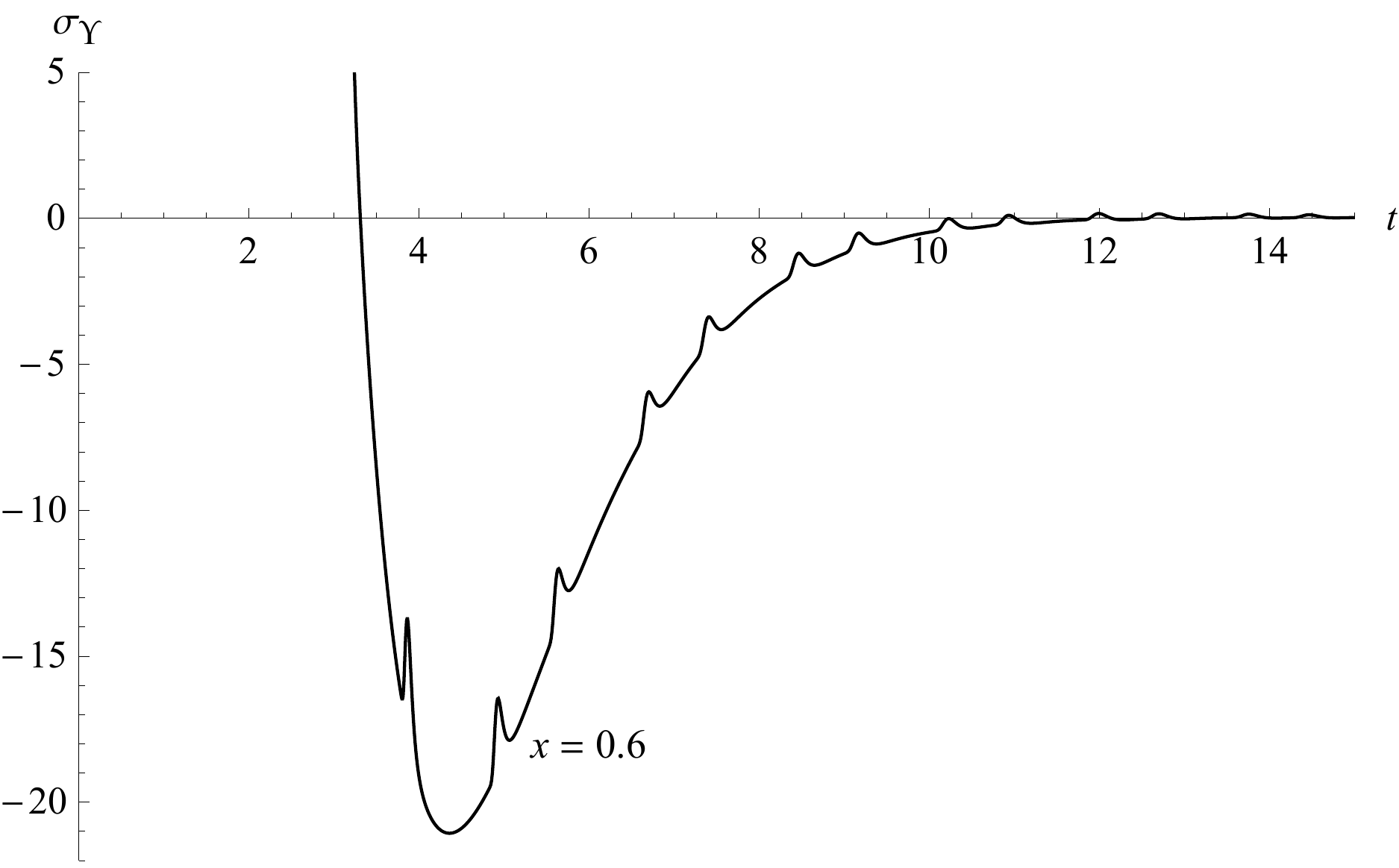}
   \label{sr-m5-sigma3-kompl-nule}}
\end{minipage}
\hfil
\begin{minipage}{0.46\columnwidth}
  \subfloat[]{
   \includegraphics[width=\columnwidth]{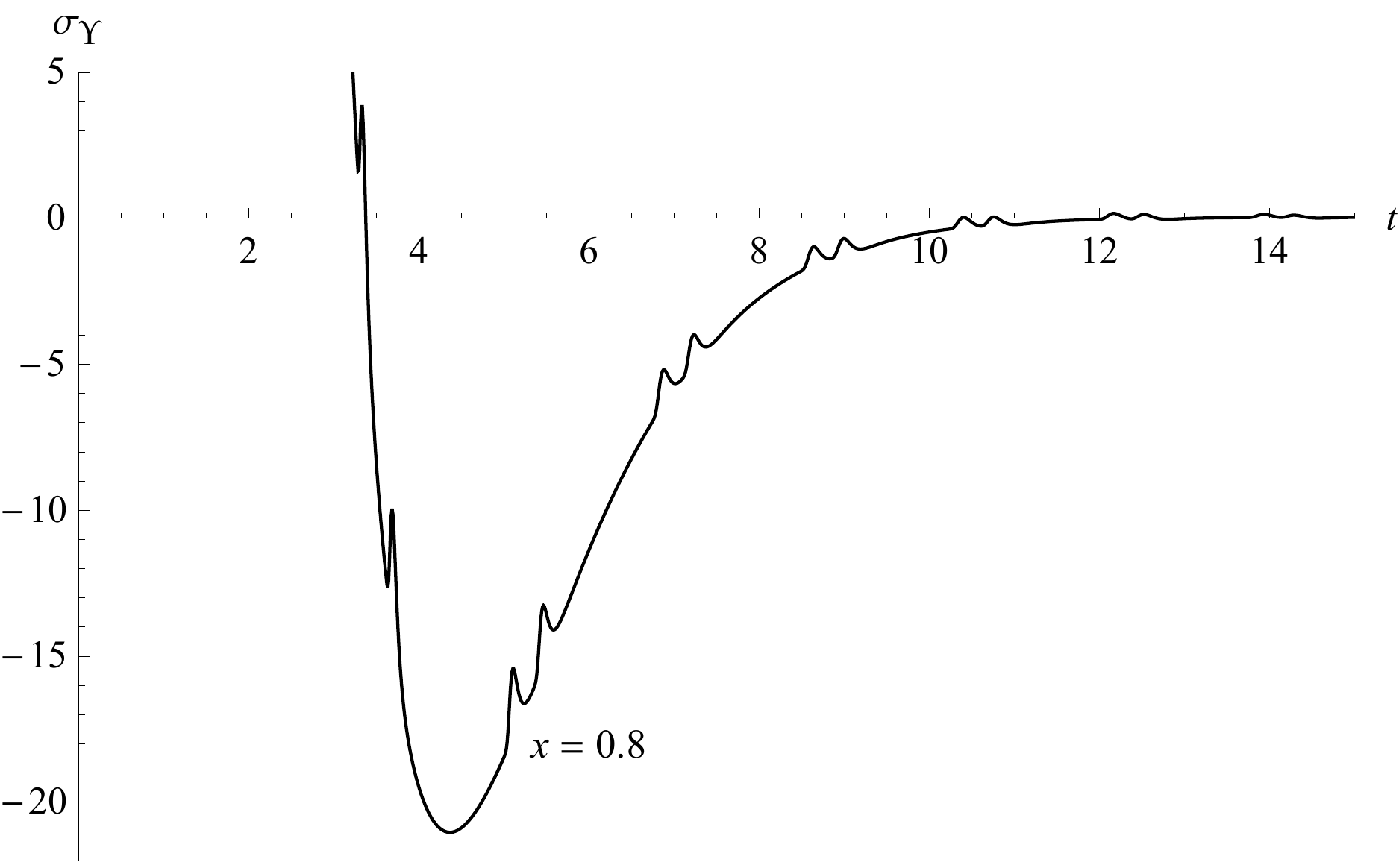}
   \label{sr-m5-sigma4-kompl-nule}}
\end{minipage}
\end{center}
\caption{Stress in a rod when the displacement of its free end is assumed as
the Heaviside function, i.e., $\Upsilon =H$, for Model V in the case of
complex conjugated branch points.}
\label{sr-sigma-kompl-nule}
\end{figure}

Figures \ref{sr-sigma} and \ref{sr-sigma-kompl-nule} present time profiles
displaying stress at several points of the rod in the case of displacement
of rod's free end taken in the form of Heaviside step function, i.e., for $%
\Upsilon =H$ as the boundary condition (\ref{bc-bd})$_{2}$, so that, by (\ref%
{sigma-ipsilon-er}), one has 
\begin{equation}
\sigma _{\Upsilon }\left( x,t\right) =H\left( t\right) \ast R\left(
x,t\right) .  \label{sigma-ipsilon}
\end{equation}

In the case of Model V, as clearly visible from Figures \ref{sr-m5-sigma-1}
- \ref{sr-m5-sigma-3}, step responses display damped oscillatory character,
taking even negative values, with the pronounced first peak whose amplitude
increases as positions are closer to rod's free end, which is expected since
the free end is subject to a sudden displacement. Note the good agreement
between curves obtained through analytical expression and through ab initio
numerical Laplace transform inversion. Time profiles in the case of Model
VII resemble to a sequence of relaxation processes interrupted by sudden
jumps decreasing in amplitude as time increases, see Figures \ref%
{sr-m7-sigma-1} and \ref{sr-m7-sigma-2}. Models V and VII differ in step
responses regarding the short-time asymptotics, since in the case of Model V
stress continuously increase from zero with significant rise depending on
point's position, as predicted by (\ref{er-te-tezi-nula}), while in the case
of Model VII stress behaves as the Dirac delta distribution for small time,
since by (\ref{er-eps-te-tezi-nula}) and (\ref{sigma-ipsilon}), as$\
t\rightarrow 0$ one has 
\begin{eqnarray*}
\sigma _{\Upsilon ,\varepsilon }\left( x,t\right) &\sim &\sqrt{\frac{b}{a_{3}%
}}\int_{0}^{t}\frac{\mathrm{d}}{\mathrm{d}t^{\prime }}\left( \frac{%
\varepsilon }{2\tau \sqrt{\pi \tau }}\mathrm{e}^{-\frac{\varepsilon ^{2}}{%
4\tau }}\right) _{\tau =t^{\prime }-\sqrt{\frac{a_{3}}{b}}\left( 1-x\right) }%
\mathrm{d}t^{\prime } \\
&\sim &\sqrt{\frac{b}{a_{3}}}\left( \frac{\varepsilon }{2\tau \sqrt{\pi \tau 
}}\mathrm{e}^{-\frac{\varepsilon ^{2}}{4\tau }}\right) _{\tau =t-\sqrt{\frac{%
a_{3}}{b}}\left( 1-x\right) } \\
&\sim &\sqrt{\frac{b}{a_{3}}}\delta \left( t-\sqrt{\frac{a_{3}}{b}}\left(
1-x\right) \right) ,\;\;\text{as}\;\;\varepsilon \rightarrow 0.
\end{eqnarray*}%
Considering the large-time asymptotic of the step response, one starts from
the Laplace transform of (\ref{sigma-ipsilon}), with the solution kernel
image $\tilde{R}$ given by (\ref{er-tilda})$_{1},$ so that%
\begin{eqnarray}
&&\tilde{\sigma}_{\Upsilon }\left( x,s\right) =s^{\frac{\xi }{2}}\frac{%
\left( 1+xs^{1-\frac{\xi }{2}}+\ldots \right) +\left( 1-xs^{1-\frac{\xi }{2}%
}+\ldots \right) }{\left( 1+s^{1-\frac{\xi }{2}}+\ldots \right) -\left(
1-s^{1-\frac{\xi }{2}}+\ldots \right) }\sim \frac{1}{s^{1-\xi }},\;\;\text{as%
}\;\;s\rightarrow 0,\;\;\text{implying}  \notag \\
&&\sigma _{\Upsilon }\left( x,t\right) \sim \frac{t^{-\xi }}{\Gamma \left(
1-\xi \right) }\rightarrow 0,\;\;\text{as}\;\;t\rightarrow \infty ,
\label{sigma-ipsilon-besk}
\end{eqnarray}%
because of the asymptotics of complex modulus $\tilde{G},$ given by (\ref%
{ge-tilde-nula}). Note, the large-time asymptotics of the step response is
exactly the same as for the relaxation modulus considered for constitutive
equation solely, see Table 2 in \cite{OZ-2}. 

Contrary to the time profiles from Figure \ref{sr-sigma}, that correspond to
solution kernel image $\tilde{R}$ having no other branch points than $s=0,$
responses presented in Figure \ref{sr-sigma-kompl-nule} correspond to the
case when solution kernel image has a pair of complex conjugated branch
points with negative real part. Plots are produced for parameters given in
Table \ref{tbl}, with regularization parameter $\varepsilon=0.25$. One
notices that step response curves from Figure \ref{sr-sigma-kompl-nule} are
superpositions of curves similar to the ones from Figures \ref{sr-m5-sigma-1}
- \ref{sr-m5-sigma-3}, whose shape does not depend on point's position, and
sequences of peaks whose position depends on point's position. For small
time, responses also resemble to the ones from Figures \ref{sr-m5-sigma-1} - %
\ref{sr-m5-sigma-3}, however they are not displayed in Figure \ref%
{sr-sigma-kompl-nule} due to the large value of their amplitudes.

\subsection{Solution for prescribed stress at rod's free end}

In the case of prescribed stress acting on rod's free end, the Laplace
transforms of displacement and stress, given by (\ref{u preko c}) and (\ref%
{sigma preko c}), are obtained as 
\begin{equation}
\tilde{u}\left( x,s\right) =\tilde{\Sigma}\left( s\right) \frac{1}{s\sqrt{%
\tilde{G}\left( s\right) }}\frac{\sinh \frac{xs}{\sqrt{\tilde{G}\left(
s\right) }}}{\cosh \frac{s}{\sqrt{\tilde{G}\left( s\right) }}},\;\;\text{with%
}\;\;\tilde{\sigma}\left( x,s\right) =\tilde{\Sigma}\left( s\right) \frac{%
\cosh \frac{xs}{\sqrt{\tilde{G}\left( s\right) }}}{\cosh \frac{s}{\sqrt{%
\tilde{G}\left( s\right) }}},  \label{u-tilda-sigma-tilda}
\end{equation}%
since the function%
\begin{equation*}
C\left( s\right) =\frac{\tilde{\Sigma}\left( s\right) }{s\sqrt{\tilde{G}%
\left( s\right) }\cosh \frac{s}{\sqrt{\tilde{G}\left( s\right) }}}
\end{equation*}%
is determined from the stress in the Laplace domain (\ref{sigma preko c})
and boundary condition (\ref{bc-bd})$_{2}$.

\subsubsection{Displacement for given $\Sigma $}

Displacement in the Laplace domain, given by (\ref{u-tilda-sigma-tilda})$%
_{1} $, can be expressed through the solution kernel image $\tilde{Q}$,
defined as%
\begin{equation}
\tilde{Q}\left( x,s\right) =\frac{1}{s\sqrt{\tilde{G}\left( s\right) }}\frac{%
\sinh \frac{xs}{\sqrt{\tilde{G}\left( s\right) }}}{\cosh \frac{s}{\sqrt{%
\tilde{G}\left( s\right) }}}.  \label{ku-ku-tilda}
\end{equation}

Considering the asymptotics of solution kernel image $\tilde{Q}$ as $%
s\rightarrow \infty ,$ one obtains%
\begin{equation}
\tilde{Q}\left( x,s\right) =\frac{1}{s\sqrt{\tilde{G}\left( s\right) }}%
\mathrm{e}^{-\left( 1-x\right) \frac{s}{\sqrt{\tilde{G}\left( s\right) }}}%
\frac{1-\mathrm{e}^{-2\frac{xs}{\sqrt{\tilde{G}\left( s\right) }}}}{1+%
\mathrm{e}^{-2\frac{s}{\sqrt{\tilde{G}\left( s\right) }}}}\sim \left\{ 
\begin{tabular}{ll}
$\sqrt{\frac{a_{3}}{b}}\frac{1}{s^{1+\frac{\delta }{2}}}\mathrm{e}^{-\sqrt{%
\frac{a_{3}}{b}}\left( 1-x\right) s^{1-\frac{\delta }{2}}},\smallskip $ & 
for models of the first class, \\ 
$\sqrt{\frac{a_{3}}{b}}\frac{1}{s}\mathrm{e}^{-\sqrt{\frac{a_{3}}{b}}\left(
1-x\right) s},$ & for models of the second class,%
\end{tabular}%
\right.  \label{ku-s-besk}
\end{equation}%
because of the asymptotics of complex modulus $\tilde{G},$ given by (\ref%
{ge-tilde-besk}), so that the short-time asymptotics of solution kernel $Q$
for models of the first class is obtained as 
\begin{equation}
Q\left( x,t\right) \sim -\frac{1}{\pi }\sqrt{\frac{a_{3}}{b}}%
\int_{0}^{\infty }\sin \left( \sqrt{\frac{a_{3}}{b}}\left( 1-x\right) \rho
^{1-\frac{\delta }{2}}\sin \frac{\delta \pi }{2}+\frac{\delta \pi }{2}%
\right) \mathrm{e}^{-\rho t+\sqrt{\frac{a_{3}}{b}}\left( 1-x\right) \rho ^{1-%
\frac{\delta }{2}}\cos \frac{\delta \pi }{2}}\frac{1}{\rho ^{1+\frac{\delta 
}{2}}}\mathrm{d}\rho ,\;\;\text{as}\;\;t\rightarrow 0,
\label{ku-short-time-asympt-m-V}
\end{equation}%
by inverting the Laplace transform of (\ref{ku-s-besk}) using the definition
and integration in the complex plane, while the short-time asymptotics of
solution kernel $Q$ corresponding to models of the second class yields%
\begin{equation}
Q\left( x,t\right) \sim \sqrt{\frac{a_{3}}{b}}H\left( t-\sqrt{\frac{a_{3}}{b}%
}\left( 1-x\right) \right) ,\;\;\text{as}\;\;t\rightarrow 0,
\label{ku-te-tezi-nula}
\end{equation}%
implying that the value of solution kernel $Q$ for small time jumps from
zero to a finite value at the time instant $t=\sqrt{\frac{a_{3}}{b}}\left(
1-x\right) $ depending on the position $x$ and material properties.

On the other hand, the asymptotics of solution kernel image $\tilde{Q}$ as $%
s\rightarrow 0,$ yields%
\begin{equation*}
\tilde{Q}\left( x,s\right) =\frac{1}{s^{1+\frac{\xi }{2}}}\frac{\left(
1+xs^{1-\frac{\xi }{2}}+\ldots \right) -\left( 1-xs^{1-\frac{\xi }{2}%
}+\ldots \right) }{\left( 1+s^{1-\frac{\xi }{2}}+\ldots \right) +\left(
1-s^{1-\frac{\xi }{2}}+\ldots \right) }\sim x\,\frac{1}{s^{\xi }},\;\;\text{%
so that}\;\;\mathcal{L}^{-1}\left[ \tilde{Q}\left( x,s\right) \right] \sim
x\,\frac{t^{-\left( 1-\xi \right) }}{\Gamma \left( \xi \right) },\;\;\text{as%
}\;\;t\rightarrow \infty ,
\end{equation*}%
because of the asymptotics of complex modulus $\tilde{G},$ given by (\ref%
{ge-tilde-nula}), implying that solution kernel $Q$ for large time
asymptotically tends to zero as the power type function, depending on the
position $x.$

The solution kernel%
\begin{align}
& Q\left( x,t\right) =\frac{1}{\pi }\int_{0}^{\infty }\func{Im}\left( \frac{1%
}{\rho \mathrm{e}^{\mathrm{i}\varphi _{0}}\sqrt{\tilde{G}\left( \rho \mathrm{%
e}^{\mathrm{i}\varphi _{0}}\right) }}\frac{\sinh \frac{x\rho \mathrm{e}_{0}^{%
\mathrm{i}\varphi _{0}}}{\sqrt{\tilde{G}\left( \rho \mathrm{e}^{\mathrm{i}%
\varphi _{0}}\right) }}}{\cosh \frac{\rho \mathrm{e}^{\mathrm{i}\varphi _{0}}%
}{\sqrt{\tilde{G}\left( \rho \mathrm{e}^{\mathrm{i}\varphi _{0}}\right) }}}%
\mathrm{e}^{\mathrm{i}\left( \varphi _{0}+\rho t\mathrm{\sin }\varphi
_{0}\right) }\right) \mathrm{e}^{-\rho t\left\vert \mathrm{\cos }\varphi
_{0}\right\vert }\mathrm{d}\rho  \notag \\
& \quad \quad \quad \quad \quad \quad +2\sum_{k=0}^{\infty }\left( -1\right)
^{k}\sin \left( \frac{\left( 2k+1\right) \pi }{2}x\right) \mathrm{e}^{-\rho
_{k}t\left\vert \cos \varphi _{k}\right\vert }\func{Re}\left( \frac{1}{s_{k}}%
\frac{1}{1+\left( \frac{\left( 2k+1\right) \pi }{2}\right) ^{2}\frac{\tilde{G%
}^{\prime }\left( s_{k}\right) }{2s_{k}}}\mathrm{e}^{\mathrm{i}\rho
_{k}t\sin \varphi _{k}}\right)  \label{ku-ku}
\end{align}%
is calculated by the definition of inverse Laplace transform similarly as
the solution kernel $P,$ see Section \ref{kalk-pe}, using the Cauchy
residues theorem, since complex valued function $\tilde{Q}$ has infinite
number of pairs of complex conjugated poles $s_{k}$ and $\bar{s}_{k},$ for
each $k\in 
\mathbb{N}
_{0},$ lying in the left complex half-plane, each of them being poles of the
first order, that are obtained as zeros of the denominator of function $%
\tilde{Q}$, i.e., as solutions of the equation%
\begin{equation}
\cosh \frac{s}{\sqrt{\tilde{G}\left( s\right) }}=0\;\;\text{implying}\;\;%
\frac{s}{\sqrt{\tilde{G}\left( s\right) }}=-\mathrm{i}\frac{\left(
2k+1\right) \pi }{2},\;\;\text{i.e.,}\;\;\frac{s^{2}}{\tilde{G}\left(
s\right) }+\left( \frac{\left( 2k+1\right) \pi }{2}\right)
^{2}=0,\;\;k=0,\pm 1,\pm 2,...,  \label{kosinus}
\end{equation}%
as proved in Section \ref{polovi}. As in the case of function $\tilde{P},$
function $\tilde{Q}$ may also has branch points other than $s=0,$ due to the
square root of function $\tilde{G}.$ The form of solution kernel $Q,$ given
by (\ref{ku-ku}), corresponds to the case of a pair of complex conjugated
branch points $s_{0}=\rho _{0}\mathrm{e}^{\mathrm{i}\varphi _{0}}$ and $\bar{%
s}_{0},$ while its form corresponding to cases of no branch points or one
negative real branch point is obtained by putting $\varphi _{0}=\pi $ in (%
\ref{ku-ku}).

Having the solution kernel calculated by (\ref{ku-ku}), the displacement in
the case of prescribed stress of rod's free end is%
\begin{equation}
u\left( x,t\right) =\Sigma \left( t\right) \ast Q\left( x,t\right) ,
\label{u-sigma-ku}
\end{equation}%
by the inverse Laplace transform of (\ref{u-tilda-sigma-tilda})$_{1}$, with
the solution kernel image $\tilde{Q}$ defined by (\ref{ku-ku-tilda}).

\subsubsection{Stress for given $\Sigma $}

Stress in the Laplace domain, given by (\ref{u-tilda-sigma-tilda})$_{2}$,
can be expressed either through the solution kernel image $\tilde{S}$ in the
case of Burgers models of the first class, or through the regularized
solution kernel $\tilde{S}_{\mathrm{reg}}$ in the case of the second class
models, that are defined by%
\begin{equation}
\tilde{S}\left( x,s\right) =\frac{\cosh \frac{xs}{\sqrt{\tilde{G}\left(
s\right) }}}{\cosh \frac{s}{\sqrt{\tilde{G}\left( s\right) }}}\;\;\text{and}%
\;\;\tilde{S}_{\mathrm{reg}}\left( x,s\right) =\frac{1}{s}\tilde{S}\left(
x,s\right) .  \label{es-tilda}
\end{equation}

Considering the asymptotics of solution kernel image $\tilde{S}$ and its
regularized version $\tilde{S}_{\mathrm{reg}}$ as $s\rightarrow \infty ,$ by
the asymptotics of complex modulus $\tilde{G},$ given by (\ref{ge-tilde-besk}%
), one obtains%
\begin{equation*}
\tilde{S}\left( x,s\right) =\mathrm{e}^{-\left( 1-x\right) \frac{s}{\sqrt{%
\tilde{G}\left( s\right) }}}\frac{1+\mathrm{e}^{-2\frac{xs}{\sqrt{\tilde{G}%
\left( s\right) }}}}{1+\mathrm{e}^{-2\frac{s}{\sqrt{\tilde{G}\left( s\right) 
}}}}\sim \mathrm{e}^{-\sqrt{\frac{a_{3}}{b}}\left( 1-x\right) s^{1-\frac{%
\delta }{2}}}\;\;\text{and}\;\;\tilde{S}_{\mathrm{reg}}\left( x,s\right)
\sim \frac{1}{s}\mathrm{e}^{-\sqrt{\frac{a_{3}}{b}}\left( 1-x\right) s},
\end{equation*}%
which has exactly the same form as the asymptotics of solution kernel image $%
\tilde{P}$ and its regularized version $\tilde{P}_{\mathrm{reg}},$ see (\ref%
{pe-s-besk}), so that the short-time asymptotics of solution kernel $S$ for
models of the first class is given by (\ref{pe-te(zi)-nula}), while the
asymptotics of $S_{\mathrm{reg}},$ corresponding models of the second class,
is given by (\ref{pe-reg-te(zi)-nula}), i.e., by%
\begin{eqnarray}
S\left( x,t\right) &\sim &\frac{1}{\pi }\int_{0}^{\infty }\sin \left( \sqrt{%
\frac{a_{3}}{b}}\left( 1-x\right) \rho ^{1-\frac{\delta }{2}}\sin \frac{%
\delta \pi }{2}\right) \mathrm{e}^{-\rho t+\sqrt{\frac{a_{3}}{b}}\left(
1-x\right) \rho ^{1-\frac{\delta }{2}}\cos \frac{\delta \pi }{2}}\mathrm{d}%
\rho ,\;\;\text{as}\;\;t\rightarrow 0,  \label{es-te(zi)-nula} \\
S_{\mathrm{reg}}\left( x,t\right) &\sim &H\left( t-\sqrt{\frac{a_{3}}{b}}%
\left( 1-x\right) \right) ,\;\;\text{as}\;\;t\rightarrow 0.
\label{es-reg-te(zi)-nula}
\end{eqnarray}

On the other hand, the asymptotics of regularized solution kernel image $%
\tilde{S}_{\mathrm{reg}}$ as $s\rightarrow 0,$ yields%
\begin{equation}
\tilde{S}_{\mathrm{reg}}\left( x,s\right) =\frac{1}{s}\frac{\left( 1+xs^{1-%
\frac{\xi }{2}}+\ldots \right) +\left( 1-xs^{1-\frac{\xi }{2}}+\ldots
\right) }{\left( 1+s^{1-\frac{\xi }{2}}+\ldots \right) +\left( 1-s^{1-\frac{%
\xi }{2}}+\ldots \right) }\sim \frac{1}{s},\;\;\text{implying}\;\;S_{\mathrm{%
reg}}\left( x,t\right) \sim H\left( t\right) =1,\;\;\text{as}%
\;\;t\rightarrow \infty ,  \label{es-reg-te(zi)-besk}
\end{equation}%
because of the asymptotics of complex modulus $\tilde{G},$ given by (\ref%
{ge-tilde-nula}).

Using the regularized solution kernel $S_{\mathrm{reg}},$ the solution
kernel reads%
\begin{eqnarray}
S(x,t) &=&\frac{\partial }{\partial t}\left( S_{\mathrm{reg}}\left(
x,t\right) \,H\left( t-\sqrt{\frac{a_{3}}{b}}\left( 1-x\right) \right)
\right)  \notag \\
&=&\frac{\partial }{\partial t}S_{\mathrm{reg}}\left( x,t\right) \,H\left( t-%
\sqrt{\frac{a_{3}}{b}}\left( 1-x\right) \right) +S_{\mathrm{reg}}\left(
x,t\right) \,\delta \left( t-\sqrt{\frac{a_{3}}{b}}\left( 1-x\right) \right)
,  \label{s-reg}
\end{eqnarray}%
similarly as in the case of solution kernel $P$ expressed through its
regularization $P_{\mathrm{reg}},$ see (\ref{pe-reg}).

The stress in the case of prescribed stress of rod's free end is obtained as%
\begin{equation}
\sigma \left( x,t\right) =\Sigma \left( t\right) \ast S\left( x,t\right) ,
\label{sigma-sigma-es}
\end{equation}
by the inverse Laplace transform of (\ref{u-tilda-sigma-tilda}), with
solution kernel image $\tilde{S}$ defined by (\ref{es-tilda}), where the
solution kernel $S$ takes the form%
\begin{align}
& S\left( x,t\right) =\frac{1}{\pi }\int_{0}^{\infty }\func{Im}\left( \frac{%
\cosh \frac{x\rho \mathrm{e}^{\mathrm{i}\varphi _{0}}}{\sqrt{\tilde{G}\left(
\rho \mathrm{e}^{\mathrm{i}\varphi _{0}}\right) }}}{\cosh \frac{\rho \mathrm{%
e}^{\mathrm{i}\varphi _{0}}}{\sqrt{\tilde{G}\left( \rho \mathrm{e}^{\mathrm{i%
}\varphi _{0}}\right) }}}\mathrm{e}^{\mathrm{i}\left( \varphi _{0}+\rho t%
\mathrm{\sin }\varphi _{0}\right) }\right) \mathrm{e}^{-\rho t\left\vert 
\mathrm{\cos }\varphi _{0}\right\vert }\mathrm{d}\rho  \notag \\
& \quad \quad \quad \quad \quad \quad +2\sum_{k=0}^{\infty }\left( -1\right)
^{k+1}\frac{\cos \left( \frac{\left( 2k+1\right) \pi }{2}x\right) }{\frac{%
\left( 2k+1\right) \pi }{2}}\mathrm{e}^{-\rho _{k}t\left\vert \cos \varphi
_{k}\right\vert }\func{Re}\left( \frac{s_{k}}{1+\left( \frac{\left(
2k+1\right) \pi }{2}\right) ^{2}\frac{\tilde{G}^{\prime }\left( s_{k}\right) 
}{2s_{k}}}\mathrm{e}^{\mathrm{i}\rho _{k}t\sin \varphi _{k}}\right) ,
\label{es}
\end{align}%
in the case of models belonging to the first class, while in the case of the
second model class, it is given by (\ref{s-reg}), with the regularized
solution kernel $S_{\mathrm{reg}}$ being of the form%
\begin{align}
& S_{\mathrm{reg}}\left( x,t\right) =\frac{\varphi _{0}}{\pi }+\frac{1}{\pi }%
\int_{0}^{\infty }\func{Im}\left( \frac{1}{\rho \mathrm{e}^{\mathrm{i}%
\varphi _{0}}}\frac{\cosh \frac{x\rho \mathrm{e}^{\mathrm{i}\varphi _{0}}}{%
\sqrt{\tilde{G}\left( \rho \mathrm{e}^{\mathrm{i}\varphi _{0}}\right) }}}{%
\cosh \frac{\rho \mathrm{e}^{\mathrm{i}\varphi _{0}}}{\sqrt{\tilde{G}\left(
\rho \mathrm{e}^{\mathrm{i}\varphi _{0}}\right) }}}\mathrm{e}^{\mathrm{i}%
\left( \varphi _{0}+\rho t\mathrm{\sin }\varphi _{0}\right) }\right) \mathrm{%
e}^{-\rho t\left\vert \mathrm{\cos }\varphi _{0}\right\vert }\mathrm{d}\rho 
\notag \\
& \quad \quad \quad \quad \quad \quad +2\sum_{k=0}^{\infty }\left( -1\right)
^{k+1}\frac{\cos \left( \frac{\left( 2k+1\right) \pi }{2}x\right) }{\frac{%
\left( 2k+1\right) \pi }{2}}\mathrm{e}^{-\rho _{k}t\left\vert \cos \varphi
_{k}\right\vert }\func{Re}\left( \frac{1}{1+\left( \frac{\left( 2k+1\right)
\pi }{2}\right) ^{2}\frac{\tilde{G}^{\prime }\left( s_{k}\right) }{2s_{k}}}%
\mathrm{e}^{\mathrm{i}\rho _{k}t\sin \varphi _{k}}\right) .  \label{es-reg}
\end{align}%
Solution kernel and its regularized version are obtained using the
definition of the inverse Laplace transform, similarly as done in Section %
\ref{kalk-pe} when calculating the solution kernel $P$. The form of solution
kernels $S$ and $S_{\mathrm{reg}}$ correspond to the case when corresponding
solution kernel image has a pair of complex conjugate branch points $%
s_{0}=\rho _{0}\mathrm{e}^{\mathrm{i}\varphi _{0}}$ and $\bar{s}_{0}$ in
addition to $s=0,$ while by putting $\varphi _{0}=\pi $ in (\ref{es}) and (%
\ref{es-reg}) one obtains their forms in cases when the image function has
either no branch points or has one negative real branch point.

\subsubsection{Numerical examples}

\begin{figure}[h]
\begin{center}
\begin{minipage}{0.46\columnwidth}
  \subfloat[Case of Model V.]{
   \includegraphics[width=\columnwidth]{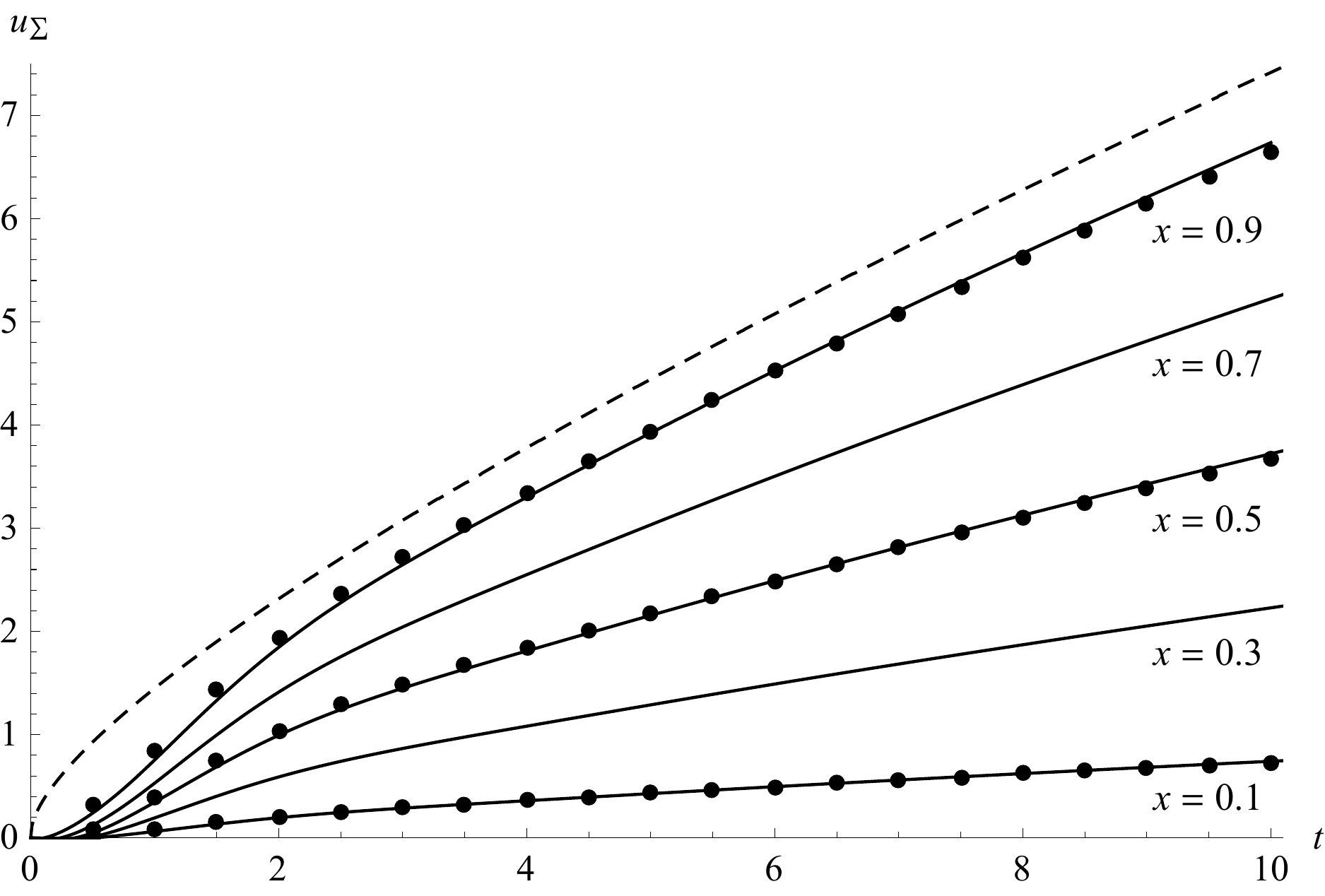}
   \label{cr-m5-u}}
  \end{minipage}
\hfil
\begin{minipage}{0.46\columnwidth}
  \subfloat[Case of Model VII.]{
   \includegraphics[width=\columnwidth]{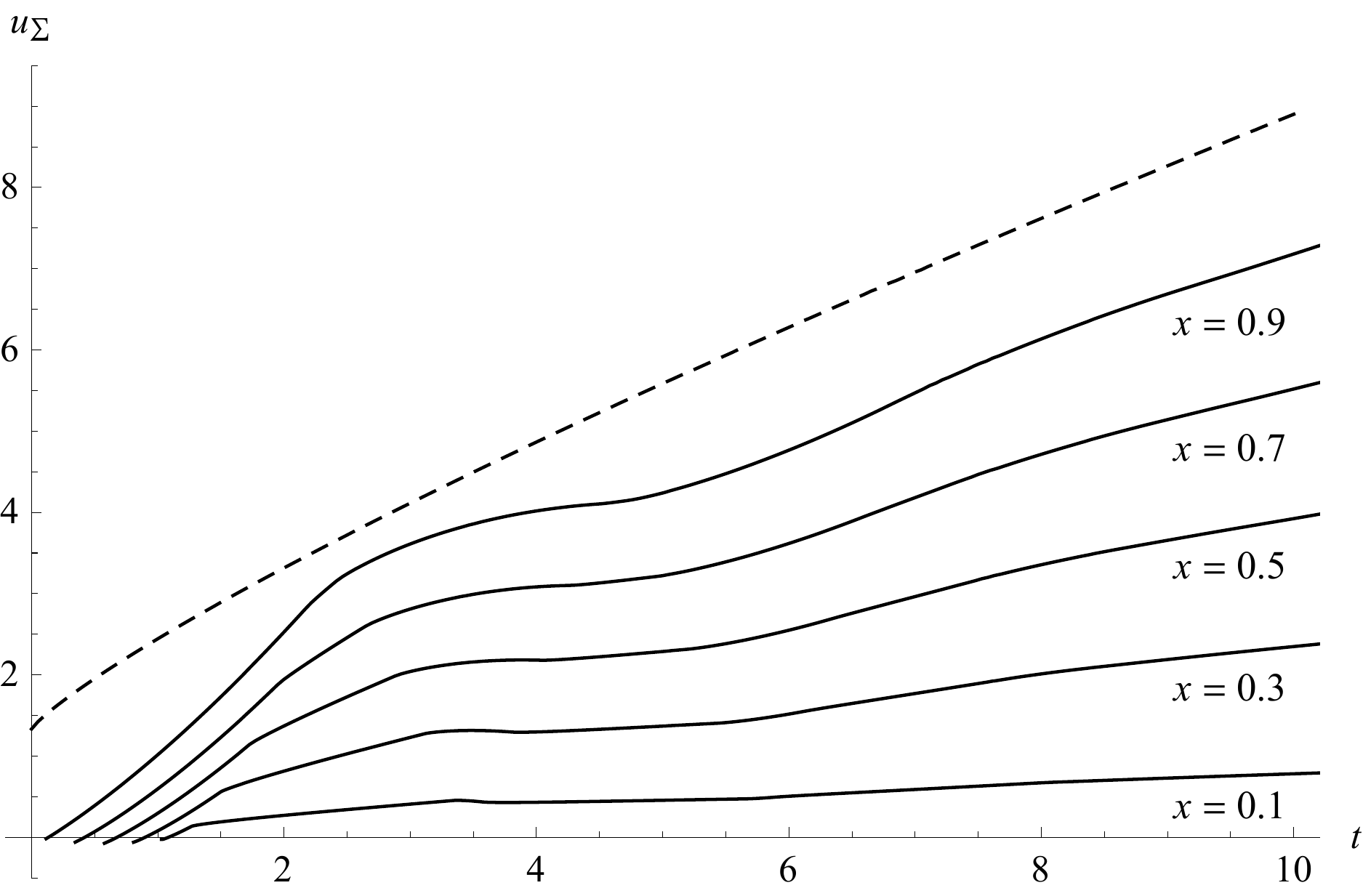}
   \label{cr-m7-u}}
  \end{minipage}
\end{center}
\caption{Displacement of a rod when the stress applied on its free end is
assumed as the Heaviside function, i.e., $\Sigma=H$, obtained according to
analytical expression (lines) and by numerical Laplace transform inversion
(dots), as well as creep curve corresponding to the constitutive equation
(dashed line). }
\label{cr-u}
\end{figure}

\begin{figure}[h]
\begin{center}
\includegraphics[width=0.6\columnwidth]{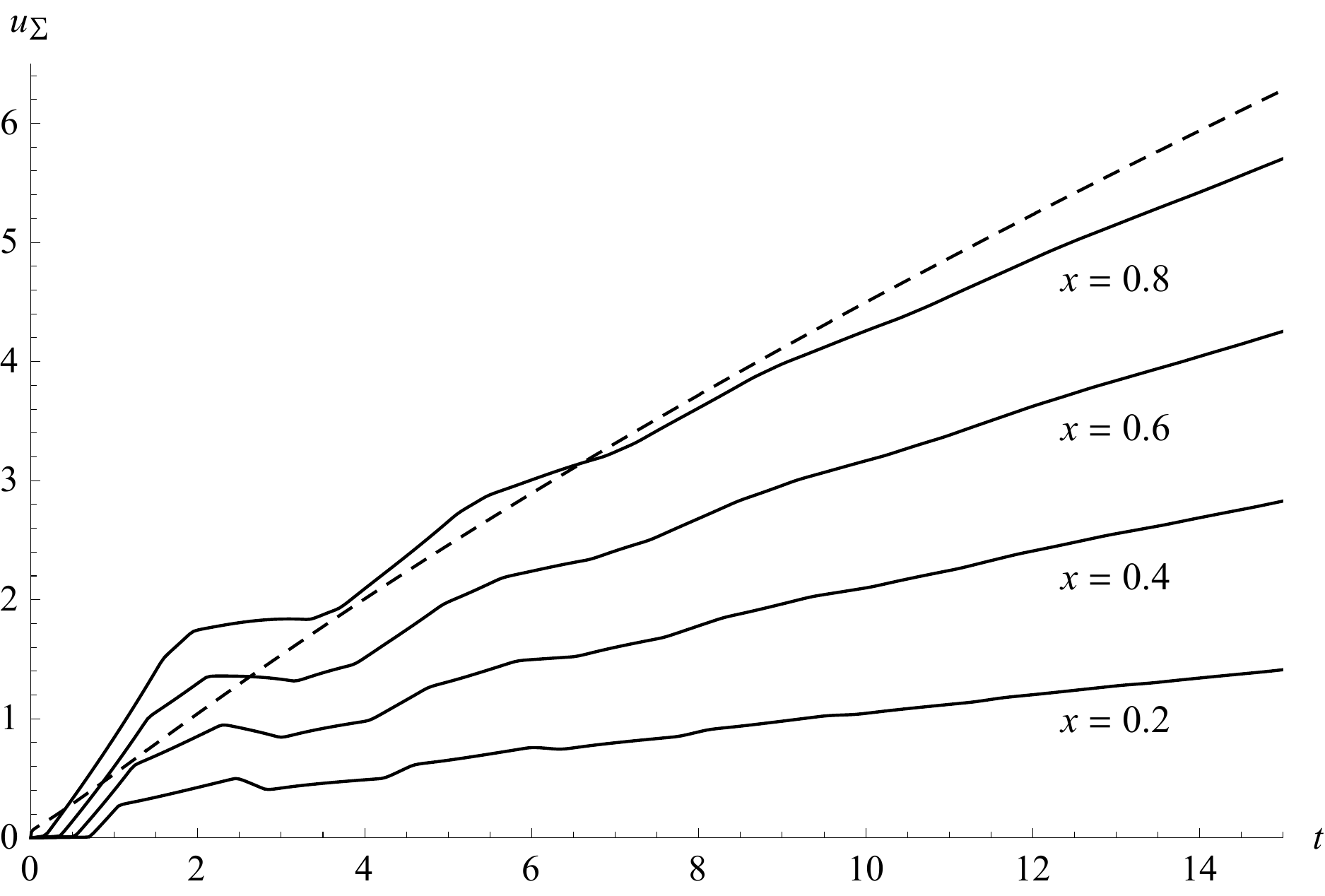}
\end{center}
\caption{Displacement of a rod when the stress applied on its free end is
assumed as the Heaviside function, i.e., $\Sigma=H$, for Model V in the case
of complex conjugated branch points, as well as creep curve corresponding to
the constitutive equation (dashed line). }
\label{cr-m5-u-kompl-nule}
\end{figure}

Figures \ref{cr-u} and \ref{cr-m5-u-kompl-nule} present time profiles of
displacement of several points of the rod for stress applied to rod's free
end assumed as the Heaviside step function, i.e., for boundary condition (%
\ref{bc-bd})$_{3}$ taken as $\Sigma =H,$ so that by (\ref{u-sigma-ku}), one
has 
\begin{equation}
u_{\Sigma }\left( x,t\right) =H\left( t\right) \ast Q\left( x,t\right) .
\label{u-sigma}
\end{equation}

The step response can be considered as a superposition of monotonically
increasing curve and oscillations having amplitudes decreasing in time, that
are quite pronounced in the case of the Model VII, see Figure \ref{cr-m7-u},
while in the case of the Model V the oscillations cannot really be noticed,
presumable due to large damping, see Figure \ref{cr-m5-u}. Note the good
agreement between curves obtained by analytical expression (\ref{u-sigma})
and by the numerical Laplace transform inversion. The displacement tends to
infinity for large time depending on point's position, since the large-time
asymptotics%
\begin{equation}
u_{\Sigma }\left( x,t\right) \sim x\frac{t^{\xi }}{\Gamma \left( 1+\xi
\right) }\rightarrow \infty ,\;\;\text{as}\;\;t\rightarrow \infty ,
\label{u-sigma-asimpt}
\end{equation}%
is obtained from the Laplace transform of (\ref{u-sigma}), with the solution
kernel image $\tilde{Q}$ given by (\ref{ku-ku-tilda}), yielding%
\begin{equation*}
\tilde{u}_{\Sigma }\left( x,s\right) =\frac{1}{s^{2+\frac{\xi }{2}}}\frac{%
\left( 1+xs^{1-\frac{\xi }{2}}+\ldots \right) -\left( 1-xs^{1-\frac{\xi }{2}%
}+\ldots \right) }{\left( 1+s^{1-\frac{\xi }{2}}+\ldots \right) +\left(
1-s^{1-\frac{\xi }{2}}+\ldots \right) }\sim x\frac{1}{s^{1+\xi }},\;\;\text{%
as}\;\;s\rightarrow 0,
\end{equation*}%
thanks to asymptotics of complex modulus $\tilde{G}$, given by (\ref%
{ge-tilde-nula}). The large-time asymptotics (\ref{u-sigma-asimpt}) implies
that, depending on their position, all points of the rod have the same time
behavior, that is exactly the same as for the creep compliance considered
for constitutive equation, see Table 2 in \cite{OZ-2}. As before, the step
response differs for Burgers models of the first and second class, since
time profiles continuously increase from zero in the case of Model V, see (%
\ref{ku-short-time-asympt-m-V}), while in the case of Model VII, according
to the asymptotics of solution kernel image (\ref{ku-s-besk}) and Laplace
transform of (\ref{u-sigma}), one has 
\begin{eqnarray*}
\tilde{u}_{\Sigma }\left( x,s\right) &\sim &\sqrt{\frac{a_{3}}{b}}\frac{1}{%
s^{2}}\mathrm{e}^{-\sqrt{\frac{a_{3}}{b}}\left( 1-x\right) s},\;\;\text{as}%
\;\;s\rightarrow \infty ,\;\;\text{i.e.,} \\
u_{\Sigma }\left( x,t\right) &\sim &\sqrt{\frac{a_{3}}{b}}\left( t-\sqrt{%
\frac{a_{3}}{b}}\left( 1-x\right) \right) ,\;\;\text{for}\;\;t>\sqrt{\frac{%
a_{3}}{b}}\left( 1-x\right) \;\;\text{as}\;\;t\rightarrow 0,
\end{eqnarray*}%
implying that the short-time asymptotics of displacement has linear trend
starting after $t=\sqrt{\frac{a_{3}}{b}}\left( 1-x\right) .$ In the case
when solution kernel image $\tilde{Q}$ has a pair of complex conjugated
branch points, the step response curves, being non-smooth, are also damped
oscillatory and superposed to monotonically increasing curves, see Figure %
\ref{cr-m5-u-kompl-nule}. As before, plots are produced using the model
parameters from Table \ref{tbl}.

\begin{figure}[p]
\begin{center}
\begin{minipage}{0.55\columnwidth}
  \subfloat[Case of Model V.]{
   \includegraphics[width=\columnwidth]{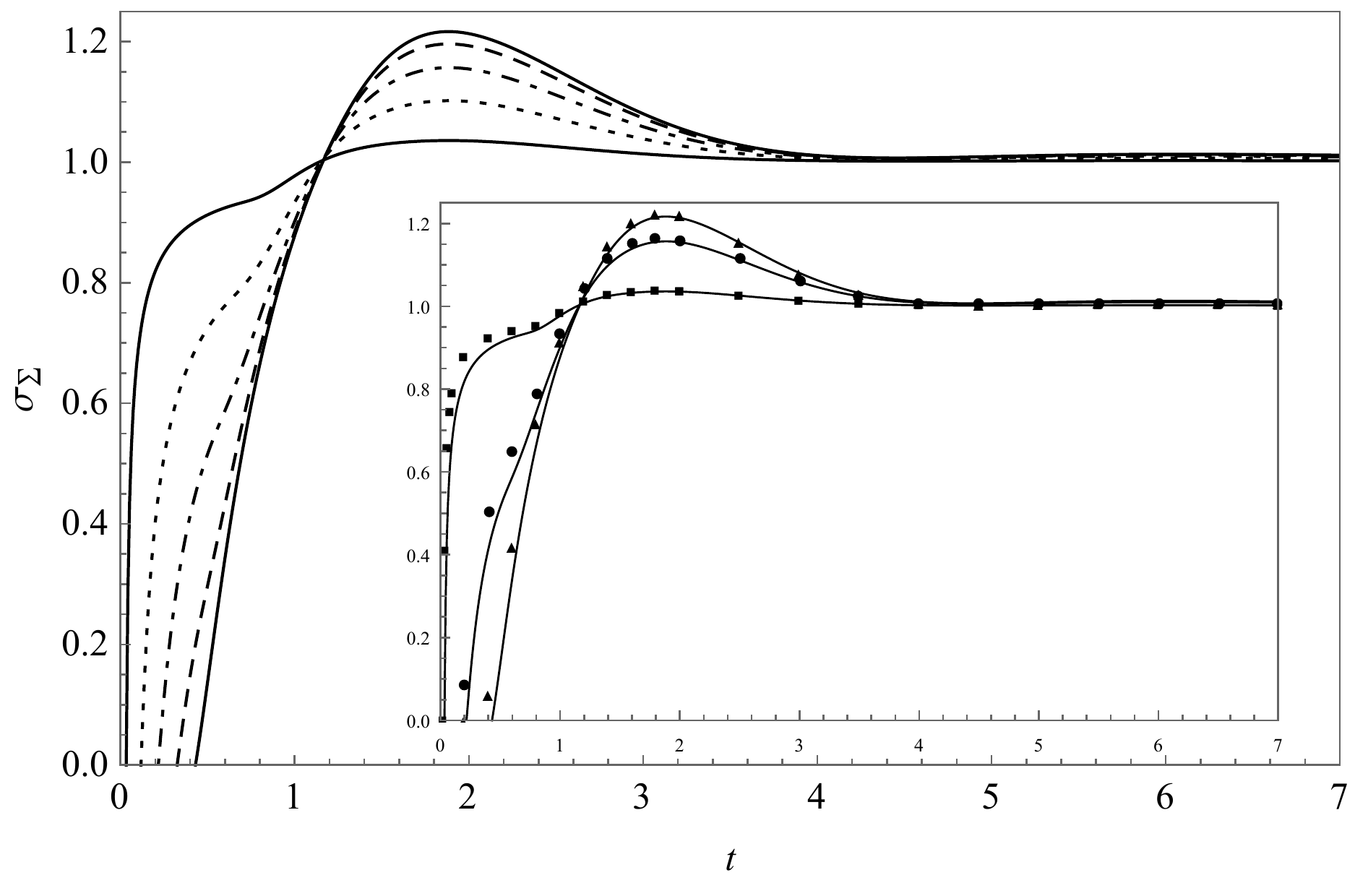}
   \label{cr-m5-sigma}}
  \end{minipage}
\medskip \vfil
\medskip 
\begin{minipage}{0.55\columnwidth}
  \subfloat[Case of Model VII.]{
   \includegraphics[width=\columnwidth]{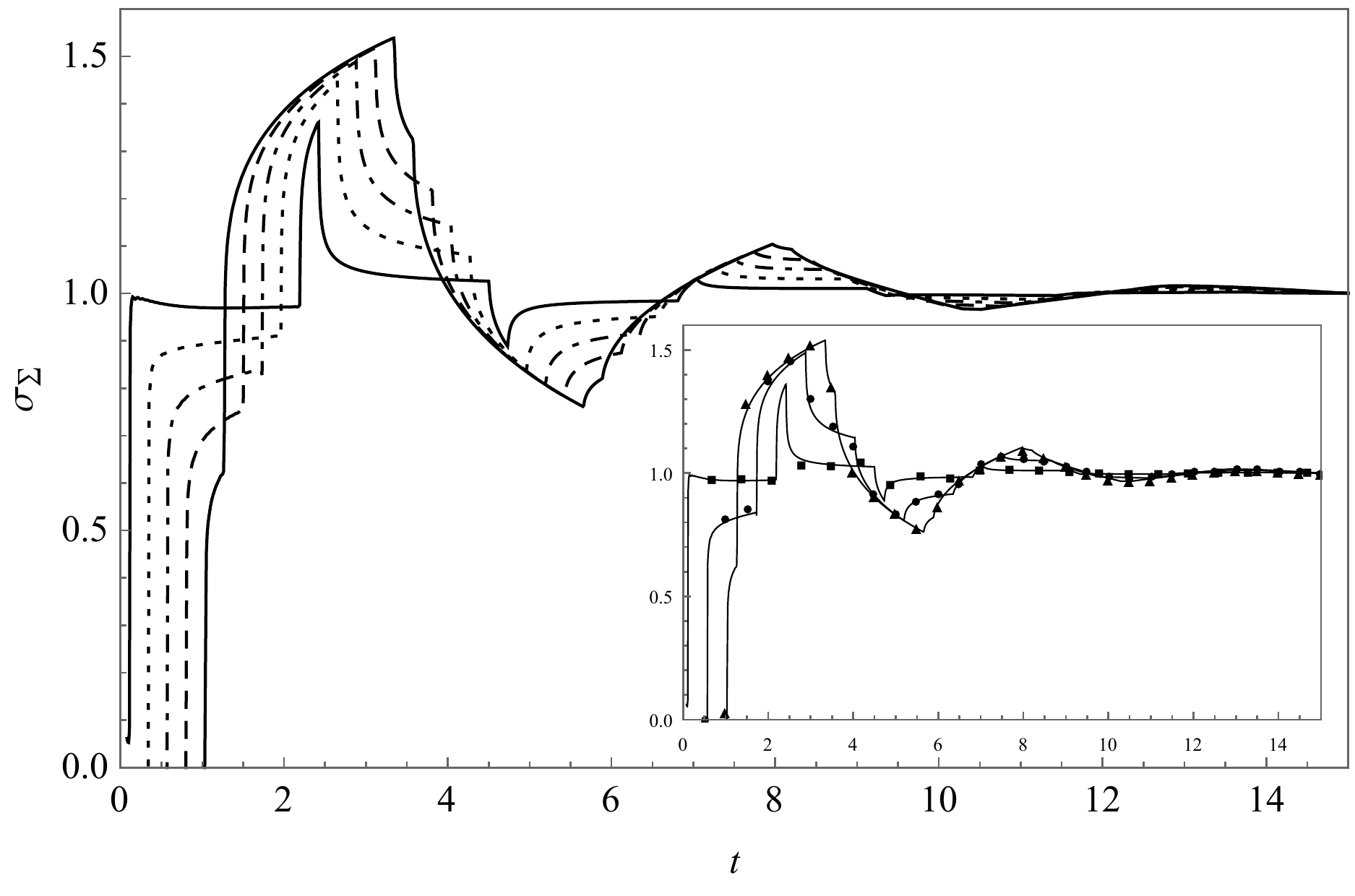}
   \label{cr-m7-sigma}}
  \end{minipage}
\end{center}
\caption{Stress in a rod when the stress applied on its free end is assumed
as the Heaviside function, i.e., $\Sigma=H$, obtained according to
analytical expression and depicted by solid, dashed, dot dashed, dotted, and
solid lines for $x\in\{0.1,0.3,0.5,0.7,0.9\}$ respectively, as well as by
numerical Laplace transform inversion and depicted by triangles, circles,
and squares for $x\in\{0.1,0.5,0.9\}$ respectively.}
\label{cr-sigma}
\end{figure}

\begin{figure}[p]
\begin{center}
\includegraphics[width=0.6\columnwidth]{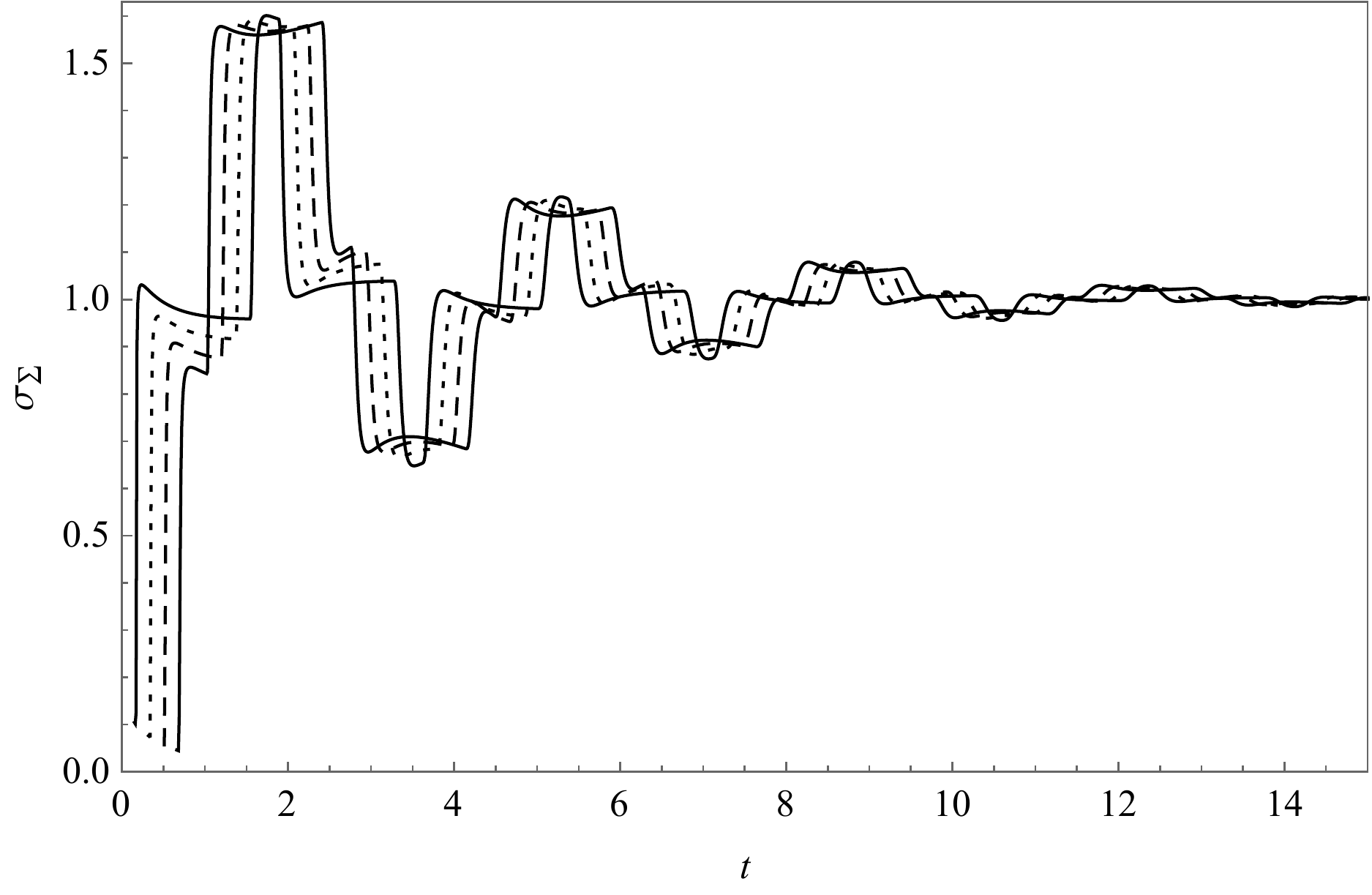}
\end{center}
\caption{Stress in a rod when the stress applied on its free end is assumed
as the Heaviside function, i.e., $\Sigma=H$, for Model V in the case of
complex conjugated branch points. }
\label{cr-m5-sigma-kompl-nule}
\end{figure}

Figures \ref{cr-sigma} and \ref{cr-m5-sigma-kompl-nule} present time
profiles displaying stress at several points of the rod for stress applied
to rod's free end assumed as the Heaviside step function, i.e., for boundary
condition (\ref{bc-bd})$_{3}$ taken as $\Sigma =H.$ The regularized solution
kernel $S_{\mathrm{reg}}$ actually represents the step response, due to
defining relation (\ref{es-tilda})$_{2}$ for regularized solution kernel
image $\tilde{S}_{\mathrm{reg}},$ that yields%
\begin{equation*}
\sigma _{\Sigma }\left( x,t\right) =S_{\mathrm{reg}}\left( x,t\right)
=H\left( t\right) \ast S\left( x,t\right)
\end{equation*}%
after performing the inverse Laplace transform, see also (\ref%
{sigma-sigma-es}).

The behavior of step response is of the damped oscillatory type, that
settles at the value of stress applied to rod's free end, i.e.,%
\begin{equation}
\lim_{t\rightarrow \infty }\sigma _{\Sigma }\left( x,t\right) =1,
\label{sigma-sigma-besk}
\end{equation}%
as predicted by the large-time asymptotics of regularized solution kernel $%
S_{\mathrm{reg}},$ given by (\ref{es-reg-te(zi)-besk}). The time profiles of
step response in the case of Model V display oscillatory behavior with the
very pronounced damping, see Figure \ref{cr-m5-sigma}. On the other hand,
the profiles in the case of Model VII, being also damped oscillatory,
resemble to the sequence of two excitation processes followed by two
relaxation processes, since profiles repeatedly change their convexity from
concave to convex, as clearly visible from Figure \ref{cr-m7-sigma}. Again,
good agreement between curves obtained analytically through (\ref{es-reg}),
using parameters from Table \ref{tbl}, and by ab initio numerical Laplace
transform inversion is observed. Step response differs for Models V and VII
regarding the short-time asymptotics, since in the case of Model V time
profiles continuously increase from zero, obtaining non-zero values
depending on point's position, see (\ref{es-te(zi)-nula}) and Figure \ref%
{cr-m5-sigma}, while for Model VII time profiles jump from zero depending on
point's position, due to the Heaviside function as the short-time
asymptotics, see (\ref{es-reg-te(zi)-nula}) and Figure \ref{cr-m7-sigma}.

Contrary to the case of plots from Figure \ref{cr-sigma}, that correspond to
the case when solution kernel image $\tilde{S}$ and its regularization $%
\tilde{S}_{\mathrm{reg}}$ have no other branch points than $s=0,$ time
profiles from Figure \ref{cr-m5-sigma-kompl-nule} correspond to the case
when kernel image additionally has a pair of complex conjugated branch
points. Time profiles presented in Figure \ref{cr-m5-sigma-kompl-nule} are
peculiarly shaped, as if several damped vibrations with different
frequencies are superposed. Responses seem to have an envelope, that is
typical for damped oscillations. 


\section{Calculation of solution kernel $P$ \label{kalk-pe}}

In order to obtain the solution kernel $P$ in the form (\ref{pe}), the
inverse Laplace transform formula is applied to solution kernel image $%
\tilde{P}$, given by (\ref{pe-tilda})$_{1}$, yielding 
\begin{equation*}
P\left( x,t\right) =\mathcal{L}^{-1}\left[ \tilde{P}\left( x,s\right) \right]
\left( x,t\right) =\frac{1}{2\pi \mathrm{i}}\int_{Br}\tilde{P}\left(
x,s\right) \mathrm{e}^{st}\mathrm{d}s,
\end{equation*}%
where the integration is performed along the Bromwich path $Br$, that is
obtained from the contour $\Gamma _{0}$ in the limit as $R\rightarrow \infty 
$, with the contour $\Gamma _{0}$ being the part of the closed contour: $%
\Gamma ^{\left( \mathrm{I}\right) }$ if function $\tilde{P}$ does not have
branch points other than $s=0,$ $\Gamma ^{\left( \mathrm{II}\right) }$ if
function $\tilde{P}$ has a negative real branch point in addition to $s=0$,
and $\Gamma ^{\left( \mathrm{III}\right) }$ if function $\tilde{P}$ has a
pair of complex conjugated branch points in addition to $s=0,$ respectively
shown in Figures \ref{nemaTG}, \ref{negativnaTG}, and \ref{komplTG}. Each of
the contours $\Gamma ^{\left( \mathrm{I}\right) },$ $\Gamma ^{\left( \mathrm{%
II}\right) },$ and $\Gamma ^{\left( \mathrm{III}\right) }$ are used in the
Cauchy residues theorem%
\begin{equation}
\lim_{\substack{ R\rightarrow \infty  \\ r\rightarrow 0}}\oint_{\Gamma
^{\left( i\right) }}\tilde{P}\left( x,s\right) \mathrm{e}^{st}\mathrm{d}%
s=2\pi \mathrm{i}\sum_{k=-\infty }^{\infty }\func{Res}\left[ \tilde{P}\left(
x,s\right) \mathrm{e}^{st},s_{k}\right] ,\;\;i\in \left\{ \mathrm{I},\mathrm{%
II},\mathrm{III}\right\} ,  \label{reziduum}
\end{equation}%
where all poles $s_{k}$ of function $\tilde{P}$ lie in the domain encircled
by contour $\Gamma ^{\left( i\right) }$ in the limit when $R\rightarrow
\infty $ and $r\rightarrow 0$.

The branch points of solution kernel image $\tilde{P}$ are due to the
complex modulus $\tilde{G},$ since, by (\ref{g-tilda}), it contains function 
$\phi _{\sigma },$ given by (\ref{fi-prva-klasa})$_{1}$ for the first class
of the Burgers models and by (\ref{fi-druga-klasa})$_{1}$ for the second
class models, that 
\begin{equation*}
\begin{tabular}{ll}
\begin{tabular}{l}
has no zeros in the complex plane%
\end{tabular}
& if $\func{Re}\phi _{\sigma }\left( \rho ^{\ast }\right) <0,$ \\ 
\begin{tabular}{l}
has one negative real zero $-\rho ^{\ast }$%
\end{tabular}
& if $\func{Re}\phi _{\sigma }\left( \rho ^{\ast }\right) =0,$ \\ 
\begin{tabular}{l}
has a pair of complex conjugated \\ 
zeros $s_{0}$ and $\bar{s}_{0}$ having negative real part%
\end{tabular}
& if $\func{Re}\phi _{\sigma }\left( \rho ^{\ast }\right) >0,$%
\end{tabular}%
\end{equation*}%
where%
\begin{equation*}
\func{Re}\phi _{\sigma }\left( \rho ^{\ast }\right) =1+a_{1}\left( \rho
^{\ast }\right) ^{\alpha }\cos \left( \alpha \pi \right) +a_{2}\left( \rho
^{\ast }\right) ^{\beta }\cos \left( \beta \pi \right) +a_{3}\left( \rho
^{\ast }\right) ^{\gamma }\cos \left( \gamma \pi \right) ,
\end{equation*}%
with $\rho ^{\ast }$ determined from 
\begin{equation*}
\frac{a_{1}\sin \left( \alpha \pi \right) }{a_{3}\left\vert \sin \left(
\gamma \pi \right) \right\vert }+\frac{a_{2}\sin \left( \beta \pi \right) }{%
a_{3}\left\vert \sin \left( \gamma \pi \right) \right\vert }\left( \rho
^{\ast }\right) ^{\beta -\alpha }=\left( \rho ^{\ast }\right) ^{\gamma
-\alpha },
\end{equation*}%
for Models I - VII, while in the case of Model VIII function $\phi _{\sigma
} $%
\begin{equation*}
\begin{tabular}{ll}
\begin{tabular}{l}
has no zeros in the complex plane%
\end{tabular}
& 
\begin{tabular}{l}
if $\left( \frac{\bar{a}_{1}}{2\bar{a}_{2}}\right) ^{2}\geq \frac{1}{\bar{a}%
_{2}},$ or \\ 
if $\left( \frac{\bar{a}_{1}}{2\bar{a}_{2}}\right) ^{2}<\frac{1}{\bar{a}_{2}}
$ and $a<b\frac{\left\vert \cos \left( \alpha \pi \right) \right\vert }{\sin
\left( \alpha \pi \right) },$%
\end{tabular}
\\ 
\begin{tabular}{l}
has one negative real zero $-\rho ^{\ast }$ \\ 
determined by $\rho ^{\ast }=\left( \frac{b}{\sin \left( \alpha \pi \right) }%
\right) ^{\frac{1}{\alpha }}$%
\end{tabular}
& 
\begin{tabular}{l}
if $\left( \frac{\bar{a}_{1}}{2\bar{a}_{2}}\right) ^{2}<\frac{1}{\bar{a}_{2}}
$ and $a=b\frac{\left\vert \cos \left( \alpha \pi \right) \right\vert }{\sin
\left( \alpha \pi \right) },$%
\end{tabular}
\\ 
\begin{tabular}{l}
has a pair of complex conjugated \\ 
zeros $s_{0}$ and $\bar{s}_{0}$ having negative real part%
\end{tabular}
& 
\begin{tabular}{l}
if $\left( \frac{\bar{a}_{1}}{2\bar{a}_{2}}\right) ^{2}<\frac{1}{\bar{a}_{2}}
$ and $a>b\frac{\left\vert \cos \left( \alpha \pi \right) \right\vert }{\sin
\left( \alpha \pi \right) },$%
\end{tabular}%
\end{tabular}%
\end{equation*}%
where $a=\frac{\bar{a}_{1}}{2\bar{a}_{2}},$ and $b=\sqrt{\frac{1}{\bar{a}_{2}%
}-\left( \frac{\bar{a}_{1}}{2\bar{a}_{2}}\right) ^{2}},$ as proved in \cite%
{OZ-2}.

\subsection{Case when function $\tilde{P}$ has no branch points other than $%
s=0$ \label{no-branch-points}}

In the case when solution kernel image $\tilde{P}$ has no branch points
other than $s=0,$ the integrals along contours $\Gamma _{0}$, $\Gamma _{3}$,
and $\Gamma _{5},$ belonging to the integration contour $\Gamma ^{\left( 
\mathrm{I}\right) }$ from Figure \ref{nemaTG} and appearing in the Cauchy
residues theorem (\ref{reziduum}), have non-zero contribution and according
to contours' parameterization given in Table \ref{nemaTG-param}, in the
limit when $R\rightarrow \infty $ and $r\rightarrow 0,$ take the form%
\begin{eqnarray}
\lim_{R\rightarrow \infty }I_{\Gamma _{0}} &=&\int_{Br}\tilde{P}\left(
x,s\right) \mathrm{e}^{st}\mathrm{d}s=2\pi \mathrm{i}P\left( x,t\right) ,
\label{i-gama-0} \\
\lim_{\substack{ R\rightarrow \infty  \\ r\rightarrow 0}}I_{\Gamma _{3}}
&=&\int_{\infty }^{0}\frac{\sinh \frac{x\rho \mathrm{e}^{\mathrm{i}\pi }}{%
\sqrt{\tilde{G}\left( \rho \mathrm{e}^{\mathrm{i}\pi }\right) }}}{\sinh 
\frac{\rho \mathrm{e}^{\mathrm{i}\pi }}{\sqrt{\tilde{G}\left( \rho \mathrm{e}%
^{\mathrm{i}\pi }\right) }}}\mathrm{e}^{\rho t\mathrm{e}^{\mathrm{i}\pi }}%
\mathrm{e}^{\mathrm{i}\pi }\mathrm{d}\rho =\int_{0}^{\infty }\frac{\sinh 
\frac{x\rho \mathrm{e}^{\mathrm{i}\pi }}{\sqrt{\tilde{G}\left( \rho \mathrm{e%
}^{\mathrm{i}\pi }\right) }}}{\sinh \frac{\rho \mathrm{e}^{\mathrm{i}\pi }}{%
\sqrt{\tilde{G}\left( \rho \mathrm{e}^{\mathrm{i}\pi }\right) }}}\mathrm{e}%
^{-\rho t}\mathrm{d}\rho ,  \label{i-gama-3} \\
\lim_{\substack{ R\rightarrow \infty  \\ r\rightarrow 0}}I_{\Gamma _{5}}
&=&\int_{0}^{\infty }\frac{\sinh \frac{x\rho \mathrm{e}^{-\mathrm{i}\pi }}{%
\sqrt{\tilde{G}\left( \rho \mathrm{e}^{-\mathrm{i}\pi }\right) }}}{\sinh 
\frac{\rho \mathrm{e}^{-\mathrm{i}\pi }}{\sqrt{\tilde{G}\left( \rho \mathrm{e%
}^{-\mathrm{i}\pi }\right) }}}\mathrm{e}^{\rho t\mathrm{e}^{-\mathrm{i}\pi }}%
\mathrm{e}^{-\mathrm{i}\pi }\mathrm{d}\rho =-\int_{0}^{\infty }\frac{\sinh 
\frac{x\rho \mathrm{e}^{-\mathrm{i}\pi }}{\sqrt{\tilde{G}\left( \rho \mathrm{%
e}^{-\mathrm{i}\pi }\right) }}}{\sinh \frac{\rho \mathrm{e}^{-\mathrm{i}\pi }%
}{\sqrt{\tilde{G}\left( \rho \mathrm{e}^{-\mathrm{i}\pi }\right) }}}\mathrm{e%
}^{-\rho t}\mathrm{d}\rho ,  \label{i-gama-5}
\end{eqnarray}%
where the notation%
\begin{equation}
I_{\Gamma _{i}}=\int_{\Gamma _{i}}\frac{\sinh \frac{xs}{\sqrt{\tilde{G}%
\left( s\right) }}}{\sinh \frac{s}{\sqrt{\tilde{G}\left( s\right) }}}\mathrm{%
e}^{st}\mathrm{d}s,\;\;i=0,\ldots ,7,  \label{i-gama-i}
\end{equation}%
is used, while the integrals along all other contours have zero contribution
when $R\rightarrow \infty $ and $r\rightarrow 0,$ as will be proved below.

\noindent 
\begin{minipage}{\columnwidth}
\begin{minipage}[c]{0.4\columnwidth}
\centering
\includegraphics[width=0.7\columnwidth]{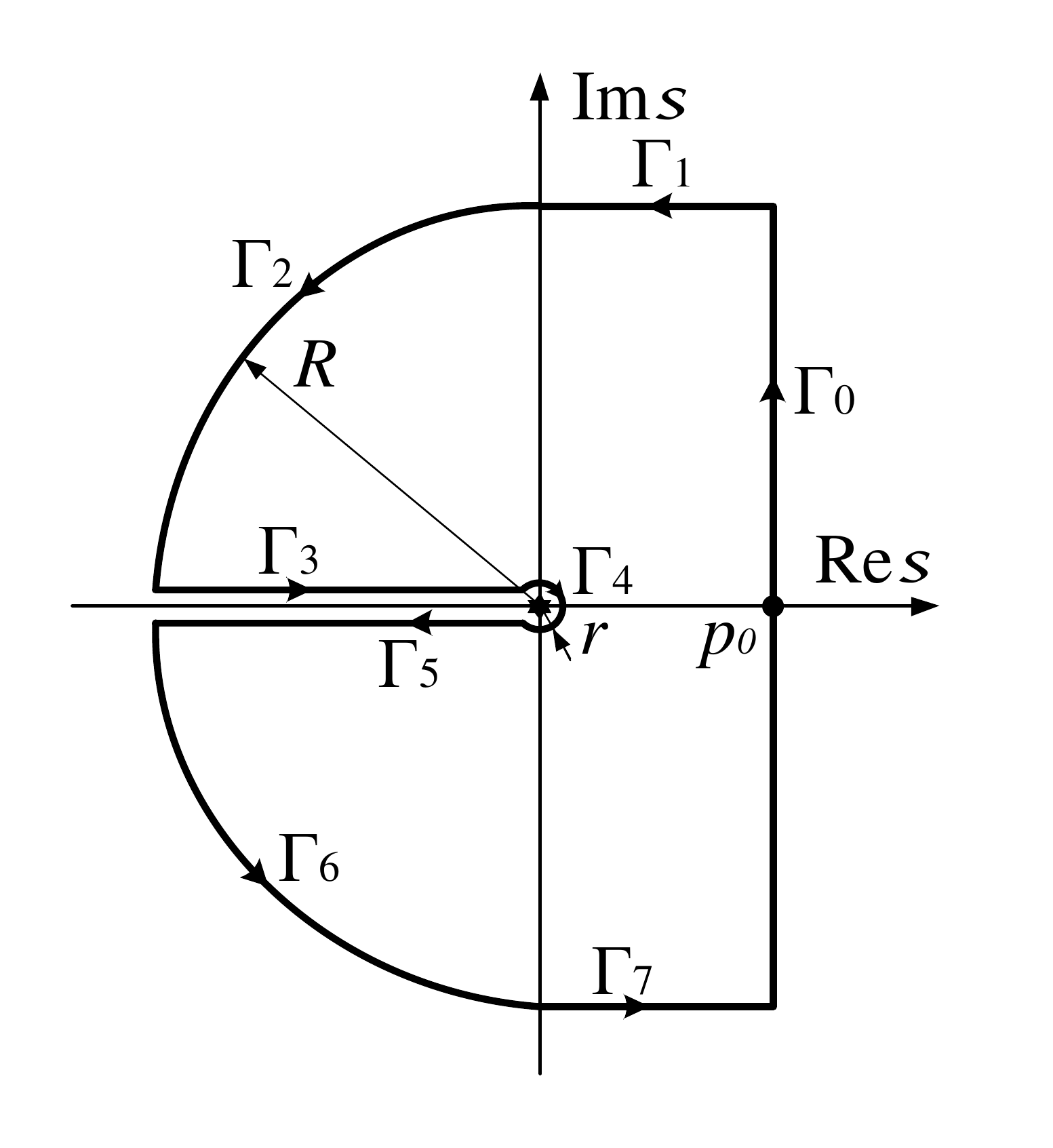}
\captionof{figure}{Integration contour $\Gamma^{(\mathrm{I})}$.}
\label{nemaTG}
\end{minipage}
\hfil
\begin{minipage}[c]{0.55\columnwidth}
\centering
\begin{tabular}{rll}
$\Gamma _{0}:$ & Bromwich path, &  \\ 
$\Gamma _{1}:$ & $s=p+\mathrm{i}R,$ & $p\in \left[ 0,p_{0}\right],\, p_0\geq 0$ arbitrary, \\ 
$\Gamma _{2}:$ & $s=R\mathrm{e}^{\mathrm{i}\varphi },$ & $\varphi \in \left[ 
\frac{\pi }{2},\pi \right] ,$ \\ 
$\Gamma _{3}:$ & $s=\rho \mathrm{e}^{\mathrm{i}\pi },$ & $\rho \in \left[ r,R%
\right] ,$ \\ 
$\Gamma _{4}:$ & $s=r\mathrm{e}^{\mathrm{i}\varphi },$ & $\varphi \in \left[ -\pi
,\pi \right] ,$ \\ 
$\Gamma _{5}:$ & $s=\rho \mathrm{e}^{-\mathrm{i}\pi },$ & $\rho \in \left[ r,R%
\right] ,$ \\
$\Gamma _{6}:$  & $s=R\mathrm{e}^{\mathrm{i}\varphi },$ & $\varphi \in \left[ 
-\pi, -\frac{\pi }{2} \right] ,$ \\
$\Gamma _{7}:$ & $s=p-\mathrm{i}R,$ & $p\in \left[ 0,p_{0}\right],\, p_0\geq 0$ arbitrary.  
\end{tabular}
\captionof{table}{Parametrization of integration contour $\Gamma^{(\mathrm{I})}$.}
\label{nemaTG-param}
\end{minipage}
\end{minipage}\smallskip

The residues in theorem (\ref{reziduum}), calculated according to%
\begin{equation}
\func{Res}\left[ \tilde{P}\left( x,s\right) \mathrm{e}^{st},s_{k}\right]
=\left. \frac{\sinh \frac{xs}{\sqrt{\tilde{G}\left( s\right) }}\mathrm{e}%
^{st}}{\frac{\mathrm{d}}{\mathrm{d}s}\sinh \frac{s}{\sqrt{\tilde{G}\left(
s\right) }}}\right\vert _{s=s_{k}}=\left. \frac{\sinh \frac{xs}{\sqrt{\tilde{%
G}\left( s\right) }}\mathrm{e}^{st}}{\left( \frac{1}{\sqrt{\tilde{G}\left(
s\right) }}-\frac{s\tilde{G}^{\prime }\left( s\right) }{2\sqrt{\tilde{G}%
\left( s\right) }^{3}}\right) \cosh \frac{s}{\sqrt{\tilde{G}\left( s\right) }%
}}\right\vert _{s=s_{k}},  \label{res-1}
\end{equation}%
where $\tilde{G}^{\prime }\left( s\right) =\frac{\mathrm{d}}{\mathrm{d}s}%
\tilde{G}\left( s\right) $ and where $s_{k}$, $k\in 
\mathbb{N}
_{0}$, are poles of the first order of function $\tilde{P}$, see (\ref%
{pe-tilda})$_{1}$, lying in the upper left complex quarter-plane as proved
in Section \ref{polovi}, become 
\begin{equation*}
\func{Res}\left[ \tilde{P}\left( x,s\right) \mathrm{e}^{st},s_{k}\right]
=\left( -1\right) ^{k}\frac{\sin \left( k\pi x\right) }{k\pi }\frac{s_{k}%
\mathrm{e}^{s_{k}t}}{1+\left( k\pi \right) ^{2}\frac{\tilde{G}^{\prime
}\left( s_{k}\right) }{2s_{k}}},
\end{equation*}%
when (\ref{sinus}) is used in (\ref{res-1}). The complex conjugation of
equation (\ref{sinus}) gives 
\begin{equation*}
\overline{\left( \frac{s}{\sqrt{\tilde{G}\left( s\right) }}\right) }=\frac{%
\bar{s}}{\sqrt{\tilde{G}\left( \bar{s}\right) }}=\mathrm{i}k\pi ,\;\;k=0,\pm
1,\pm 2,...
\end{equation*}%
implying $s_{-k}=\bar{s}_{k}$, where $\bar{s}_{k}$ is the complex conjugate
of $s_{k}$, so that the right hand side of the Cauchy residues theorem (\ref%
{reziduum}) becomes%
\begin{align}
& \sum_{k=-\infty }^{\infty }\func{Res}\left[ \tilde{P}\left( x,s\right) 
\mathrm{e}^{st},s_{k}\right]  \notag \\
& \qquad \qquad \qquad =\sum_{k=1}^{\infty }\left( -1\right) ^{k}\frac{\sin
\left( k\pi x\right) }{k\pi }\left( \frac{s_{k}\mathrm{e}^{s_{k}t}}{1+\left(
k\pi \right) ^{2}\frac{\tilde{G}^{\prime }\left( s_{k}\right) }{2s_{k}}}+%
\frac{\bar{s}_{k}\mathrm{e}^{\bar{s}_{k}t}}{1+\left( k\pi \right) ^{2}\frac{%
\tilde{G}^{\prime }\left( \bar{s}_{k}\right) }{2\bar{s}_{k}}}\right)
+x\lim_{k\rightarrow 0}\frac{s_{k}\mathrm{e}^{s_{k}t}}{1+\left( k\pi \right)
^{2}\frac{\tilde{G}^{\prime }\left( s_{k}\right) }{2s_{k}}}  \label{res-2} \\
& \qquad \qquad \qquad =2\sum_{k=1}^{\infty }\left( -1\right) ^{k}\frac{\sin
\left( k\pi x\right) }{k\pi }\mathrm{e}^{-\rho _{k}t\left\vert \cos \varphi
_{k}\right\vert }\func{Re}\left( \frac{s_{k}\mathrm{e}^{\mathrm{i}\rho
_{k}t\sin \varphi _{k}}}{1+\left( k\pi \right) ^{2}\frac{\tilde{G}^{\prime
}\left( s_{k}\right) }{2s_{k}}}\right) ,  \label{res-3}
\end{align}%
where $s_{k}=\rho _{k}\mathrm{e}^{\mathrm{i}\varphi _{k}}$, $\varphi _{k}\in
\left( \frac{\pi }{2},\pi \right) $. Note, the last term in (\ref{res-2}) is
zero since%
\begin{equation*}
\frac{s_{k}\mathrm{e}^{s_{k}t}}{1+\left( k\pi \right) ^{2}\frac{\tilde{G}%
^{\prime }\left( s_{k}\right) }{2s_{k}}}\sim \frac{\mathrm{i}k\pi \mathrm{e}%
^{\mathrm{i}k\pi t}}{1+k\pi \frac{\tilde{G}^{\prime }\left( \mathrm{i}k\pi
\right) }{2\mathrm{i}}}\sim \mathrm{i}k\pi \rightarrow 0\;\;\text{as}%
\;\;k\rightarrow 0,
\end{equation*}%
due to $s_{k}\sim \mathrm{i}k\pi $ as $k\rightarrow 0$, according to the
equation (\ref{sinus}).

Summing up, the integrals having non-zero contribution (\ref{i-gama-0}), (%
\ref{i-gama-3}), and (\ref{i-gama-5}), as well as the residues (\ref{res-3}%
), according to the Cauchy residues theorem (\ref{reziduum}) yield%
\begin{align}
& 2\pi \mathrm{i}P\left( x,t\right) +2\mathrm{i}\int_{0}^{\infty }\func{Im}%
\left( \frac{\sinh \frac{x\rho \mathrm{e}^{\mathrm{i}\pi }}{\sqrt{\tilde{G}%
\left( \rho \mathrm{e}^{\mathrm{i}\pi }\right) }}}{\sinh \frac{\rho \mathrm{e%
}^{\mathrm{i}\pi }}{\sqrt{\tilde{G}\left( \rho \mathrm{e}^{\mathrm{i}\pi
}\right) }}}\right) \mathrm{e}^{-\rho t}\mathrm{d}\rho  \notag \\
& \quad \quad \quad \quad \quad \quad =4\pi \mathrm{i}\sum_{k=1}^{\infty
}\left( -1\right) ^{k}\frac{\sin \left( k\pi x\right) }{k\pi }\mathrm{e}%
^{-\rho _{k}t\left\vert \cos \varphi _{k}\right\vert }\func{Re}\left( \frac{%
s_{k}\mathrm{e}^{\mathrm{i}\rho _{k}t\sin \varphi _{k}}}{1+\left( k\pi
\right) ^{2}\frac{\tilde{G}^{\prime }\left( s_{k}\right) }{2s_{k}}}\right) 
\notag \\
& P\left( x,t\right) =-\frac{1}{\pi }\int_{0}^{\infty }\func{Im}\left( \frac{%
\sinh \frac{x\rho \mathrm{e}^{\mathrm{i}\pi }}{\sqrt{\tilde{G}\left( \rho 
\mathrm{e}^{\mathrm{i}\pi }\right) }}}{\sinh \frac{\rho \mathrm{e}^{\mathrm{i%
}\pi }}{\sqrt{\tilde{G}\left( \rho \mathrm{e}^{\mathrm{i}\pi }\right) }}}%
\right) \mathrm{e}^{-\rho t}\mathrm{d}\rho +2\sum_{k=1}^{\infty }\left(
-1\right) ^{k}\frac{\sin \left( k\pi x\right) }{k\pi }\mathrm{e}^{-\rho
_{k}t\left\vert \cos \varphi _{k}\right\vert }\func{Re}\left( \frac{s_{k}%
\mathrm{e}^{\mathrm{i}\rho _{k}t\sin \varphi _{k}}}{1+\left( k\pi \right)
^{2}\frac{\tilde{G}^{\prime }\left( s_{k}\right) }{2s_{k}}}\right) .
\label{pee}
\end{align}

It is left to prove that the integrals along contours $\Gamma _{1},$ $\Gamma
_{2},$ $\Gamma _{4},$ $\Gamma _{6},$ and $\Gamma _{7}$ tend to zero in the
limit $R\rightarrow \infty $ and $r\rightarrow 0$. Regardless of the fact
that the form (\ref{pee}) of solution kernel $P$ is only valid for models of
the first class, in showing zero contributions of the mentioned integrals,
models of the second class will be also addressed, since the regularized
solution kernel $P_{\mathrm{reg}}$ is treated analogously.

The integral along the contour $\Gamma _{1},$ according to its
parameterization from Table \ref{nemaTG-param}, is%
\begin{equation*}
I_{\Gamma _{1}}=\int_{p_{0}}^{0}\frac{\sinh \frac{x\left( p+\mathrm{i}%
R\right) }{\sqrt{\tilde{G}\left( p+\mathrm{i}R\right) }}}{\sinh \frac{p+%
\mathrm{i}R}{\sqrt{\tilde{G}\left( p+\mathrm{i}R\right) }}}\mathrm{e}^{pt}%
\mathrm{e}^{\mathrm{i}Rt}\mathrm{d}p,
\end{equation*}%
so that%
\begin{eqnarray}
\left\vert I_{\Gamma _{1}}\right\vert &\leqslant &\int_{0}^{p_{0}}\left\vert 
\mathrm{e}^{-\left( 1-x\right) \frac{p+\mathrm{i}R}{\sqrt{\tilde{G}\left( p+%
\mathrm{i}R\right) }}}\right\vert \frac{\left\vert 1-\mathrm{e}^{-\frac{%
2x\left( p+\mathrm{i}R\right) }{\sqrt{\tilde{G}\left( p+\mathrm{i}R\right) }}%
}\right\vert }{\left\vert 1-\mathrm{e}^{-\frac{2\left( p+\mathrm{i}R\right) 
}{\sqrt{\tilde{G}\left( p+\mathrm{i}R\right) }}}\right\vert }\mathrm{e}^{pt}%
\mathrm{d}p  \notag \\
&\leqslant &\int_{0}^{p_{0}}\mathrm{e}^{-\left( 1-x\right) \func{Re}\frac{p+%
\mathrm{i}R}{\sqrt{\tilde{G}\left( p+\mathrm{i}R\right) }}}\frac{1+\mathrm{e}%
^{-2x\func{Re}\frac{p+\mathrm{i}R}{\sqrt{\tilde{G}\left( p+\mathrm{i}%
R\right) }}}}{\left\vert 1-\mathrm{e}^{-2\func{Re}\frac{p+\mathrm{i}R}{\sqrt{%
\tilde{G}\left( p+\mathrm{i}R\right) }}}\right\vert }\mathrm{e}^{pt}\mathrm{d%
}p  \label{i-gama-1} \\
&\leqslant &\int_{0}^{p_{0}}\mathrm{e}^{-\left( 1-x\right) \sqrt{\frac{a_{3}%
}{b}}R^{\frac{2-\delta }{2}}\cos \frac{\left( 2-\delta \right) \pi }{4}}%
\frac{1+\mathrm{e}^{-2x\sqrt{\frac{a_{3}}{b}}R^{\frac{2-\delta }{2}}\cos 
\frac{\left( 2-\delta \right) \pi }{4}}}{\left\vert 1-\mathrm{e}^{-2\sqrt{%
\frac{a_{3}}{b}}R^{\frac{2-\delta }{2}}\cos \frac{\left( 2-\delta \right)
\pi }{4}}\right\vert }\mathrm{e}^{pt}\mathrm{d}p\rightarrow 0\;\;\text{as}%
\;\;R\rightarrow \infty ,  \notag
\end{eqnarray}%
due to the asymptotic behavior 
\begin{equation*}
\frac{p+\mathrm{i}R}{\sqrt{\tilde{G}\left( p+\mathrm{i}R\right) }}\sim \sqrt{%
\frac{a_{3}}{b}}R^{\frac{2-\delta }{2}}\mathrm{e}^{\mathrm{i}\frac{\left(
2-\delta \right) \pi }{4}},\;\;\text{as}\;\;R\rightarrow \infty ,
\end{equation*}%
with $\delta =\mu +\eta -\gamma \in \left( 0,1\right) $, implying $\frac{%
\left( 2-\delta \right) \pi }{4}\in \left( \frac{\pi }{4},\frac{\pi }{2}%
\right) \ $in the case of the first class of Burgers models, while in the
case of the second class of Burgers models, the expression (\ref{i-gama-1})
becomes 
\begin{equation*}
\left\vert I_{\Gamma _{1}}\right\vert \leqslant \int_{0}^{p_{0}}\mathrm{e}%
^{-\left( 1-x\right) \frac{1}{2}\sqrt{\frac{b}{a_{3}}}\frac{a_{2}}{b^{2}}%
\left( b-\frac{a_{3}}{a_{2}}\right) \sin \frac{\eta \pi }{2}R^{1-\eta }}%
\frac{1+\mathrm{e}^{-2x\frac{1}{2}\sqrt{\frac{b}{a_{3}}}\frac{a_{2}}{b^{2}}%
\left( b-\frac{a_{3}}{a_{2}}\right) \sin \frac{\eta \pi }{2}R^{1-\eta }}}{1-%
\mathrm{e}^{-2\frac{1}{2}\sqrt{\frac{b}{a_{3}}}\frac{a_{2}}{b^{2}}\left( b-%
\frac{a_{3}}{a_{2}}\right) \sin \frac{\eta \pi }{2}R^{1-\eta }}}\mathrm{e}%
^{pt}\mathrm{d}p\rightarrow 0\;\;\text{as}\;\;R\rightarrow \infty ,
\end{equation*}%
since $\eta \in \left( 0,1\right) ,$ due to the asymptotics%
\begin{eqnarray*}
\func{Re}\frac{p+\mathrm{i}R}{\sqrt{\tilde{G}\left( p+\mathrm{i}R\right) }}
&\sim &\frac{1}{2}\sqrt{\frac{b}{a_{3}}}\frac{a_{2}}{b^{2}}\left( b-\frac{%
a_{3}}{a_{2}}\right) \sin \frac{\eta \pi }{2}R^{1-\eta }\rightarrow \infty
\;\;\text{as}\;\;R\rightarrow \infty , \\
\func{Im}\frac{p+\mathrm{i}R}{\sqrt{\tilde{G}\left( p+\mathrm{i}R\right) }}
&\sim &\sqrt{\frac{a_{3}}{b}}R\rightarrow \infty \;\;\text{as}%
\;\;R\rightarrow \infty ,
\end{eqnarray*}%
found by applying formulae $\func{Re}z=\sqrt{\frac{\left\vert
z^{2}\right\vert +\func{Re}z^{2}}{2}}$ and $\func{Im}z=\sqrt{\frac{%
\left\vert z^{2}\right\vert -\func{Re}z^{2}}{2}}\mathrm{sgn}\func{Im}z^{2}$
to 
\begin{eqnarray*}
\func{Re}\left( \frac{p+\mathrm{i}R}{\sqrt{\tilde{G}\left( p+\mathrm{i}%
R\right) }}\right) ^{2} &\sim &\func{Re}\psi \left( R,\frac{\pi }{2}\right)
\sim -\frac{a_{3}}{b}R^{2}\;\;\text{as}\;\;R\rightarrow \infty , \\
\func{Im}\left( \frac{p+\mathrm{i}R}{\sqrt{\tilde{G}\left( p+\mathrm{i}%
R\right) }}\right) ^{2} &\sim &\func{Im}\psi \left( R,\frac{\pi }{2}\right)
\sim \frac{a_{2}}{b^{2}}\left( b-\frac{a_{3}}{a_{2}}\right) R^{2-\eta }\sin 
\frac{\eta \pi }{2}\;\;\text{as}\;\;R\rightarrow \infty ,
\end{eqnarray*}%
that are obtained by (\ref{repsi-gama1-pi-pola-II-klas}) and (\ref%
{impsi-gama1-pi-pola-II-klas}), since by (\ref{psi}) it holds $\psi \left(
s\right) \sim \frac{s^{2}}{\tilde{G}\left( s\right) }$ as $\left\vert
s\right\vert \rightarrow \infty $. Therefore, the integral $I_{\Gamma
_{1}}\rightarrow 0$ when $R\rightarrow \infty $, and by similar arguments
integral $I_{\Gamma _{7}}\rightarrow 0$ when $R\rightarrow \infty $.

Using the parameterization from Table \ref{nemaTG-param}, the integral along
contour $\Gamma _{2}$ takes the form%
\begin{equation*}
I_{\Gamma _{2}}=\int_{\frac{\pi }{2}}^{\pi }\frac{\sinh \frac{xR\mathrm{e}^{%
\mathrm{i}\varphi }}{\sqrt{\tilde{G}\left( R\mathrm{e}^{\mathrm{i}\varphi
}\right) }}}{\sinh \frac{R\mathrm{e}^{\mathrm{i}\varphi }}{\sqrt{\tilde{G}%
\left( R\mathrm{e}^{\mathrm{i}\varphi }\right) }}}\mathrm{e}^{Rt\mathrm{e}^{%
\mathrm{i}\varphi }}R\mathrm{ie}^{\mathrm{i}\varphi }\mathrm{d}\varphi ,
\end{equation*}%
so that 
\begin{eqnarray}
\left\vert I_{\Gamma _{2}}\right\vert &\leqslant &\int_{\frac{\pi }{2}}^{\pi
}\left\vert \mathrm{e}^{\left( 1-x\right) \frac{R\mathrm{e}^{\mathrm{i}%
\varphi }}{\sqrt{\tilde{G}\left( R\mathrm{e}^{\mathrm{i}\varphi }\right) }}%
}\right\vert \frac{\left\vert \mathrm{e}^{\frac{2xR\mathrm{e}^{\mathrm{i}%
\varphi }}{\sqrt{\tilde{G}\left( R\mathrm{e}^{\mathrm{i}\varphi }\right) }}%
}-1\right\vert }{\left\vert \mathrm{e}^{\frac{2R\mathrm{e}^{\mathrm{i}%
\varphi }}{\sqrt{\tilde{G}\left( R\mathrm{e}^{\mathrm{i}\varphi }\right) }}%
}-1\right\vert }\mathrm{e}^{Rt\mathrm{\cos \varphi }}R\mathrm{d}\varphi 
\notag \\
&\leqslant &\int_{\frac{\pi }{2}}^{\pi }\mathrm{e}^{\left( 1-x\right) \func{%
Re}\frac{R\mathrm{e}^{\mathrm{i}\varphi }}{\sqrt{\tilde{G}\left( R\mathrm{e}%
^{\mathrm{i}\varphi }\right) }}}\frac{\mathrm{e}^{2x\func{Re}\frac{R\mathrm{e%
}^{\mathrm{i}\varphi }}{\sqrt{\tilde{G}\left( R\mathrm{e}^{\mathrm{i}\varphi
}\right) }}}+1}{\left\vert \mathrm{e}^{2\func{Re}\frac{R\mathrm{e}^{\mathrm{i%
}\varphi }}{\sqrt{\tilde{G}\left( R\mathrm{e}^{\mathrm{i}\varphi }\right) }}%
}-1\right\vert }\mathrm{e}^{Rt\mathrm{\cos \varphi }}R\mathrm{d}\varphi
\label{i-gama-2}
\end{eqnarray}%
In the case of the first class of Burgers models, the asymptotic behavior of 
$\frac{s}{\sqrt{\tilde{G}\left( s\right) }}$ on contour $\Gamma _{2}$ is%
\begin{equation*}
\frac{R\mathrm{e}^{\mathrm{i}\varphi }}{\sqrt{\tilde{G}\left( R\mathrm{e}^{%
\mathrm{i}\varphi }\right) }}\sim \sqrt{\frac{a_{3}}{b}}R^{\frac{2-\delta }{2%
}}\mathrm{e}^{\mathrm{i}\frac{\left( 2-\delta \right) \varphi }{2}}\;\;\text{%
as}\;\;R\rightarrow \infty ,
\end{equation*}%
where $\delta =\mu +\eta -\gamma \in \left( 0,1\right) $, implying $\frac{%
\left( 2-\delta \right) \varphi }{2}\in \left( \frac{\pi }{4},\pi \right) $,
so it is necessary to consider two intervals: $\varphi \in \left( \frac{\pi 
}{2},\varphi _{\delta }\right) $ since then $\cos \frac{\left( 2-\delta
\right) \varphi }{2}>0$ and $\varphi \in \left( \varphi _{\delta },\pi
\right) $ since then $\cos \frac{\left( 2-\delta \right) \varphi }{2}<0,$
with $\varphi _{\delta }=\frac{\pi }{2-\delta }.$ Therefore, the expression (%
\ref{i-gama-2}) become%
\begin{eqnarray*}
\left\vert I_{\Gamma _{2}}\right\vert &\leqslant &\int_{\frac{\pi }{2}%
}^{\varphi _{\delta }}\mathrm{e}^{-2\left( 1-x\right) \sqrt{\frac{a_{3}}{b}}%
R^{\frac{2-\delta }{2}}\cos \frac{\left( 2-\delta \right) \varphi }{2}}\frac{%
1+\mathrm{e}^{-2x\sqrt{\frac{a_{3}}{b}}R^{\frac{2-\delta }{2}}\cos \frac{%
\left( 2-\delta \right) \varphi }{2}}}{\left\vert 1-\mathrm{e}^{-2\sqrt{%
\frac{a_{3}}{b}}R^{\frac{2-\delta }{2}}\cos \frac{\left( 2-\delta \right)
\varphi }{2}}\right\vert }\mathrm{e}^{Rt\mathrm{\cos \varphi }}R\mathrm{d}%
\varphi \\
&&+\int_{\varphi _{\delta }}^{\pi }\mathrm{e}^{-2\left( 1-x\right) \sqrt{%
\frac{a_{3}}{b}}R^{\frac{2-\delta }{2}}\left\vert \cos \frac{\left( 2-\delta
\right) \varphi }{2}\right\vert }\frac{\mathrm{e}^{-2x\sqrt{\frac{a_{3}}{b}}%
R^{\frac{2-\delta }{2}}\left\vert \cos \frac{\left( 2-\delta \right) \varphi 
}{2}\right\vert }+1}{\left\vert \mathrm{e}^{-2\sqrt{\frac{a_{3}}{b}}R^{\frac{%
2-\delta }{2}}\left\vert \cos \frac{\left( 2-\delta \right) \varphi }{2}%
\right\vert }-1\right\vert }\mathrm{e}^{Rt\mathrm{\cos \varphi }}R\mathrm{d}%
\varphi \\
&\leqslant &\int_{\frac{\pi }{2}}^{\varphi _{\delta }}\mathrm{e}%
^{-Rt\left\vert \mathrm{\cos }\varphi \right\vert -2\left( 1-x\right) \sqrt{%
\frac{a_{3}}{b}}R^{\frac{2-\delta }{2}}\cos \frac{\left( 2-\delta \right)
\varphi }{2}}R\mathrm{d}\varphi \\
&&+\int_{\varphi _{\delta }}^{\pi }\mathrm{e}^{-Rt\left\vert \mathrm{\cos }%
\varphi \right\vert -2\left( 1-x\right) \sqrt{\frac{a_{3}}{b}}R^{\frac{%
2-\delta }{2}}\left\vert \cos \frac{\left( 2-\delta \right) \varphi }{2}%
\right\vert }R\mathrm{d}\varphi 
\begin{tabular}{l}
$\rightarrow $%
\end{tabular}%
0\;\;\text{as}\;\;R%
\begin{tabular}{l}
$\rightarrow $%
\end{tabular}%
\infty .
\end{eqnarray*}%
For the Burgers models of the second class the expression (\ref{i-gama-2})
gives%
\begin{equation*}
\left\vert I_{\Gamma _{2}}\right\vert \leqslant \int_{\frac{\pi }{2}}^{\pi }%
\mathrm{e}^{-R\left\vert \mathrm{\cos \varphi }\right\vert \left( t+\left(
1-x\right) \sqrt{\frac{a_{3}}{b}}\right) }\frac{\mathrm{e}^{-2x\sqrt{\frac{%
a_{3}}{b}}R\left\vert \cos \varphi \right\vert }+1}{\left\vert \mathrm{e}^{-2%
\sqrt{\frac{a_{3}}{b}}R\left\vert \cos \varphi \right\vert }-1\right\vert }R%
\mathrm{d}\varphi \rightarrow 0\;\;\text{as}\;\;R\rightarrow \infty ,
\end{equation*}%
due to the asymptotics%
\begin{equation*}
\frac{R\mathrm{e}^{\mathrm{i}\varphi }}{\sqrt{\tilde{G}\left( R\mathrm{e}^{%
\mathrm{i}\varphi }\right) }}\sim \sqrt{\frac{a_{3}}{b}}R\mathrm{e}^{\mathrm{%
i}\varphi }\;\;\text{as}\;\;R\rightarrow \infty .
\end{equation*}%
So, $I_{\Gamma _{2}}\rightarrow 0$ when $R\rightarrow \infty $, and using a
similar argumentation $I_{\Gamma _{6}}\rightarrow 0$ when $R\rightarrow
\infty .$

Integral on the contour $\Gamma _{4},$ parameterized using parameterization
given in Table \ref{nemaTG-param}, is%
\begin{equation*}
I_{\Gamma _{4}}=\int_{\pi }^{-\pi }\frac{\sinh \frac{xr\mathrm{e}^{\mathrm{i}%
\varphi }}{\sqrt{\tilde{G}\left( r\mathrm{e}^{\mathrm{i}\varphi }\right) }}}{%
\sinh \frac{r\mathrm{e}^{\mathrm{i}\varphi }}{\sqrt{\tilde{G}\left( r\mathrm{%
e}^{\mathrm{i}\varphi }\right) }}}\mathrm{e}^{rt\mathrm{e}^{\mathrm{i}%
\varphi }}r\mathrm{ie}^{\mathrm{i}\varphi }\mathrm{d}\varphi ,
\end{equation*}%
so that%
\begin{eqnarray*}
\left\vert I_{\Gamma _{4}}\right\vert &\leqslant &\int_{-\pi }^{\pi
}\left\vert \mathrm{e}^{\left( 1-x\right) \frac{r\mathrm{e}^{\mathrm{i}%
\varphi }}{\sqrt{\tilde{G}\left( r\mathrm{e}^{\mathrm{i}\varphi }\right) }}%
}\right\vert \frac{\left\vert \mathrm{e}^{\frac{2xr\mathrm{e}^{\mathrm{i}%
\varphi }}{\sqrt{\tilde{G}\left( r\mathrm{e}^{\mathrm{i}\varphi }\right) }}%
}-1\right\vert }{\left\vert \mathrm{e}^{\frac{2r\mathrm{e}^{\mathrm{i}%
\varphi }}{\sqrt{\tilde{G}\left( r\mathrm{e}^{\mathrm{i}\varphi }\right) }}%
}-1\right\vert }\mathrm{e}^{rt\mathrm{\cos \varphi }}r\mathrm{d}\varphi \\
&\leqslant &\int_{-\pi }^{\pi }\mathrm{e}^{\left( 1-x\right) r^{\frac{2-\xi 
}{2}}\cos \frac{\left( 2-\xi \right) \varphi }{2}+rt\cos \varphi }\frac{%
\mathrm{e}^{2xr^{\frac{2-\xi }{2}}\cos \frac{\left( 2-\xi \right) \varphi }{2%
}}+1}{\left\vert \mathrm{e}^{2r^{\frac{2-\xi }{2}}\cos \frac{\left( 2-\xi
\right) \varphi }{2}}-1\right\vert }r\mathrm{d}\varphi \\
&\leqslant &\frac{1}{2}\int_{-\pi }^{\pi }\frac{r^{\frac{\xi }{2}}}{%
\left\vert \cos \frac{\left( 2-\xi \right) \varphi }{2}\right\vert }\mathrm{d%
}\varphi \rightarrow 0\;\;\text{as}\;\;r\rightarrow 0,
\end{eqnarray*}%
with $\xi \in \left\{ \mu ,\beta \right\} $, since for the Burgers models of
both classes the following asymptotics holds%
\begin{equation*}
\frac{r\mathrm{e}^{\mathrm{i}\varphi }}{\sqrt{\tilde{G}\left( r\mathrm{e}^{%
\mathrm{i}\varphi }\right) }}\sim r^{\frac{2-\xi }{2}}\mathrm{e}^{\mathrm{i}%
\frac{\left( 2-\xi \right) \varphi }{2}}\;\;\text{as}\;\;r\rightarrow 0,
\end{equation*}%
where $\mathrm{e}^{2r^{\frac{2-\xi }{2}}\cos \frac{\left( 2-\xi \right)
\varphi }{2}}-1\approx 2r^{\frac{2-\xi }{2}}\cos \frac{\left( 2-\xi \right)
\varphi }{2}$ for small $r$ is additionally used.

\subsection{Case when function $\tilde{P}$ has negative real branch point in
addition to $s=0$}

In the case when solution kernel image $\tilde{P}$ has a negative real
branch point $s=-\rho ^{\ast }$ in addition to $s=0,$ the integration in the
Cauchy residues theorem (\ref{reziduum}) is performed along the contour $%
\Gamma ^{\left( \mathrm{II}\right) },$ shown in Figure \ref{negativnaTG},
and, as in the previous case, the integrals along contours $\Gamma _{0}$, $%
\Gamma _{3a}\cup \Gamma _{3b}$, and $\Gamma _{5a}\cup \Gamma _{5b}$ have
non-zero contribution and according to contours' parameterization given in
Table \ref{negativnaTG-param} take the same form as given by (\ref{i-gama-0}%
), (\ref{i-gama-3}), and (\ref{i-gama-5}). In addition to the integrals, the
residues are also already calculated in Section \ref{no-branch-points} and
given by (\ref{res-3}). Solution kernel $P$ takes the form (\ref{pee}),
since, as in the case when function $\tilde{P}$ has no other branch points
than $s=0$, the integrals along contours $\Gamma _{1},$ $\Gamma _{2},$ $%
\Gamma _{6},$ and $\Gamma _{7},$ have zero contribution, as already proved
in Section \ref{no-branch-points}, so it is left to prove that the integrals
along the contours $\Gamma _{8}$ and $\Gamma _{9}$ have zero contributions.

\noindent 
\begin{minipage}{\columnwidth}
\begin{minipage}[c]{0.4\columnwidth}
\centering
\includegraphics[width=0.7\columnwidth]{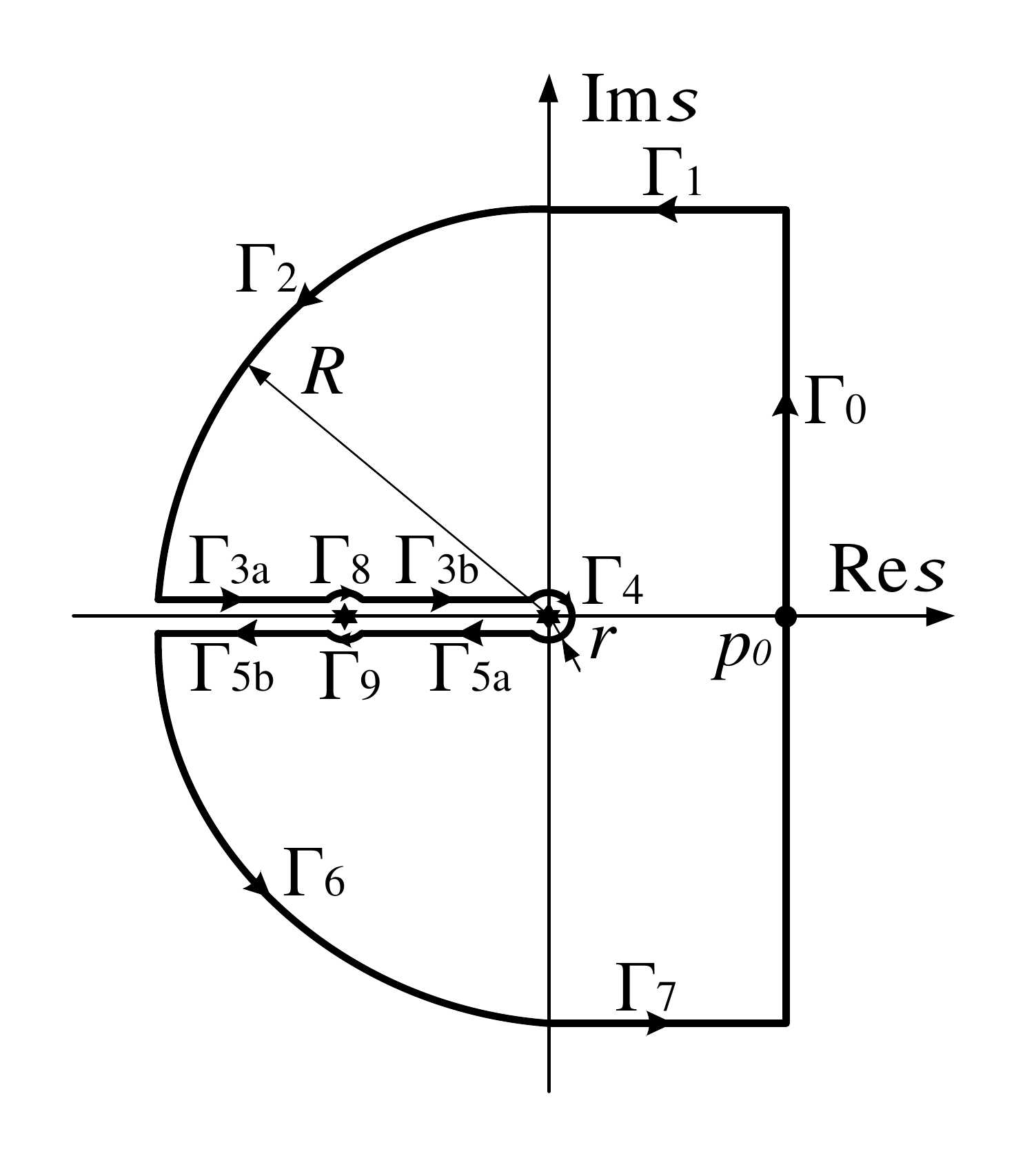}
\captionof{figure}{Integration contour $\Gamma^{(\mathrm{II})}$.}
\label{negativnaTG}
\end{minipage}
\hfil
\begin{minipage}[c]{0.55\columnwidth}
\centering
\begin{tabular}{rll}
$\Gamma _{0}:$ & Bromwich path, &  \\ 
$\Gamma _{1}:$ & $s=p+\mathrm{i}R,$ & $p\in \left[ 0,p_{0}\right],\, p_0\geq 0$ arbitrary, \\ 
$\Gamma _{2}:$ & $s=R\mathrm{e}^{\mathrm{i}\varphi },$ & $\varphi \in \left[ 
\frac{\pi }{2},\pi \right] ,$ \\ 
$\Gamma _{3a}\cup\Gamma _{3b}:$ & $s=\rho \mathrm{e}^{\mathrm{i}\pi },$ & $\rho \in \left[ r,R%
\right] ,$ \\ 
$\Gamma _{4}:$ & $s=r\mathrm{e}^{\mathrm{i}\varphi },$ & $\varphi \in \left[ -\pi
,\pi \right] ,$ \\ 
$\Gamma _{5a}\cup\Gamma _{5b}:$ & $s=\rho \mathrm{e}^{-\mathrm{i}\pi },$ & $\rho \in \left[ r,R%
\right] ,$ \\
$\Gamma _{6}:$  & $s=R\mathrm{e}^{\mathrm{i}\varphi },$ & $\varphi \in \left[ 
-\pi, -\frac{\pi }{2} \right] ,$ \\
$\Gamma _{7}:$ & $s=p-\mathrm{i}R,$ & $p\in \left[ 0,p_{0}\right],\, p_0\geq 0$ arbitrary,\\
$\Gamma _{8}:$  & $s=-\rho^*+r\mathrm{e}^{\mathrm{i}\varphi },$ & $\varphi \in \left[0,
\pi\right] ,$ \\
$\Gamma _{9}:$  & $s=-\rho^*+r\mathrm{e}^{\mathrm{i}\varphi },$ & $\varphi \in \left[ -\pi,
0 \right]$.  
\end{tabular}
\captionof{table}{Parametrization of integration contour $\Gamma^{(\mathrm{II})}$.}
\label{negativnaTG-param}
\end{minipage}
\end{minipage}\smallskip

Integral on the contour $\Gamma _{8},$ parameterized using parameterization
given in Table \ref{negativnaTG-param}, by (\ref{i-gama-i}) is%
\begin{equation*}
I_{\Gamma _{8}}=\int_{\pi }^{0}\frac{\sinh \frac{x\left( \rho ^{\ast }%
\mathrm{e}^{\mathrm{i}\pi }+r\mathrm{e}^{\mathrm{i}\varphi }\right) }{\sqrt{%
\tilde{G}\left( \rho ^{\ast }\mathrm{e}^{\mathrm{i}\pi }+r\mathrm{e}^{%
\mathrm{i}\varphi }\right) }}}{\sinh \frac{\left( \rho ^{\ast }\mathrm{e}^{%
\mathrm{i}\pi }+r\mathrm{e}^{\mathrm{i}\varphi }\right) }{\sqrt{\tilde{G}%
\left( \rho ^{\ast }\mathrm{e}^{\mathrm{i}\pi }+r\mathrm{e}^{\mathrm{i}%
\varphi }\right) }}}\mathrm{e}^{\left( \rho ^{\ast }\mathrm{e}^{\mathrm{i}%
\pi }+r\mathrm{e}^{\mathrm{i}\varphi }\right) t}r\mathrm{ie}^{\mathrm{i}%
\varphi }\mathrm{d}\varphi ,
\end{equation*}%
so that%
\begin{eqnarray*}
\left\vert I_{\Gamma _{8}}\right\vert &\leqslant &\mathrm{e}^{-\rho ^{\ast
}t}\int_{0}^{\pi }\left\vert \mathrm{e}^{\left( 1-x\right) \sqrt{r}\left(
X\left( \varphi \right) +\mathrm{i}Y\left( \varphi \right) \right)
}\right\vert \frac{\left\vert \mathrm{e}^{2x\sqrt{r}\left( X\left( \varphi
\right) +\mathrm{i}Y\left( \varphi \right) \right) }-1\right\vert }{%
\left\vert \mathrm{e}^{2\sqrt{r}\left( X\left( \varphi \right) +\mathrm{i}%
Y\left( \varphi \right) \right) }-1\right\vert }r\mathrm{d}\varphi \\
&\leqslant &\mathrm{e}^{-\rho ^{\ast }t}\int_{0}^{\pi }\mathrm{e}^{\left(
1-x\right) \sqrt{r}X\left( \varphi \right) }\frac{\mathrm{e}^{2x\sqrt{r}%
X\left( \varphi \right) }+1}{\left\vert \mathrm{e}^{2\sqrt{r}X\left( \varphi
\right) }-1\right\vert }r\mathrm{d}\varphi \\
&\leqslant &\mathrm{e}^{-\rho ^{\ast }t}\int_{0}^{\pi }\frac{\sqrt{r}}{%
\left\vert X\left( \varphi \right) \right\vert }\mathrm{d}\varphi
\rightarrow 0\;\;\text{as}\;\;r\rightarrow 0,
\end{eqnarray*}%
where $\mathrm{e}^{2\sqrt{r}X\left( \varphi \right) }-1\approx 2\sqrt{r}%
X\left( \varphi \right) $ for small $r$ is additionally used, since the
asymptotics%
\begin{eqnarray*}
\frac{\rho ^{\ast }\mathrm{e}^{\mathrm{i}\pi }+r\mathrm{e}^{\mathrm{i}%
\varphi }}{\sqrt{\tilde{G}\left( \rho ^{\ast }\mathrm{e}^{\mathrm{i}\pi }+r%
\mathrm{e}^{\mathrm{i}\varphi }\right) }} &\sim &\mathrm{i}\sqrt{r}\left(
\rho ^{\ast }\right) ^{\frac{1-\xi }{2}}\mathrm{e}^{\mathrm{i}\frac{\left(
2-\xi \right) \pi +\varphi }{2}}\sqrt{\frac{\alpha a_{1}\left( \rho ^{\ast
}\right) ^{\alpha }\mathrm{e}^{\mathrm{i}\alpha \pi }+\beta a_{2}\left( \rho
^{\ast }\right) ^{\beta }\mathrm{e}^{\mathrm{i}\beta \pi }+\gamma
a_{3}\left( \rho ^{\ast }\right) ^{\gamma }\mathrm{e}^{\mathrm{i}\gamma \pi }%
}{1+b\left( \rho ^{\ast }\right) ^{\eta }\mathrm{e}^{\mathrm{i}\eta \pi }}}%
\;\;\text{i.e.,} \\
&\sim &\sqrt{r}\left( X\left( \varphi \right) +\mathrm{i}Y\left( \varphi
\right) \right) \;\;\text{as}\;\;r\rightarrow 0,
\end{eqnarray*}%
holds for the Burgers models of both classes because%
\begin{equation}
\phi _{\sigma }\left( \rho ^{\ast }\mathrm{e}^{\mathrm{i}\pi }+r\mathrm{e}^{%
\mathrm{i}\varphi }\right) \sim -\frac{r\mathrm{e}^{\mathrm{i}\varphi }}{%
\rho ^{\ast }}\left( \alpha a_{1}\left( \rho ^{\ast }\right) ^{\alpha }%
\mathrm{e}^{\mathrm{i}\alpha \pi }+\beta a_{2}\left( \rho ^{\ast }\right)
^{\beta }\mathrm{e}^{\mathrm{i}\beta \pi }+\gamma a_{3}\left( \rho ^{\ast
}\right) ^{\gamma }\mathrm{e}^{\mathrm{i}\gamma \pi }\right) \;\;\text{as}%
\;\;r\rightarrow 0,  \label{fi-sigma-gama-8}
\end{equation}%
see equation (B.31) in \cite{OZ-2}, implying%
\begin{equation*}
\tilde{G}\left( \rho ^{\ast }\mathrm{e}^{\mathrm{i}\pi }+r\mathrm{e}^{%
\mathrm{i}\varphi }\right) \sim -\frac{1}{r\mathrm{e}^{\mathrm{i}\varphi }}%
\left( \rho ^{\ast }\right) ^{1+\xi }\mathrm{e}^{\mathrm{i}\xi \pi }\frac{%
1+b\left( \rho ^{\ast }\right) ^{\eta }\mathrm{e}^{\mathrm{i}\eta \pi }}{%
\alpha a_{1}\left( \rho ^{\ast }\right) ^{\alpha }\mathrm{e}^{\mathrm{i}%
\alpha \pi }+\beta a_{2}\left( \rho ^{\ast }\right) ^{\beta }\mathrm{e}^{%
\mathrm{i}\beta \pi }+\gamma a_{3}\left( \rho ^{\ast }\right) ^{\gamma }%
\mathrm{e}^{\mathrm{i}\gamma \pi }}
\end{equation*}%
with $\xi \in \left\{ \mu ,\beta \right\} .$ By the similar arguments the
integral $I_{\Gamma _{9}}$ also tends to zero as $r\rightarrow 0$.

\subsection{Case when function $\tilde{P}$ has a pair of complex conjugated
branch points in addition to $s=0$}

In the case when solution kernel image $\tilde{P},$ in addition to $s=0,$
has a pair of complex conjugated branch points $s_{0}$ and $\bar{s}_{0}$
having negative real part, with $s_{0}=\rho _{0}\mathrm{e}^{\mathrm{i}%
\varphi _{0}},$ $\varphi _{0}\in \left( \frac{\pi }{2},\pi \right) ,$
assuming that the absolute value of poles' argument is less then the
argument of branch points, the integration in the Cauchy residues theorem (%
\ref{reziduum}) is performed along the contour $\Gamma ^{\left( \mathrm{III}%
\right) },$ shown in Figure \ref{komplTG}, so that the integrals along
contours $\Gamma _{0}$, $\Gamma _{3a}\cup \Gamma _{3b}$, and $\Gamma
_{5a}\cup \Gamma _{5b}$ have non-zero contribution and according to
contours' parameterization given in Table \ref{komplTG-param} by (\ref%
{i-gama-i}) take the following forms%
\begin{eqnarray*}
\lim_{R\rightarrow \infty }I_{\Gamma _{0}} &=&\int_{Br}\tilde{P}\left(
x,s\right) \mathrm{e}^{st}\mathrm{d}s=2\pi \mathrm{i}P\left( x,t\right) , \\
\lim_{\substack{ R\rightarrow \infty  \\ r\rightarrow 0}}I_{\Gamma _{3a}\cup
\Gamma _{3b}} &=&\int_{\infty }^{0}\frac{\sinh \frac{x\rho \mathrm{e}^{%
\mathrm{i}\varphi _{0}}}{\sqrt{\tilde{G}\left( \rho \mathrm{e}^{\mathrm{i}%
\varphi _{0}}\right) }}}{\sinh \frac{\rho \mathrm{e}^{\mathrm{i}\varphi _{0}}%
}{\sqrt{\tilde{G}\left( \rho \mathrm{e}^{\mathrm{i}\varphi _{0}}\right) }}}%
\mathrm{e}^{\rho t\mathrm{e}^{\mathrm{i}\varphi _{0}}}\mathrm{e}^{\mathrm{i}%
\varphi _{0}}\mathrm{d}\rho =-\int_{0}^{\infty }\frac{\sinh \frac{x\rho 
\mathrm{e}^{\mathrm{i}\varphi _{0}}}{\sqrt{\tilde{G}\left( \rho \mathrm{e}^{%
\mathrm{i}\varphi _{0}}\right) }}}{\sinh \frac{\rho \mathrm{e}^{\mathrm{i}%
\varphi _{0}}}{\sqrt{\tilde{G}\left( \rho \mathrm{e}^{\mathrm{i}\varphi
_{0}}\right) }}}\mathrm{e}^{\mathrm{i}\left( \rho t\mathrm{\sin }\varphi
_{0}+\varphi _{0}\right) }\mathrm{e}^{-\rho t\left\vert \mathrm{\cos }%
\varphi _{0}\right\vert }\mathrm{d}\rho , \\
\lim_{\substack{ R\rightarrow \infty  \\ r\rightarrow 0}}I_{\Gamma _{5a}\cup
\Gamma _{5b}} &=&\int_{0}^{\infty }\frac{\sinh \frac{x\rho \mathrm{e}^{-%
\mathrm{i}\varphi _{0}}}{\sqrt{\tilde{G}\left( \rho \mathrm{e}^{-\mathrm{i}%
\varphi _{0}}\right) }}}{\sinh \frac{\rho \mathrm{e}^{-\mathrm{i}\varphi
_{0}}}{\sqrt{\tilde{G}\left( \rho \mathrm{e}^{-\mathrm{i}\varphi
_{0}}\right) }}}\mathrm{e}^{\rho t\mathrm{e}^{-\mathrm{i}\varphi _{0}}}%
\mathrm{e}^{-\mathrm{i}\varphi _{0}}\mathrm{d}\rho =\int_{0}^{\infty }\frac{%
\sinh \frac{x\rho \mathrm{e}^{-\mathrm{i}\varphi _{0}}}{\sqrt{\tilde{G}%
\left( \rho \mathrm{e}^{-\mathrm{i}\varphi _{0}}\right) }}}{\sinh \frac{\rho 
\mathrm{e}^{-\mathrm{i}\varphi _{0}}}{\sqrt{\tilde{G}\left( \rho \mathrm{e}%
^{-\mathrm{i}\varphi _{0}}\right) }}}\mathrm{e}^{-\mathrm{i}\left( \rho t%
\mathrm{\sin }\varphi _{0}+\varphi _{0}\right) }\mathrm{e}^{-\rho
t\left\vert \mathrm{\cos }\varphi _{0}\right\vert }\mathrm{d}\rho ,
\end{eqnarray*}%
while the residues are already calculated in Section \ref{no-branch-points}
and given by (\ref{res-3}), so that the solution kernel $P$ takes the form 
\begin{align}
& 2\pi \mathrm{i}P\left( x,t\right) -2\mathrm{i}\int_{0}^{\infty }\func{Im}%
\left( \frac{\sinh \frac{x\rho \mathrm{e}^{\mathrm{i}\varphi _{0}}}{\sqrt{%
\tilde{G}\left( \rho \mathrm{e}^{\mathrm{i}\varphi _{0}}\right) }}}{\sinh 
\frac{\rho \mathrm{e}^{\mathrm{i}\varphi _{0}}}{\sqrt{\tilde{G}\left( \rho 
\mathrm{e}^{\mathrm{i}\varphi _{0}}\right) }}}\mathrm{e}^{\mathrm{i}\left(
\rho t\mathrm{\sin }\varphi _{0}+\varphi _{0}\right) }\right) \mathrm{e}%
^{-\rho t\left\vert \mathrm{\cos }\varphi _{0}\right\vert }\mathrm{d}\rho 
\notag \\
& \quad \quad \quad \quad \quad \quad =4\pi \mathrm{i}\sum_{k=1}^{\infty
}\left( -1\right) ^{k}\frac{\sin \left( k\pi x\right) }{k\pi }\mathrm{e}%
^{-\rho _{k}t\left\vert \cos \varphi _{k}\right\vert }\func{Re}\left( \frac{%
s_{k}\mathrm{e}^{\mathrm{i}\rho _{k}t\sin \varphi _{k}}}{1+\left( k\pi
\right) ^{2}\frac{\tilde{G}^{\prime }\left( s_{k}\right) }{2s_{k}}}\right) 
\notag \\
& P\left( x,t\right) =\frac{1}{\pi }\int_{0}^{\infty }\func{Im}\left( \frac{%
\sinh \frac{x\rho \mathrm{e}^{\mathrm{i}\varphi _{0}}}{\sqrt{\tilde{G}\left(
\rho \mathrm{e}^{\mathrm{i}\varphi _{0}}\right) }}}{\sinh \frac{\rho \mathrm{%
e}^{\mathrm{i}\varphi _{0}}}{\sqrt{\tilde{G}\left( \rho \mathrm{e}^{\mathrm{i%
}\varphi _{0}}\right) }}}\mathrm{e}^{\mathrm{i}\left( \rho t\mathrm{\sin }%
\varphi _{0}+\varphi _{0}\right) }\right) \mathrm{e}^{-\rho t\left\vert 
\mathrm{\cos }\varphi _{0}\right\vert }\mathrm{d}\rho  \notag \\
& \quad \quad \quad \quad \quad \quad +2\sum_{k=1}^{\infty }\left( -1\right)
^{k}\frac{\sin \left( k\pi x\right) }{k\pi }\mathrm{e}^{-\rho
_{k}t\left\vert \cos \varphi _{k}\right\vert }\func{Re}\left( \frac{s_{k}%
\mathrm{e}^{\mathrm{i}\rho _{k}t\sin \varphi _{k}}}{1+\left( k\pi \right)
^{2}\frac{\tilde{G}^{\prime }\left( s_{k}\right) }{2s_{k}}}\right) ,
\label{p3e}
\end{align}%
since the integrals along all other contours are zero. It is already proved
in Section \ref{no-branch-points} that the integrals along contours $\Gamma
_{1},$ $\Gamma _{2},$ $\Gamma _{6},$ and $\Gamma _{7}$ have zero
contribution, so it is left to prove that the integrals along the contours $%
\Gamma _{8}$ and $\Gamma _{9}$ have zero contributions as well.

\noindent 
\begin{minipage}{\columnwidth}
\begin{minipage}[c]{0.4\columnwidth}
\centering
\includegraphics[width=0.7\columnwidth]{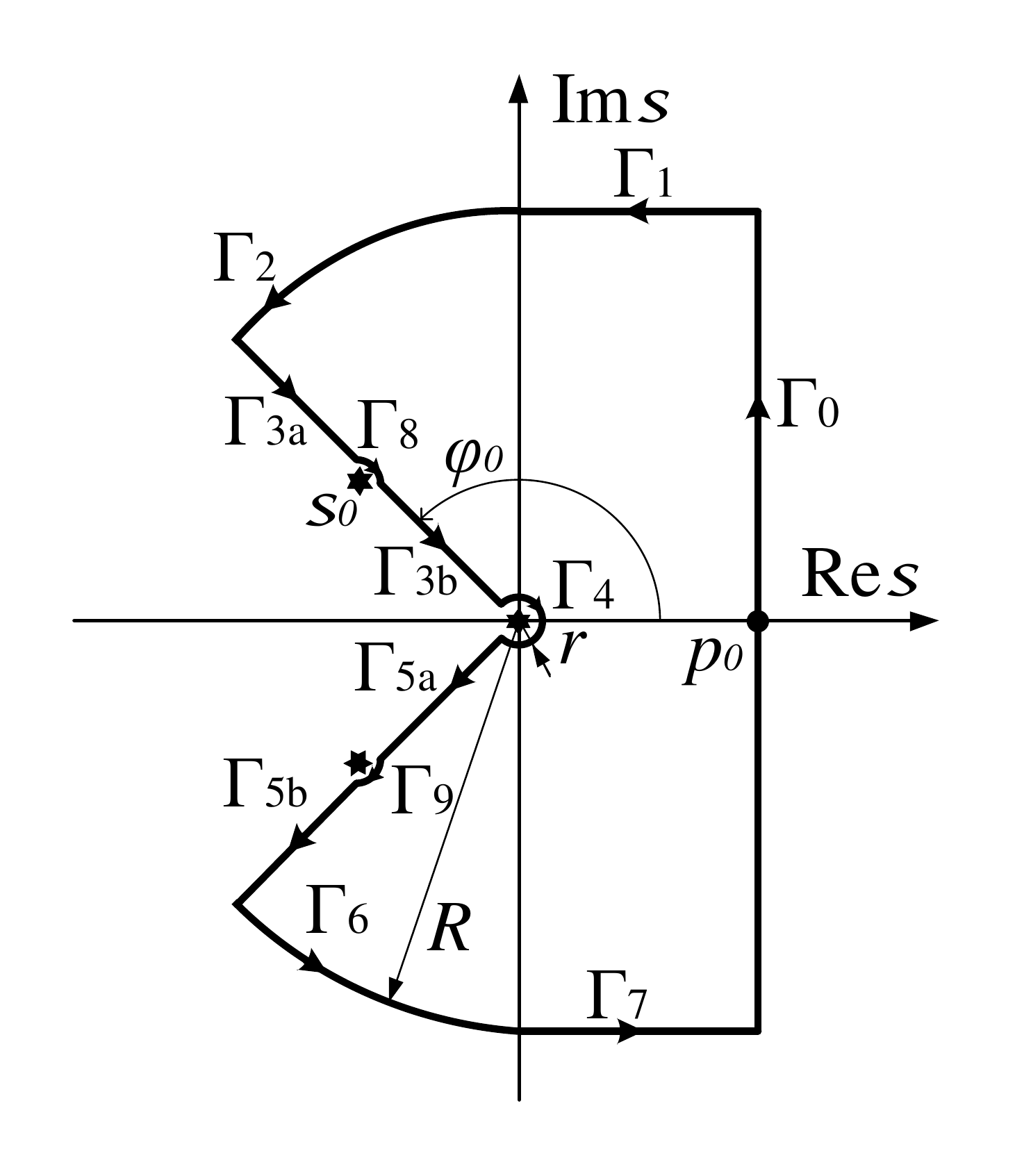}
\captionof{figure}{Integration contour $\Gamma^{(\mathrm{III})}$.}
\label{komplTG}
\end{minipage}
\hfil
\begin{minipage}[c]{0.55\columnwidth}
\centering
\begin{tabular}{rll}
$\Gamma _{0}:$ & Bromwich path, &  \\ 
$\Gamma _{1}:$ & $s=p+\mathrm{i}R,$ & $p\in \left[ 0,p_{0}\right],\, p_0\geq 0$ arbitrary, \\ 
$\Gamma _{2}:$ & $s=R\mathrm{e}^{\mathrm{i}\varphi },$ & $\varphi \in \left[ 
\frac{\pi }{2},\varphi_0 \right] ,$ \\ 
$\Gamma _{3a}\cup\Gamma _{3b}:$ & $s=\rho \mathrm{e}^{\mathrm{i}\varphi_0},$ & $\rho \in \left[ r,R%
\right] ,$ \\ 
$\Gamma _{4}:$ & $s=r\mathrm{e}^{\mathrm{i}\varphi },$ & $\varphi \in \left[ -\varphi_0,\varphi_0\right] ,$ \\ 
$\Gamma _{5a}\cup\Gamma _{5b}:$ & $s=\rho \mathrm{e}^{-\mathrm{i}\varphi_0},$ & $\rho \in \left[ r,R%
\right] ,$ \\
$\Gamma _{6}:$  & $s=R\mathrm{e}^{\mathrm{i}\varphi },$ & $\varphi \in \left[ 
-\varphi_0 , -\frac{\pi }{2} \right] ,$ \\
$\Gamma _{7}:$ & $s=p-\mathrm{i}R,$ & $p\in \left[ 0,p_{0}\right],\, p_0\geq 0$ arbitrary,\\
$\Gamma _{8}:$  & $s=s_0+r\mathrm{e}^{\mathrm{i}\varphi },$ & $\varphi \in \left[ 
-\pi+\varphi_0, \varphi_0 \right] ,$ \\
$\Gamma _{9}:$  & $s=\bar{s}_0+r\mathrm{e}^{\mathrm{i}\varphi },$ & $\varphi \in \left[ -\varphi_0,
\pi-\varphi_0 \right]$.   
\end{tabular}
\captionof{table}{Parametrization of integration contour $\Gamma^{(\mathrm{III})}$.}
\label{komplTG-param}
\end{minipage}
\end{minipage}\smallskip

Integral on the contour $\Gamma _{8},$ parameterized using parameterization
given in Table \ref{komplTG-param}, by (\ref{i-gama-i}) is%
\begin{equation*}
I_{\Gamma _{8}}=\int_{-\varphi _{0}}^{\pi -\varphi _{0}}\frac{\sinh \frac{%
x\left( s_{0}+r\mathrm{e}^{\mathrm{i}\varphi }\right) }{\sqrt{\tilde{G}%
\left( s_{0}+r\mathrm{e}^{\mathrm{i}\varphi }\right) }}}{\sinh \frac{\left(
s_{0}+r\mathrm{e}^{\mathrm{i}\varphi }\right) }{\sqrt{\tilde{G}\left( s_{0}+r%
\mathrm{e}^{\mathrm{i}\varphi }\right) }}}\mathrm{e}^{\left( s_{0}+r\mathrm{e%
}^{\mathrm{i}\varphi }\right) t}r\mathrm{ie}^{\mathrm{i}\varphi }\mathrm{d}%
\varphi ,
\end{equation*}%
so that%
\begin{eqnarray*}
\left\vert I_{\Gamma _{8}}\right\vert &\leqslant &\mathrm{e}%
^{s_{0}t}\int_{-\varphi _{0}}^{\pi -\varphi _{0}}\left\vert \mathrm{e}%
^{\left( 1-x\right) \sqrt{r}\left( X\left( \varphi \right) +\mathrm{i}%
Y\left( \varphi \right) \right) }\right\vert \frac{\left\vert \mathrm{e}^{2x%
\sqrt{r}\left( X\left( \varphi \right) +\mathrm{i}Y\left( \varphi \right)
\right) }-1\right\vert }{\left\vert \mathrm{e}^{2\sqrt{r}\left( X\left(
\varphi \right) +\mathrm{i}Y\left( \varphi \right) \right) }-1\right\vert }r%
\mathrm{d}\varphi \\
&\leqslant &\mathrm{e}^{s_{0}t}\int_{-\varphi _{0}}^{\pi -\varphi _{0}}%
\mathrm{e}^{\left( 1-x\right) \sqrt{r}X\left( \varphi \right) }\frac{\mathrm{%
e}^{2x\sqrt{r}X\left( \varphi \right) }+1}{\left\vert \mathrm{e}^{2\sqrt{r}%
X\left( \varphi \right) }-1\right\vert }r\mathrm{d}\varphi \\
&\leqslant &\mathrm{e}^{s_{0}t}\int_{-\varphi _{0}}^{\pi -\varphi _{0}}\frac{%
\sqrt{r}}{\left\vert X\left( \varphi \right) \right\vert }\mathrm{d}\varphi
\rightarrow 0\;\;\text{as}\;\;r\rightarrow 0,
\end{eqnarray*}%
where $\mathrm{e}^{2\sqrt{r}X\left( \varphi \right) }-1\approx 2\sqrt{r}%
X\left( \varphi \right) $ for small $r$ is additionally used, since the
asymptotics%
\begin{eqnarray*}
\frac{s_{0}+r\mathrm{e}^{\mathrm{i}\varphi }}{\sqrt{\tilde{G}\left( s_{0}+r%
\mathrm{e}^{\mathrm{i}\varphi }\right) }} &\sim &\sqrt{r}\mathrm{e}^{\mathrm{%
i}\frac{\varphi }{2}}s_{0}^{\frac{1-\xi }{2}}\sqrt{\frac{\alpha
a_{1}s_{0}^{\alpha }+\beta a_{2}s_{0}^{\beta }+\gamma a_{3}s_{0}^{\gamma }}{%
1+bs_{0}^{\eta }}}\;\;\text{i.e.,} \\
&\sim &\sqrt{r}\left( X\left( \varphi \right) +\mathrm{i}Y\left( \varphi
\right) \right) \;\;\text{as}\;\;r\rightarrow 0,
\end{eqnarray*}%
holds for the Burgers models of both classes, because%
\begin{equation*}
\phi _{\sigma }\left( s_{0}+r\mathrm{e}^{\mathrm{i}\varphi }\right) \sim 
\frac{r\mathrm{e}^{\mathrm{i}\varphi }}{s_{0}}\left( \alpha
a_{1}s_{0}^{\alpha }+\beta a_{2}s_{0}^{\beta }+\gamma a_{3}s_{0}^{\gamma
}\right) \;\;\text{as}\;\;r\rightarrow 0,
\end{equation*}%
that is obtained similarly as (\ref{fi-sigma-gama-8}) implies%
\begin{equation*}
\tilde{G}\left( s_{0}+r\mathrm{e}^{\mathrm{i}\varphi }\right) \sim \frac{1}{r%
\mathrm{e}^{\mathrm{i}\varphi }}s_{0}^{1+\xi }\frac{1+bs_{0}^{\eta }}{\alpha
a_{1}s_{0}^{\alpha }+\beta a_{2}s_{0}^{\beta }+\gamma a_{3}s_{0}^{\gamma }}
\end{equation*}%
with $\xi \in \left\{ \mu ,\beta \right\} .$ By the similar arguments the
integral $I_{\Gamma _{9}}$ also tends to zero as $r\rightarrow 0$.

\section{Zeros of function $\protect\psi $ \label{polovi}}

In order to obtain poles of functions $\tilde{P},$ $\tilde{R},$ $\tilde{Q},$
and $\tilde{S},$ according to equations (\ref{sinus}) and (\ref{kosinus}),
one needs to examine the position, number, and multiplicity of zeros of
function $\psi $, defined by%
\begin{equation}
\psi \left( s\right) =\frac{s^{2}}{\tilde{G}\left( s\right) }+\vartheta
=0,\;\;\text{with}\;\;\vartheta =\left\{ 
\begin{tabular}{l}
$\left( k\pi \right) ^{2},$ \\ 
$\left( \frac{2k+1}{2}\pi \right) ^{2}.$%
\end{tabular}%
\right.  \label{psi}
\end{equation}%
Although equation (\ref{psi}) cannot be solved analytically, it will be
proved that function $\psi $, defined by (\ref{psi}), for each $k$ has a
pair of complex conjugated zeros having negative real part, representing
poles of functions $\tilde{P},$ $\tilde{R},$ $\tilde{Q},$ and $\tilde{S}.$

Introducing the substitution $s=\rho \mathrm{e}^{\mathrm{i}\varphi }$ into
function $\psi ,$ given by (\ref{psi}), its real and imaginary parts become%
\begin{eqnarray}
\func{Re}\psi \left( \rho ,\varphi \right)  &=&\frac{\rho ^{2-\xi }}{%
\left\vert 1+b\rho ^{\eta }\mathrm{e}^{\mathrm{i}\eta \varphi }\right\vert
^{2}}\left( g^{\left( \mathrm{I},\mathrm{II}\right) }\left( \rho ,\varphi
\right) \cos \left( 2\varphi \right) +f^{\left( \mathrm{I},\mathrm{II}%
\right) }\left( \rho ,\varphi \right) \sin \left( 2\varphi \right) \right)
+\vartheta ,  \label{repsi} \\
\func{Im}\psi \left( \rho ,\varphi \right)  &=&\frac{\rho ^{2-\xi }}{%
\left\vert 1+b\rho ^{\eta }\mathrm{e}^{\mathrm{i}\eta \varphi }\right\vert
^{2}}\left( g^{\left( \mathrm{I},\mathrm{II}\right) }\left( \rho ,\varphi
\right) \sin \left( 2\varphi \right) -f^{\left( \mathrm{I},\mathrm{II}%
\right) }\left( \rho ,\varphi \right) \cos \left( 2\varphi \right) \right) ,
\label{impsi}
\end{eqnarray}%
with $\xi =\left\{ \mu ,\beta \right\} $ for the first, respectively for the
second class of Burgers models, where functions $g^{\left( \mathrm{I},%
\mathrm{II}\right) }$ and $f^{\left( \mathrm{I},\mathrm{II}\right) }$ are
given by%
\begin{eqnarray}
g^{\left( \mathrm{I}\right) }\left( \rho ,\varphi \right)  &=&\cos \left(
\mu \varphi \right) +a_{1}\rho ^{\alpha }\cos \left( \left( \mu -\alpha
\right) \varphi \right) +a_{2}\rho ^{\beta }\cos \left( \left( \mu -\beta
\right) \varphi \right) +a_{3}\rho ^{\gamma }\cos \left( \left( \mu -\gamma
\right) \varphi \right) +b\rho ^{\eta }\cos \left( \left( \mu +\eta \right)
\varphi \right)   \notag \\
&&+a_{1}b\rho ^{\alpha +\eta }\cos \left( \left( \mu +\eta -\alpha \right)
\varphi \right) +a_{2}b\rho ^{\beta +\eta }\cos \left( \left( \mu +\eta
-\beta \right) \varphi \right) +a_{3}b\rho ^{\gamma +\eta }\cos \left(
\left( \mu +\eta -\gamma \right) \varphi \right) ,  \notag \\
f^{\left( \mathrm{I}\right) }\left( \rho ,\varphi \right)  &=&\sin \left(
\mu \varphi \right) +a_{1}\rho ^{\alpha }\sin \left( \left( \mu -\alpha
\right) \varphi \right) +a_{2}\rho ^{\beta }\sin \left( \left( \mu -\beta
\right) \varphi \right) +a_{3}\rho ^{\gamma }\sin \left( \left( \mu -\gamma
\right) \varphi \right) +b\rho ^{\eta }\sin \left( \left( \mu +\eta \right)
\varphi \right)   \notag \\
&&+a_{1}b\rho ^{\alpha +\eta }\sin \left( \left( \mu +\eta -\alpha \right)
\varphi \right) +a_{2}b\rho ^{\beta +\eta }\sin \left( \left( \mu +\eta
-\beta \right) \varphi \right) +a_{3}b\rho ^{\gamma +\eta }\sin \left(
\left( \mu +\eta -\gamma \right) \varphi \right) ,  \label{ef1}
\end{eqnarray}%
for the first class and by 
\begin{eqnarray}
g^{\left( \mathrm{II}\right) }\left( \rho ,\varphi \right)  &=&\cos \left(
\beta \varphi \right) +a_{1}\rho ^{\alpha }\cos \left( \left( \beta -\alpha
\right) \varphi \right) +a_{2}\rho ^{\beta }+b\rho ^{\eta }\cos \left(
\left( \beta +\eta \right) \varphi \right)   \notag \\
&&+a_{1}b\rho ^{\alpha +\eta }\cos \left( \left( \eta +\beta -\alpha \right)
\varphi \right) +\left( a_{2}b+a_{3}\right) \rho ^{\beta +\eta }\cos \left(
\eta \varphi \right) +a_{3}b\rho ^{\beta +2\eta },  \notag \\
f^{\left( \mathrm{II}\right) }\left( \rho ,\varphi \right)  &=&\sin \left(
\beta \varphi \right) +a_{1}\rho ^{\alpha }\sin \left( \left( \beta -\alpha
\right) \varphi \right) +b\rho ^{\eta }\sin \left( \left( \beta +\eta
\right) \varphi \right)   \notag \\
&&+a_{1}b\rho ^{\alpha +\eta }\sin \left( \left( \eta +\beta -\alpha \right)
\varphi \right) +a_{2}\left( b-\frac{a_{3}}{a_{2}}\right) \rho ^{\beta +\eta
}\sin \left( \eta \varphi \right) ,  \label{ef2}
\end{eqnarray}%
for the second class. Note that $b-\frac{a_{3}}{a_{2}}>0$ according to the
thermodynamical restrictions.

If $s_{0}=\rho _{0}\mathrm{e}^{\mathrm{i}\varphi _{0}}$ is a zero of
function $\psi ,$ then its complex conjugate $\bar{s}_{0}=\rho _{0}\mathrm{e}%
^{-\mathrm{i}\varphi _{0}}$ is a zero as well, since $\func{Im}\psi \left(
\rho ,-\varphi \right) =-\func{Im}\psi \left( \rho ,\varphi \right) ,$
according to (\ref{impsi}), hence it is sufficient to seek for zeros of
function $\psi $ in the upper complex half-plane only. Moreover, as proved
in Appendix A of \cite{OZO}, function $\psi $ does not have zeros in the
right complex half-plane and therefore in order to have a pair of complex
conjugated zeros of $\psi $ lying in the left half-plane it is left to prove
using the argument principle and contour $\gamma ,$ shown in Figure \ref%
{fig-gama} along with its parameterization given in Table \ref%
{fig-gama-param}, that the zeros of $\psi $ lie in the upper left complex
quarter-plane. Recall, the argument principle claims that if the variable $s$
changes along the contour $\gamma $ closed in the complex plane, then the
number of zeros $N$ of function $\psi $ in the domain encircled by contour $%
\gamma $ is given by $\Delta \arg \psi \left( s\right) =2\pi N,$ provided
that function $\psi $ does not have poles in the mentioned domain.

\noindent 
\begin{minipage}{\columnwidth}
\begin{minipage}[c]{0.4\columnwidth}
\centering
\includegraphics[width=0.7\columnwidth]{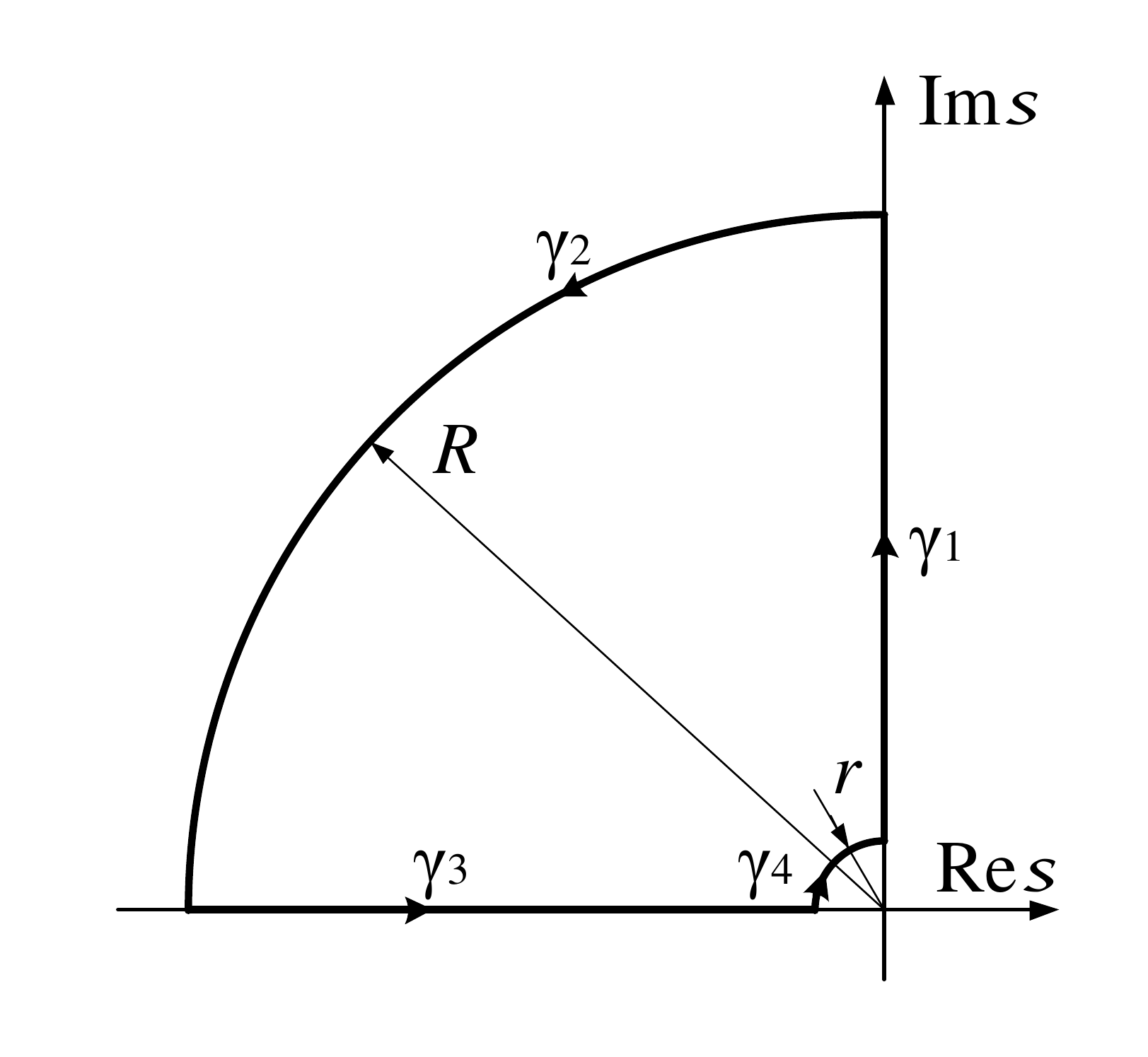}
\captionof{figure}{Contour $\gamma$.}
\label{fig-gama}
\end{minipage}
\hfil
\begin{minipage}[c]{0.55\columnwidth}
\centering
\begin{tabular}{rll}
$\gamma _{1}:$ & $s=\rho \mathrm{e}^{\mathrm{i}\frac{\pi}{2}},$ & $\rho \in \left[ r,R%
\right] ,$ \\
$\gamma _{2}:$ & $s=R\mathrm{e}^{\mathrm{i}\varphi },$ & $\varphi \in \left[ 
\frac{\pi }{2},\pi \right] ,$ \\ 
$\gamma _{3}:$ & $s=\rho \mathrm{e}^{\mathrm{i}\pi },$ & $\rho \in \left[ r,R%
\right] ,$ \\ 
$\gamma _{4}:$ & $s=r\mathrm{e}^{\mathrm{i}\varphi },$ & $\varphi \in \left[ 
\frac{\pi }{2},\pi \right] .$ \\ 
\end{tabular}
\captionof{table}{Parametrization of contour $\gamma$.}
\label{fig-gama-param}
\end{minipage}
\end{minipage}\smallskip

The argument of complex numbers belonging to the contour $\gamma _{1}$ has a
fixed value $\varphi =\frac{\pi }{2},$ while their modulus changes so that $%
\rho \in \left( 0,\infty \right) ,$ implying the real and imaginary parts of
function $\psi ,$ given by (\ref{repsi}) and (\ref{impsi}), in the form%
\begin{eqnarray}
\func{Re}\psi \left( \rho ,\frac{\pi }{2}\right)  &=&-\frac{\rho ^{2-\xi }}{%
\left\vert 1+b\rho ^{\eta }\mathrm{e}^{\mathrm{i}\eta \varphi }\right\vert
^{2}}g^{\left( \mathrm{I},\mathrm{II}\right) }\left( \rho ,\frac{\pi }{2}%
\right) +\vartheta ,  \label{repsi-gama1} \\
\func{Im}\psi \left( \rho ,\frac{\pi }{2}\right)  &=&\frac{\rho ^{2-\xi }}{%
\left\vert 1+b\rho ^{\eta }\mathrm{e}^{\mathrm{i}\eta \varphi }\right\vert
^{2}}f^{\left( \mathrm{I},\mathrm{II}\right) }\left( \rho ,\frac{\pi }{2}%
\right) >0,  \label{impsi-gama1}
\end{eqnarray}%
since all terms in $f^{\left( \mathrm{I},\mathrm{II}\right) }$ are positive,
see (\ref{ef1}) and (\ref{ef2}). The asymptotics of (\ref{repsi-gama1}) and (%
\ref{impsi-gama1}) for both model classes as $\rho =r\rightarrow 0$ yields 
\begin{eqnarray*}
\func{Re}\psi \left( r,\frac{\pi }{2}\right)  &\sim &-r^{2-\xi }\cos \frac{%
\xi \pi }{2}+\vartheta \rightarrow \vartheta \;\;\text{and} \\
\func{Im}\psi \left( r,\frac{\pi }{2}\right)  &\sim &r^{2-\xi }\sin \frac{%
\xi \pi }{2}\rightarrow 0^{+},
\end{eqnarray*}%
while the asymptotics as $\rho =R\rightarrow \infty $ for the first model
class gives%
\begin{eqnarray}
\func{Re}\psi \left( R,\frac{\pi }{2}\right)  &\sim &-\frac{a_{3}}{b}%
R^{2-\mu -\eta +\gamma }\cos \frac{\left( \mu +\eta -\gamma \right) \pi }{2}%
\rightarrow -\infty \;\;\text{and}  \label{repsi-gama1-pi-pola} \\
\func{Im}\psi \left( R,\frac{\pi }{2}\right)  &\sim &\frac{a_{3}}{b}R^{2-\mu
-\eta +\gamma }\sin \frac{\left( \mu +\eta -\gamma \right) \pi }{2}%
\rightarrow \infty ,  \label{impsi-gama1-pi-pola}
\end{eqnarray}%
since by the thermodynamic requirements $0\leq \mu +\eta -\gamma \leq 1,$ as
well as%
\begin{eqnarray}
\func{Re}\psi \left( R,\frac{\pi }{2}\right)  &\sim &-\frac{a_{3}}{b}%
R^{2}\rightarrow -\infty \;\;\text{and}  \label{repsi-gama1-pi-pola-II-klas}
\\
\func{Im}\psi \left( R,\frac{\pi }{2}\right)  &\sim &\frac{a_{2}}{b^{2}}%
\left( b-\frac{a_{3}}{a_{2}}\right) R^{2-\eta }\sin \frac{\eta \pi }{2}%
\rightarrow \infty   \label{impsi-gama1-pi-pola-II-klas}
\end{eqnarray}%
for the second model class. From the asymptotics (\ref{repsi-gama1-pi-pola})
and (\ref{impsi-gama1-pi-pola}) it follows that $\left\vert \psi \left( R,%
\frac{\pi }{2}\right) \right\vert \sim \frac{a_{3}}{b}R^{2-\mu -\eta +\gamma
}\rightarrow \infty $ and $\tan \arg \psi \left( R,\frac{\pi }{2}\right)
\sim -\tan \frac{\left( \mu +\eta -\gamma \right) \pi }{2}$ implying $\arg
\psi \left( R,\frac{\pi }{2}\right) \sim \frac{\left( 2-\mu -\eta +\gamma
\right) \pi }{2},$ with $\frac{\pi }{2}\leq \frac{\left( 2-\mu -\eta +\gamma
\right) \pi }{2}\leq \pi ,$ while form the asymptotics (\ref%
{repsi-gama1-pi-pola-II-klas}) and (\ref{impsi-gama1-pi-pola-II-klas}) it
follows that $\left\vert \psi \left( R,\frac{\pi }{2}\right) \right\vert
\sim \frac{a_{3}}{b}R^{2}\rightarrow \infty $ and $\tan \arg \psi \left( R,%
\frac{\pi }{2}\right) \sim -\frac{a_{2}}{a_{3}b}\left( b-\frac{a_{3}}{a_{2}}%
\right) \frac{1}{R^{\eta }}\sin \frac{\eta \pi }{2}$ implying $\arg \psi
\left( R,\frac{\pi }{2}\right) \sim \pi $. In conclusion, as $s$ changes
along the contour $\gamma _{1},$ the argument of function $\psi $ changes
from zero either to $\arg \psi \left( R,\frac{\pi }{2}\right) \sim \frac{%
\left( 2-\mu -\eta +\gamma \right) \pi }{2}$ in the case of Burgers models
of the first class, or to $\arg \psi \left( R,\frac{\pi }{2}\right) \sim \pi 
$ in the case of models of the second class.

Complex number lying on the contour $\gamma _{2}$ has large but fixed
modulus $\rho =R,$ while its argument changes along the contour $\gamma _{2}$
taking the values $\varphi \in \left[ \frac{\pi }{2},\pi \right] ,$ so that
the asymptotics of (\ref{repsi}) and (\ref{impsi}) as $R\rightarrow \infty $
in the case of Burgers models belonging to the first class gives%
\begin{eqnarray}
\func{Re}\psi \left( R,\varphi \right) &\sim &\frac{a_{3}}{b}R^{2-\mu -\eta
+\gamma }\cos \left( \left( 2-\mu -\eta +\gamma \right) \varphi \right) \;\;%
\text{and}  \label{repsi-gama2} \\
\func{Im}\psi \left( R,\varphi \right) &\sim &\frac{a_{3}}{b}R^{2-\mu -\eta
+\gamma }\sin \left( \left( 2-\mu -\eta +\gamma \right) \varphi \right) ,
\label{impsi-gama2}
\end{eqnarray}%
as well as 
\begin{eqnarray}
\func{Re}\psi \left( R,\varphi \right) &\sim &\frac{a_{3}}{b}R^{2}\cos
\left( 2\varphi \right) \;\;\text{and}  \label{repsi-gama2-II-klas} \\
\func{Im}\psi \left( R,\varphi \right) &\sim &R^{2}\left( \frac{a_{3}}{b}%
\sin \left( 2\varphi \right) -\frac{1}{R^{\eta }}\frac{a_{2}}{b^{2}}\left( b-%
\frac{a_{3}}{a_{2}}\right) \sin \left( \eta \varphi \right) \cos \left(
2\varphi \right) \right) ,  \label{impsi-gama2-II-klas}
\end{eqnarray}%
for the Burgers models belonging to the second class, where the second term
in $\func{Im}\psi $ disappears for $\varphi \in \left( \frac{\pi }{2},\pi
\right) ,$ but has a dominant role if $\varphi =\frac{\pi }{2}$ or $\varphi
=\pi .$ Therefore, real and imaginary parts of function $\psi $\ cannot be
simultaneously positive for $\varphi \in \left( \frac{\pi }{2},\pi \right) ,$%
\ since $\frac{\pi }{2}\leq \left( 2-\mu -\eta +\gamma \right) \varphi \leq
2\pi $\ in the case of Models I, III, and IV, as well as $\frac{3\pi }{4}%
\leq \left( 2-\mu -\eta +\gamma \right) \varphi \leq 2\pi $\ in the case of
Models II and V, and obviously for models of the second class. If $\varphi =%
\frac{\pi }{2},$ then (\ref{repsi-gama2}) and (\ref{impsi-gama2}) reduce to (%
\ref{repsi-gama1-pi-pola}) and (\ref{impsi-gama1-pi-pola}), so as (\ref%
{repsi-gama2-II-klas}) and (\ref{impsi-gama2-II-klas}) to (\ref%
{repsi-gama1-pi-pola-II-klas}) and (\ref{impsi-gama1-pi-pola-II-klas}),
while if $\varphi =\pi ,$ then (\ref{repsi-gama2}) and (\ref{impsi-gama2})
become%
\begin{eqnarray}
\func{Re}\psi \left( R,\pi \right) &\sim &\frac{a_{3}}{b}R^{2-\mu -\eta
+\gamma }\cos \left( \left( \mu +\eta -\gamma \right) \pi \right)
\rightarrow \pm \infty \;\;\text{and}  \label{repsi-gama2-pi} \\
\func{Im}\psi \left( R,\pi \right) &\sim &-\frac{a_{3}}{b}R^{2-\mu -\eta
+\gamma }\sin \left( \left( \mu +\eta -\gamma \right) \pi \right)
\rightarrow -\infty ,  \label{impsi-gama2-pi}
\end{eqnarray}%
while (\ref{repsi-gama2-II-klas}) and (\ref{impsi-gama2-II-klas}) become%
\begin{eqnarray}
\func{Re}\psi \left( R,\pi \right) &\sim &\frac{a_{3}}{b}R^{2}\rightarrow
\infty \;\;\text{and}  \label{repsi-gama2-pi-II-klas} \\
\func{Im}\psi \left( R,\pi \right) &\sim &-\frac{a_{2}}{b^{2}}\left( b-\frac{%
a_{3}}{a_{2}}\right) R^{2-\eta }\sin \left( \eta \pi \right) \rightarrow
-\infty .  \label{impsi-gama2-pi-II-klas}
\end{eqnarray}%
Therefore, as $s$ changes along contour $\gamma _{2}$, the argument of
function $\psi $ changes through the second, third, and possibly fourth
quadrant.

The argument of complex numbers lying on the contour $\gamma _{3}$ has a
fixed value $\varphi =\pi ,$ while their modulus changes in the interval $%
\rho \in \left( 0,\infty \right) ,$ yielding%
\begin{eqnarray}
\func{Re}\psi \left( \rho ,\pi \right)  &=&\frac{\rho ^{2-\xi }}{\left\vert
1+b\rho ^{\eta }\mathrm{e}^{\mathrm{i}\eta \varphi }\right\vert ^{2}}%
g^{\left( \mathrm{I},\mathrm{II}\right) }\left( \rho ,\pi \right) +\vartheta
,  \label{repsi-gama3} \\
\func{Im}\psi \left( \rho ,\pi \right)  &=&-\frac{\rho ^{2-\xi }}{\left\vert
1+b\rho ^{\eta }\mathrm{e}^{\mathrm{i}\eta \varphi }\right\vert ^{2}}%
f^{\left( \mathrm{I},\mathrm{II}\right) }\left( \rho ,\pi \right) <0,
\label{impsi-gama3}
\end{eqnarray}%
for the real and imaginary parts of the function $\psi ,$ since $f^{\left( 
\mathrm{I},\mathrm{II}\right) }\left( \rho ,\pi \right) >0$ for $\rho \in
\left( 0,\infty \right) ,$ due to the conditions narrowing thermodynamical
restrictions in the case of all fractional Burgers models, which is proved
for each model separately. In the case of Model I, function $f^{\left( 
\mathrm{I}\right) }$ obtained in the form 
\begin{eqnarray*}
f^{\left( \mathrm{I}\right) }\left( \rho ,\pi \right)  &=&\sin \left( \mu
\pi \right)  \\
&&+\left\{ 
\begin{tabular}{l}
$\rho ^{\alpha }\left\vert \sin \left( \left( \mu +\alpha \right) \pi
\right) \right\vert \left( a_{1}\frac{\sin \left( \left( \mu -\alpha \right)
\pi \right) }{\left\vert \sin \left( \left( \mu +\alpha \right) \pi \right)
\right\vert }-b\right) +a_{2}\rho ^{\beta }\sin \left( \left( \mu -\beta
\right) \pi \right) +a_{3}\rho ^{\gamma }\sin \left( \left( \mu -\gamma
\right) \pi \right) $ \\ 
$\quad +a_{1}b\rho ^{2\alpha }\sin \left( \mu \pi \right) +a_{2}b\rho
^{\alpha +\beta }\sin \left( \left( \mu +\alpha -\beta \right) \pi \right)
+a_{3}b\rho ^{\alpha +\gamma }\sin \left( \left( \mu +\alpha -\gamma \right)
\pi \right) ,$ \\ 
$a_{1}\rho ^{\alpha }\sin \left( \left( \mu -\alpha \right) \pi \right)
+\rho ^{\beta }\left\vert \sin \left( \left( \mu +\beta \right) \pi \right)
\right\vert \left( a_{2}\frac{\sin \left( \left( \mu -\beta \right) \pi
\right) }{\left\vert \sin \left( \left( \mu +\beta \right) \pi \right)
\right\vert }-b\right) +a_{3}\rho ^{\gamma }\sin \left( \left( \mu -\gamma
\right) \pi \right) $ \\ 
$\quad +a_{1}b\rho ^{\alpha +\beta }\sin \left( \left( \mu +\beta -\alpha
\right) \pi \right) +a_{2}b\rho ^{2\beta }\sin \left( \mu \pi \right)
+a_{3}b\rho ^{\gamma +\beta }\sin \left( \left( \mu +\beta -\gamma \right)
\pi \right) ,$ \\ 
$a_{1}\rho ^{\alpha }\sin \left( \left( \mu -\alpha \right) \pi \right)
+a_{2}\rho ^{\beta }\sin \left( \left( \mu -\beta \right) \pi \right) +\rho
^{\gamma }\left\vert \sin \left( \left( \mu +\gamma \right) \pi \right)
\right\vert \left( a_{3}\frac{\sin \left( \left( \mu -\gamma \right) \pi
\right) }{\left\vert \sin \left( \left( \mu +\gamma \right) \pi \right)
\right\vert }-b\right) $ \\ 
$\quad +a_{1}b\rho ^{\alpha +\gamma }\sin \left( \left( \mu +\gamma -\alpha
\right) \pi \right) +a_{2}b\rho ^{\beta +\gamma }\sin \left( \left( \mu
+\gamma -\beta \right) \pi \right) +a_{3}b\rho ^{2\gamma }\sin \left( \mu
\pi \right) ,$%
\end{tabular}%
\right. 
\end{eqnarray*}%
by (\ref{ef1}), is positive due to the narrowed thermodynamical restrictions
(\ref{model 1 str}). In the case of Model II, function $f^{\left( \mathrm{I}%
\right) }$ obtained as 
\begin{eqnarray*}
f^{\left( \mathrm{I}\right) }\left( \rho ,\pi \right)  &=&\sin \left( \mu
\pi \right) +\rho ^{\alpha }\left( a_{1}\frac{\sin \left( \left( \mu -\alpha
\right) \pi \right) }{\left\vert \sin \left( \left( \mu +\alpha \right) \pi
\right) \right\vert }-b\right) +a_{2}\rho ^{\beta }\sin \left( \left( \mu
-\beta \right) \pi \right)  \\
&&+a_{1}\rho ^{2\alpha }\sin \left( \mu \pi \right) \left( b-\frac{a_{3}}{%
a_{1}}\frac{\left\vert \sin \left( \left( \mu -2\alpha \right) \pi \right)
\right\vert }{\sin \left( \mu \pi \right) }\right) +a_{2}b\rho ^{\alpha
+\beta }\sin \left( \left( \mu +\alpha -\beta \right) \pi \right)  \\
&&+a_{3}b\rho ^{3\alpha }\sin \left( \left( \mu -\alpha \right) \pi \right) ,
\end{eqnarray*}%
is positive due to the narrowed thermodynamical restrictions (\ref{model 2
str}). In the case of Model III, function $f^{\left( \mathrm{I}\right) }$
obtained as%
\begin{eqnarray*}
f^{\left( \mathrm{I}\right) }\left( \rho ,\pi \right)  &=&\sin \left( \mu
\pi \right) +\rho ^{\alpha }\left\vert \sin \left( \left( \mu +\alpha
\right) \pi \right) \right\vert \left( a_{1}\frac{\sin \left( \left( \mu
-\alpha \right) \pi \right) }{\left\vert \sin \left( \left( \mu +\alpha
\right) \pi \right) \right\vert }-b\right) +a_{2}\rho ^{\beta }\sin \left(
\left( \mu -\beta \right) \pi \right)  \\
&&+a_{1}b\rho ^{2\alpha }\sin \left( \mu \pi \right) +a_{2}\rho ^{\alpha
+\beta }\sin \left( \left( \mu +\alpha -\beta \right) \pi \right) \left( b-%
\frac{a_{3}}{a_{2}}\frac{\left\vert \sin \left( \left( \mu -\beta -\alpha
\right) \pi \right) \right\vert }{\sin \left( \left( \mu -\beta +\alpha
\right) \pi \right) }\right)  \\
&&+a_{3}b\rho ^{2\alpha +\beta }\sin \left( \left( \mu -\beta \right) \pi
\right) ,
\end{eqnarray*}%
by (\ref{ef1}), is positive due to the narrowed thermodynamical restrictions
(\ref{model 3 str}). In the case of Model IV, function $f^{\left( \mathrm{I}%
\right) }$ obtained as%
\begin{eqnarray*}
f^{\left( \mathrm{I}\right) }\left( \rho ,\pi \right)  &=&\sin \left( \mu
\pi \right) +a_{1}\rho ^{\alpha }\sin \left( \left( \mu -\alpha \right) \pi
\right) +\rho ^{\beta }\left\vert \sin \left( \left( \mu +\beta \right) \pi
\right) \right\vert \left( a_{2}\frac{\sin \left( \left( \mu -\beta \right)
\pi \right) }{\left\vert \sin \left( \left( \mu +\beta \right) \pi \right)
\right\vert }-b\right)  \\
&&+a_{1}\rho ^{\alpha +\beta }\sin \left( \left( \mu -\alpha +\beta \right)
\pi \right) \left( b-\frac{a_{3}}{a_{1}}\frac{\left\vert \sin \left( \left(
\mu -\alpha -\beta \right) \pi \right) \right\vert }{\sin \left( \left( \mu
-\alpha +\beta \right) \pi \right) }\right) +a_{2}b\rho ^{2\beta }\sin
\left( \mu \pi \right)  \\
&&+a_{3}b\rho ^{\alpha +2\beta }\sin \left( \left( \mu -\alpha \right) \pi
\right) ,
\end{eqnarray*}%
by (\ref{ef1}), is positive due to the narrowed thermodynamical restrictions
(\ref{model 4 str}). In the case of Model V, function $f^{\left( \mathrm{I}%
\right) }$ obtained as%
\begin{eqnarray*}
f^{\left( \mathrm{I}\right) }\left( \rho ,\pi \right)  &=&\sin \left( \mu
\pi \right) +a_{1}\rho ^{\alpha }\sin \left( \left( \mu -\alpha \right) \pi
\right) +\rho ^{\beta }\left\vert \sin \left( \left( \mu +\beta \right) \pi
\right) \right\vert \left( a_{2}\frac{\sin \left( \left( \mu -\beta \right)
\pi \right) }{\left\vert \sin \left( \left( \mu +\beta \right) \pi \right)
\right\vert }-b\right)  \\
&&+a_{1}b\rho ^{\alpha +\beta }\sin \left( \left( \mu +\beta -\alpha \right)
\pi \right) +a_{2}\rho ^{2\beta }\sin \left( \mu \pi \right) \left( b-\frac{%
a_{3}}{a_{2}}\frac{\left\vert \sin \left( \left( \mu -2\beta \right) \pi
\right) \right\vert }{\sin \left( \mu \pi \right) }\right)  \\
&&+a_{3}b\rho ^{3\beta }\sin \left( \left( \mu -\beta \right) \pi \right) ,
\end{eqnarray*}%
by (\ref{ef1}), is positive due to the narrowed thermodynamical restrictions
(\ref{model 5 str}). In the case of Model VI, function $f^{\left( \mathrm{II}%
\right) }$ obtained as%
\begin{eqnarray*}
f^{\left( \mathrm{II}\right) }\left( \rho ,\pi \right)  &=&\sin \left( \beta
\pi \right) +\rho ^{\alpha }\left\vert \sin \left( \left( \alpha +\beta
\right) \pi \right) \right\vert \left( a_{1}\frac{\sin \left( \left( \beta
-\alpha \right) \pi \right) }{\left\vert \sin \left( \left( \beta +\alpha
\right) \pi \right) \right\vert }-b\right)  \\
&&+a_{1}b\rho ^{2\alpha }\sin \left( \beta \pi \right) +a_{2}\left( b-\frac{%
a_{3}}{a_{2}}\right) \rho ^{\alpha +\beta }\sin \left( \alpha \pi \right) ,
\end{eqnarray*}%
by (\ref{ef2}), is positive due to the narrowed thermodynamical restrictions
(\ref{model 6 str}). In the case of Model VII, the positivity of function $%
f^{\left( \mathrm{II}\right) },$ obtained as%
\begin{equation*}
f^{\left( \mathrm{II}\right) }\left( \rho ,\pi \right) =a_{1}\rho ^{\alpha
}\sin \left( \left( \beta -\alpha \right) \pi \right) +a_{1}b\rho ^{\alpha
+\beta }\sin \left( \left( 2\beta -\alpha \right) \pi \right) +\left(
1+2b\rho ^{\beta }\cos \left( \beta \pi \right) +a_{2}\left( b-\frac{a_{3}}{%
a_{2}}\right) \rho ^{2\beta }\right) \sin \left( \beta \pi \right) ,
\end{equation*}%
by (\ref{ef2}), is achieved by requiring the positivity of the trinomial
quadratic in $\rho ^{\beta }$ that is guaranteed by the narrowed
thermodynamical restrictions (\ref{model 7 str}). In the case of Model VIII,
the positivity of function $f^{\left( \mathrm{II}\right) },$ obtained as%
\begin{equation*}
f^{\left( \mathrm{II}\right) }\left( \rho ,\pi \right) =\left( 1+2b\rho
^{\alpha }\cos \left( \alpha \pi \right) +\bar{a}_{1}\left( b-\frac{\bar{a}%
_{2}}{\bar{a}_{1}}\right) \rho ^{2\alpha }\right) \sin \left( \alpha \pi
\right) ,
\end{equation*}%
by (\ref{ef2}), is achieved by requiring the positivity of the trinomial
quadratic in $\rho ^{\beta }$ that is guaranteed by the narrowed
thermodynamical restrictions (\ref{model 8 str}). The asymptotics as $\rho
=R\rightarrow \infty $ of the real and imaginary parts of function $\psi ,$
given by (\ref{repsi-gama3}) and (\ref{impsi-gama3}), reduce to (\ref%
{repsi-gama2-pi}) and (\ref{impsi-gama2-pi}) for the Burgers models of the
first class, as well as to (\ref{repsi-gama2-pi-II-klas}) and (\ref%
{impsi-gama2-pi-II-klas}) for the Burgers models of the second class, while
for $\rho =r\rightarrow 0$ the real and imaginary parts of function $\psi $
become 
\begin{eqnarray}
\func{Re}\psi \left( \rho ,\pi \right)  &\sim &\rho ^{2-\xi }\cos \left( \xi
\varphi \right) +\vartheta \rightarrow \vartheta \;\;\text{and}
\label{repsi-gama3-ro-0} \\
\func{Im}\psi \left( \rho ,\pi \right)  &\sim &-\rho ^{2-\xi }\sin \left(
\xi \varphi \right) \rightarrow 0^{-}.  \label{impsi-gama3-ro-0}
\end{eqnarray}%
Hence, as $s$ changes along contour $\gamma _{3}$, by (\ref{impsi-gama3})
and (\ref{impsi-gama3-ro-0}), imaginary part of function $\psi $ is negative
and as $\rho \rightarrow 0,$ by (\ref{repsi-gama3-ro-0}), the real part of
function $\psi $ tends to $\vartheta .$

The modulus of complex numbers belonging to the contour $\gamma _{4}$ has a
fixed but small value $\rho =r,$ while their argument changes in the
interval $\varphi \in \left[ \frac{\pi }{2},\pi \right] ,$ so that (\ref%
{repsi}) and (\ref{impsi}) become%
\begin{eqnarray*}
\func{Re}\psi \left( r,\varphi \right) &\sim &r^{2-\xi }\cos \left( \left(
2-\xi \right) \varphi \right) +\vartheta \sim \vartheta , \\
\func{Im}\psi \left( r,\varphi \right) &\sim &r^{2-\xi }\sin \left( \left(
2-\xi \right) \varphi \right) ,
\end{eqnarray*}%
implying that regardless of the sign of $\func{Im}\psi $, the function $\psi 
$ remains in the neighborhood of a finite positive number $\vartheta $.

Summing up, the change of argument of function $\psi $ is $\Delta \arg \psi
\left( s\right) =2\pi $ as $s$ changes along the contour $\gamma ,$ implying
that function $\psi $ has one zero in the upper left complex quarter-plane
for each $k\in 
\mathbb{N}
.$

\section{Conclusion}

The fractional Burgers wave equation, written as the system of equations
consisting of the equation of motion and strain (\ref{eq-motion}), that are
coupled either with Burgers models of the first class (\ref{UCE-1-5}) or
with models of the second class (\ref{UCE-6-8}), is used to model the
dynamic response of the initially undisturbed one-dimensional viscoelastic
rod of finite length having one end fixed and the other subject to
prescribed either displacement or stress, according to boundary conditions (%
\ref{bc}). Laplace transform method is used in order to express the
displacement and stress of an arbitrary rod's point in terms of boundary
condition convoluted with the solution kernel.

The short-time asymptotics of solution kernels implied that their time
profiles continuously increase from zero as time increases, with the
significant rise depending on the point's position in the case of Burgers
models of the first class, see short-time asymptotics for solution kernels $%
P,$ $R,$ $Q,$ and $S,$ given by (\ref{pe-reg-te(zi)-nula}), (\ref%
{er-eps-te-tezi-nula}), (\ref{ku-short-time-asympt-m-V}), and (\ref%
{es-te(zi)-nula}), respectively, implying the infinite wave propagation
speed. On the other hand, solution kernels $P,$ $R,$ and $S,$ corresponding
to the Burgers models of the second class, have to be regularized, so that
the short-time asymptotics of $P_{\mathrm{reg}}$ and $S_{\mathrm{reg}}$ is
the Heaviside function of the argument $t-\sqrt{\frac{a_{3}}{b}}\left(
1-x\right) ,$ see (\ref{pe-reg-te(zi)-nula}) and (\ref{es-reg-te(zi)-nula}),
implying the finite wave propagation speed $c=\sqrt{\frac{b}{a_{3}}}$, due
to the sudden jump in the value of $P_{\mathrm{reg}}$ and $S_{\mathrm{reg}}$
at $t=\frac{1-x}{c}.$ Solution kernel $R$ is regularized differently,
implying that for small time its regularization behaves as the time
derivative of the Dirac delta regularization, calculated at $t-\sqrt{\frac{%
a_{3}}{b}}\left( 1-x\right) ,$ see (\ref{er-eps-te-tezi-nula}). It is
noteworthy that the regularization of solution kernel $Q$ is not necessary,
since its short-time asymptotics is the Heaviside function of the argument $%
t-\sqrt{\frac{a_{3}}{b}}\left( 1-x\right) ,$ see (\ref{ku-te-tezi-nula}).

All solution kernels consist of two terms: the integral one is at most
non-monotonic in both space and time and the one expressed either through
the sine or cosine Fourier series represents a superposition of standing
waves, each of them oscillating with amplitude decreasing in time, having
the damping and angular frequency determined by the pole of the solution
kernel image. Moreover, the form of solution kernel also depends on
occurrence of branch points of solution kernel image.

Time profiles of the displacement and stress step responses are of quite
classical shapes corresponding to the damped oscillatory behavior in the
case of Burgers models of the first class, while in the case of the second
class models, the time profiles are peculiarly shaped resembling to the
sequence of excitation and relaxation processes. Nevertheless, the
large-time asymptotics of the step response for prescribed displacement of
rod's free end yielded%
\begin{equation*}
\varepsilon _{\Upsilon }\left( x,t\right) \sim 1\;\;\text{and}\;\;\sigma
_{\Upsilon }\left( x,t\right) \sim \frac{t^{-\xi }}{\Gamma \left( 1-\xi
\right) },\;\;\text{as}\;\;t\rightarrow \infty ,
\end{equation*}%
while in the case of prescribed stress acting on rod's free end, one has%
\begin{equation*}
\varepsilon _{\Sigma }\left( x,t\right) \sim \frac{t^{\xi }}{\Gamma \left(
1+\xi \right) },\;\;\text{and}\;\;\sigma _{\Sigma }\left( x,t\right) \sim
1\;\;\text{as}\;\;t\rightarrow \infty ,
\end{equation*}%
see (\ref{btj-0})$_{2}$, (\ref{u-ipsilon-besk}), (\ref{sigma-ipsilon-besk}),
(\ref{u-sigma-asimpt}), and (\ref{sigma-sigma-besk}), respectively, that is
in a perfect accordance with the behavior of relaxation modulus and creep
compliance, studied in \cite{OZ-2} for the thermodynamically consistent
fractional Burgers models.

\section*{Acknowledgment}

This work is supported by the Serbian Ministry of Science, Education and
Technological Development under grant 451-03-9/2021-14/200125 (DZ).

\appendix

\section{Fractional Burgers models \label{FBMS}}

Thermodynamically consistent fractional Burgers models are listed below,
along with corresponding thermodynamical constraints, as well as with the
constraints on monotonicity of relaxation modulus and creep compliance,
narrowing down the thermodynamical requirements and guaranteeing that
relaxation modulus is completely monotonic, while creep compliance is
Bernstein function.

\noindent \textbf{Model I}: 
\begin{gather}
\left( 1+a_{1}\,{}_{0}\mathrm{D}_{t}^{\alpha }+a_{2}\,{}_{0}\mathrm{D}%
_{t}^{\beta }+a_{3}\,{}_{0}\mathrm{D}_{t}^{\gamma }\right) \sigma \left(
t\right) =\left( b_{1}\,{}_{0}\mathrm{D}_{t}^{\mu }+b_{2}\,{}_{0}\mathrm{D}%
_{t}^{\mu +\eta }\right) \varepsilon \left( t\right) ,  \label{Model 1} \\
0\leq \alpha \leq \beta \leq \gamma \leq \mu \leq 1,\;\;1\leq \mu +\eta \leq
1+\alpha ,\;\;\frac{b_{2}}{b_{1}}\leq a_{i}\frac{\cos \frac{\left( \mu -\eta
\right) \pi }{2}}{\left\vert \cos \frac{\left( \mu +\eta \right) \pi }{2}%
\right\vert },  \notag \\
\frac{b_{2}}{b_{1}}\leq a_{i}\frac{\sin \frac{\left( \mu -\eta \right) \pi }{%
2}}{\sin \frac{\left( \mu +\eta \right) \pi }{2}}\frac{\cos \frac{\left( \mu
-\eta \right) \pi }{2}}{\left\vert \cos \frac{\left( \mu +\eta \right) \pi }{%
2}\right\vert },  \label{model 1 str}
\end{gather}%
with $\left( \eta ,i\right) \in \left\{ \left( \alpha ,1\right) ,\left(
\beta ,2\right) ,\left( \gamma ,3\right) \right\} ;$

\noindent \textbf{Model II}:%
\begin{gather}
\left( 1+a_{1}\,{}_{0}\mathrm{D}_{t}^{\alpha }+a_{2}\,{}_{0}\mathrm{D}%
_{t}^{\beta }+a_{3}\,{}_{0}\mathrm{D}_{t}^{2\alpha }\right) \sigma \left(
t\right) =\left( b_{1}\,{}_{0}\mathrm{D}_{t}^{\mu }+b_{2}\,{}_{0}\mathrm{D}%
_{t}^{\mu +\alpha }\right) \varepsilon \left( t\right) ,  \label{Model 2} \\
\frac{1}{2}\leq \alpha \leq \beta \leq \mu \leq 1,\;\;\frac{a_{3}}{a_{1}}%
\frac{\left\vert \sin \frac{\left( \mu -2\alpha \right) \pi }{2}\right\vert 
}{\sin \frac{\mu \pi }{2}}\leq \frac{b_{2}}{b_{1}}\leq a_{1}\frac{\cos \frac{%
\left( \mu -\alpha \right) \pi }{2}}{\left\vert \cos \frac{\left( \mu
+\alpha \right) \pi }{2}\right\vert },  \notag \\
\frac{a_{3}}{a_{1}}\frac{\left\vert \sin \frac{\left( \mu -2\alpha \right)
\pi }{2}\right\vert }{\sin \frac{\mu \pi }{2}}\frac{\cos \frac{\left( \mu
-2\alpha \right) \pi }{2}}{\cos \frac{\mu \pi }{2}}\leq \frac{b_{2}}{b_{1}}%
\leq a_{1}\frac{\sin \frac{\left( \mu -\alpha \right) \pi }{2}}{\sin \frac{%
\left( \mu +\alpha \right) \pi }{2}}\frac{\cos \frac{\left( \mu -\alpha
\right) \pi }{2}}{\left\vert \cos \frac{\left( \mu +\alpha \right) \pi }{2}%
\right\vert };  \label{model 2 str}
\end{gather}

\noindent \textbf{Model III}:%
\begin{gather}
\left( 1+a_{1}\,{}_{0}\mathrm{D}_{t}^{\alpha }+a_{2}\,{}_{0}\mathrm{D}%
_{t}^{\beta }+a_{3}\,{}_{0}\mathrm{D}_{t}^{\alpha +\beta }\right) \sigma
\left( t\right) =\left( b_{1}\,{}_{0}\mathrm{D}_{t}^{\mu }+b_{2}\,{}_{0}%
\mathrm{D}_{t}^{\mu +\alpha }\right) \varepsilon \left( t\right) ,
\label{Model 3} \\
0\leq \alpha \leq \beta \leq \mu \leq 1,\;\;\alpha +\beta \geq 1,\;\;\frac{%
a_{3}}{a_{2}}\frac{\left\vert \sin \frac{\left( \mu -\beta -\alpha \right)
\pi }{2}\right\vert }{\sin \frac{\left( \mu -\beta +\alpha \right) \pi }{2}}%
\leq \frac{b_{2}}{b_{1}}\leq a_{1}\frac{\cos \frac{\left( \mu -\alpha
\right) \pi }{2}}{\left\vert \cos \frac{\left( \mu +\alpha \right) \pi }{2}%
\right\vert },  \notag \\
\frac{a_{3}}{a_{2}}\frac{\left\vert \sin \frac{\left( \mu -\beta -\alpha
\right) \pi }{2}\right\vert }{\sin \frac{\left( \mu -\beta +\alpha \right)
\pi }{2}}\frac{\cos \frac{\left( \mu -\beta -\alpha \right) \pi }{2}}{\cos 
\frac{\left( \mu -\beta +\alpha \right) \pi }{2}}\leq \frac{b_{2}}{b_{1}}%
\leq a_{1}\frac{\sin \frac{\left( \mu -\alpha \right) \pi }{2}}{\sin \frac{%
\left( \mu +\alpha \right) \pi }{2}}\frac{\cos \frac{\left( \mu -\alpha
\right) \pi }{2}}{\left\vert \cos \frac{\left( \mu +\alpha \right) \pi }{2}%
\right\vert };  \label{model 3 str}
\end{gather}

\noindent \textbf{Model IV}:%
\begin{gather}
\left( 1+a_{1}\,{}_{0}\mathrm{D}_{t}^{\alpha }+a_{2}\,{}_{0}\mathrm{D}%
_{t}^{\beta }+a_{3}\,{}_{0}\mathrm{D}_{t}^{\alpha +\beta }\right) \sigma
\left( t\right) =\left( b_{1}\,{}_{0}\mathrm{D}_{t}^{\mu }+b_{2}\,{}_{0}%
\mathrm{D}_{t}^{\mu +\beta }\right) \varepsilon \left( t\right) ,
\label{Model 4} \\
0\leq \alpha \leq \beta \leq \mu \leq 1,\;\;1-\alpha \leq \beta \leq
1-\left( \mu -\alpha \right) ,\;\;\frac{a_{3}}{a_{1}}\frac{\left\vert \sin 
\frac{\left( \mu -\alpha -\beta \right) \pi }{2}\right\vert }{\sin \frac{%
\left( \mu -\alpha +\beta \right) \pi }{2}}\leq \frac{b_{2}}{b_{1}}\leq a_{2}%
\frac{\cos \frac{\left( \mu -\beta \right) \pi }{2}}{\left\vert \cos \frac{%
\left( \mu +\beta \right) \pi }{2}\right\vert },  \notag \\
\frac{a_{3}}{a_{1}}\frac{\left\vert \sin \frac{\left( \mu -\alpha -\beta
\right) \pi }{2}\right\vert }{\sin \frac{\left( \mu -\alpha +\beta \right)
\pi }{2}}\frac{\cos \frac{\left( \mu -\alpha -\beta \right) \pi }{2}}{\cos 
\frac{\left( \mu -\alpha +\beta \right) \pi }{2}}\leq \frac{b_{2}}{b_{1}}%
\leq a_{2}\frac{\sin \frac{\left( \mu -\beta \right) \pi }{2}}{\sin \frac{%
\left( \mu +\beta \right) \pi }{2}}\frac{\cos \frac{\left( \mu -\beta
\right) \pi }{2}}{\left\vert \cos \frac{\left( \mu +\beta \right) \pi }{2}%
\right\vert };  \label{model 4 str}
\end{gather}

\noindent \textbf{Model V}:%
\begin{gather}
\left( 1+a_{1}\,{}_{0}\mathrm{D}_{t}^{\alpha }+a_{2}\,{}_{0}\mathrm{D}%
_{t}^{\beta }+a_{3}\,{}_{0}\mathrm{D}_{t}^{2\beta }\right) \sigma \left(
t\right) =\left( b_{1}\,{}_{0}\mathrm{D}_{t}^{\mu }+b_{2}\,{}_{0}\mathrm{D}%
_{t}^{\mu +\beta }\right) \varepsilon \left( t\right) ,  \label{Model 5} \\
0\leq \alpha \leq \beta \leq \mu \leq 1,\;\;\frac{1}{2}\leq \beta \leq
1-\left( \mu -\alpha \right) ,\;\;\frac{a_{3}}{a_{2}}\frac{\left\vert \sin 
\frac{\left( \mu -2\beta \right) \pi }{2}\right\vert }{\sin \frac{\mu \pi }{2%
}}\leq \frac{b_{2}}{b_{1}}\leq a_{2}\frac{\cos \frac{\left( \mu -\beta
\right) \pi }{2}}{\left\vert \cos \frac{\left( \mu +\beta \right) \pi }{2}%
\right\vert },  \notag \\
\frac{a_{3}}{a_{2}}\frac{\left\vert \sin \frac{\left( \mu -2\beta \right)
\pi }{2}\right\vert }{\sin \frac{\mu \pi }{2}}\frac{\cos \frac{\left( \mu
-2\beta \right) \pi }{2}}{\cos \frac{\mu \pi }{2}}\leq \frac{b_{2}}{b_{1}}%
\leq a_{2}\frac{\sin \frac{\left( \mu -\beta \right) \pi }{2}}{\sin \frac{%
\left( \mu +\beta \right) \pi }{2}}\frac{\cos \frac{\left( \mu -\beta
\right) \pi }{2}}{\left\vert \cos \frac{\left( \mu +\beta \right) \pi }{2}%
\right\vert };  \label{model 5 str}
\end{gather}

\noindent \textbf{Model VI}:%
\begin{gather}
\left( 1+a_{1}\,{}_{0}\mathrm{D}_{t}^{\alpha }+a_{2}\,{}_{0}\mathrm{D}%
_{t}^{\beta }+a_{3}\,{}_{0}\mathrm{D}_{t}^{\alpha +\beta }\right) \sigma
\left( t\right) =\left( b_{1}\,{}_{0}\mathrm{D}_{t}^{\beta }+b_{2}\,{}_{0}%
\mathrm{D}_{t}^{\alpha +\beta }\right) \varepsilon \left( t\right) ,
\label{Model 6} \\
0\leq \alpha \leq \beta \leq 1,\;\;\alpha +\beta \geq 1,\;\;\frac{a_{3}}{%
a_{2}}\leq \frac{b_{2}}{b_{1}}\leq a_{1}\frac{\cos \frac{\left( \beta
-\alpha \right) \pi }{2}}{\left\vert \cos \frac{\left( \beta +\alpha \right)
\pi }{2}\right\vert },  \notag \\
\frac{a_{3}}{a_{2}}\leq \frac{b_{2}}{b_{1}}\leq a_{1}\frac{\sin \frac{\left(
\beta -\alpha \right) \pi }{2}}{\sin \frac{\left( \beta +\alpha \right) \pi 
}{2}}\frac{\cos \frac{\left( \beta -\alpha \right) \pi }{2}}{\left\vert \cos 
\frac{\left( \beta +\alpha \right) \pi }{2}\right\vert }\leq a_{1}\frac{\cos 
\frac{\left( \beta -\alpha \right) \pi }{2}}{\left\vert \cos \frac{\left(
\beta +\alpha \right) \pi }{2}\right\vert };  \label{model 6 str}
\end{gather}

\noindent \textbf{Model VII}:%
\begin{gather}
\left( 1+a_{1}\,{}_{0}\mathrm{D}_{t}^{\alpha }+a_{2}\,{}_{0}\mathrm{D}%
_{t}^{\beta }+a_{3}\,{}_{0}\mathrm{D}_{t}^{2\beta }\right) \sigma \left(
t\right) =\left( b_{1}\,{}_{0}\mathrm{D}_{t}^{\beta }+b_{2}\,{}_{0}\mathrm{D}%
_{t}^{2\beta }\right) \varepsilon \left( t\right) ,  \label{Model 7} \\
0\leq \alpha \leq \beta \leq 1,\;\;\frac{1}{2}\leq \beta \leq \frac{1+\alpha 
}{2},\;\;\frac{a_{3}}{a_{2}}\leq \frac{b_{2}}{b_{1}}\leq a_{2}\frac{1}{%
\left\vert \cos \left( \beta \pi \right) \right\vert },  \notag \\
\frac{a_{3}}{a_{2}}\leq \frac{a_{2}}{2\cos ^{2}\left( \beta \pi \right) }%
\left( 1-\sqrt{1-\frac{4a_{3}\cos ^{2}\left( \beta \pi \right) }{a_{2}^{2}}}%
\right) \leq \frac{b_{2}}{b_{1}}\leq \frac{a_{2}}{\left\vert \cos \left(
\beta \pi \right) \right\vert };  \label{model 7 str}
\end{gather}

\noindent \textbf{Model VIII}:%
\begin{gather}
\left( 1+\bar{a}_{1}\,{}_{0}\mathrm{D}_{t}^{\alpha }+\bar{a}_{2}\,{}_{0}%
\mathrm{D}_{t}^{2\alpha }\right) \sigma \left( t\right) =\left( b_{1}\,{}_{0}%
\mathrm{D}_{t}^{\alpha }+b_{2}\,{}_{0}\mathrm{D}_{t}^{2\alpha }\right)
\varepsilon \left( t\right) ,  \label{Model 8} \\
\frac{1}{2}\leq \alpha \leq 1,\;\;\frac{\bar{a}_{2}}{\bar{a}_{1}}\leq \frac{%
b_{2}}{b_{1}}\leq \bar{a}_{1}\frac{1}{\left\vert \cos \left( \alpha \pi
\right) \right\vert },  \notag \\
\frac{\bar{a}_{2}}{\bar{a}_{1}}\leq \frac{\bar{a}_{1}}{2\cos ^{2}\left(
\alpha \pi \right) }\left( 1-\sqrt{1-\frac{4\bar{a}_{2}\cos ^{2}\left(
\alpha \pi \right) }{\bar{a}_{1}^{2}}}\right) \leq \frac{b_{2}}{b_{1}}\leq 
\frac{\bar{a}_{1}}{\left\vert \cos \left( \alpha \pi \right) \right\vert }.
\label{model 8 str}
\end{gather}


\end{document}